\documentclass[
  journal=largetwo,
  manuscript=article-type,
  year=2024,
  volume=?,
]{cup-journal}

\usepackage{amsmath}
\usepackage{amssymb}
\usepackage[stretch=10,shrink=10]{microtype}
\usepackage{booktabs}
\usepackage[normalem]{ulem}

\newcommand{\Msun}{\,{\rm M}_{\odot}}

\newcommand{\iso}[2]{\hbox{${}^{#1}{\rm #2}$}}

\title{Using Binary Population Synthesis to Examine the Impact of Binary Evolution on the C, N, O, and $S$-Process Yields of Solar-Metallicity Low- and Intermediate-Mass Stars}

\author{Zara Osborn}
\affiliation{School of Physics \& Astronomy, Monash University, Clayton 3800, VIC, Australia}
\alsoaffiliation{Centre of Excellence for Astrophysics in Three Dimensions (ASTRO-3D), Melbourne, VIC, Australia}
\email[Z. Osborn]{zara.osborn@monash.edu}

\author{Amanda I. Karakas}
\affiliation{School of Physics \& Astronomy, Monash University, Clayton 3800, VIC, Australia}
\alsoaffiliation{Centre of Excellence for Astrophysics in Three Dimensions (ASTRO-3D), Melbourne, VIC, Australia}

\author{Alex J. Kemp}
\affiliation{Institute of Astronomy, KU Leuven, Celestijnenlaan 200D, 3001 Leuven, Belgium}

\author{Robert Izzard}
\affiliation{Astrophysics Research Group, University of Surrey, Guildford, Surrey GU2 7XH, UK}

\author{Devika Kamath}
\affiliation{School of Mathematical and Physical Sciences, Macquarie University, Balaclava Road, Sydney, NSW 2109, Australia}
\alsoaffiliation{Centre of Excellence for Astrophysics in Three Dimensions (ASTRO-3D), Melbourne, VIC, Australia}

\author{Maria Lugaro}
\affiliation{Konkoly Observatory, HUN-REN Research Centre for Astronomy and Earth Sciences, Konkoly Thege Mikl\'os út 15-17., H-1121, Hungary}
\alsoaffiliation{CSFK, MTA Centre of Excellence, Budapest, Konkoly Thege Mikl\'os út 15-17., H-1121, Hungary}
\alsoaffiliation{ELTE E\"{o}tv\"{o}s Lor\'and University, Institute of Physics andAstronomy, Budapest 1117, P\'azm\'any P\'eter s\'et\'any 1/A, Hungary}
\alsoaffiliation{School of Physics \& Astronomy, Monash University, Clayton 3800, VIC, Australia}

\defcitealias{Cristallo2009_2}{Cristallo, Piersanti et al. 2019}
\defcitealias{Lugaro2012}{Lugaro et al. 2012}


\begin{document}
\emergencystretch 10em

\begin{abstract}
Asymptotic giant branch (AGB) stars play a significant role in our understanding of the origin of the elements. They contribute to the abundances of C, N, and approximately 50\% of the abundances of the elements heavier than iron. An aspect often neglected in studies of AGB stars is the impact of a stellar companion on AGB stellar evolution and nucleosynthesis. In this study, we update the stellar abundances of AGB stars in the binary population synthesis code \textsc{binary\_c} and calibrate our treatment of the third dredge-up using observations of Galactic carbon stars. We model stellar populations of low- to intermediate-mass stars at solar-metallicity and examine the stellar wind contributions to C, N, O, Sr, Ba, and Pb yields at binary fractions between 0 and 1. For a stellar population with a binary fraction of 0.7, we find $\sim 20-25\%$ less C and $s$-process elements ejected than from a population composed of only single stars, and we find little change in the N and O yields. We also compare our models with observed abundances from Ba stars and find our models can reproduce most Ba star abundances, but our population estimates a higher frequency of Ba stars with a surface [Ce/Y] > $+0.2\,$dex. Our models also predict the rare existence of Ba stars with masses $> 10\Msun$.
\end{abstract}

\section{Introduction}
Most elements, except for hydrogen, helium, and trace amounts of lithium, beryllium, and boron, are forged by stars, including asymptotic giant branch (AGB) stars. AGB stars are giant stars of low- to intermediate-mass ($\sim 1-8\Msun$) that have completed core He burning. The unique nucleosynthesis that occurs during the AGB is thought to be responsible for producing significant fractions of carbon, nitrogen, fluorine, and about half of the elements heavier than iron \citep[for example, see][]{Renda2004, Bensby2006, Vangioni2018, Prantzos2020, Kobayashi2020}. 

The total mass of a given isotope or element ejected by a star over its lifetime is the stellar yield \citep[e.g.][]{Karakas2010}. The stellar yield of single low- and intermediate-mass stars originate from the ejection of the stellar envelopes via stellar winds, primarily during the AGB phase. Over the lifetime of a star, the stellar surface becomes enriched with the products of nuclear-burning forged deep within the stellar interior. These nuclear burning products are mixed to the stellar surface through convective processes known as dredge-ups. The first and second dredge-ups occur during the first giant branch (GB) and early-AGB (E-AGB), respectively, and mix products of partial H-burning to the stellar surface. The third dredge-up occurs repeatedly on the thermally pulsing AGB (TP-AGB). 

TP-AGB stars are sites of complex stellar nucleosynthesis driven by periodic unstable shell He burning (thermal pulses). TP-AGB stars can synthesize carbon via partial He burning and elements heavier than iron through the slow neutron capture process ($s$-process). These heavy nuclides are transported to the stellar surface through recurrent third dredge-up events \citep{Gallino1998, Busso2001}. Furthermore, in TP-AGB stars with mass $\gtrsim 5\Msun$, temperatures at the bottom of the convective envelope are sufficient to sustain H burning ($\sim 10^8 \,$K), in a process known as hot-bottom burning \citep{Boothroyd1995}. The stellar evolution and yield of AGB stars have been researched extensively for single stars \citep{Herwig2005, Cristallo2011, Karakas2014, Ventura2020, Cinquegrana2022_2}.  However, all these are single-star models, whereas observations show that at least 40-75\% of low- and intermediate-mass stars are in a binary \citep{Raghavan2010, Moe2017}. 

The presence of a stellar companion introduces processes that can potentially alter the evolutions of the stars. These processes include, for example, mass transfer via Roche-robe overflow \citep{Eggleton1983}, stellar wind accretion \citep{Bondi1944, Abate2013}, and mergers (for reviews on binary evolution, see \citealp{Iben1991} and \citealp{DeMarco2017}). Stellar companions have been shown to interact with AGB stars through the shaping of their stellar winds and planetary nebulae \citep{Jones2012, DeMarco2022}. Objects such as post-RGB stars \citep{Kamath2015}, and barium stars \citep{McClure1983, Jorissen2019}, can only be formed by interacting binaries. Although the existence of such objects implies that binary evolution has the potential to alter the stellar evolution and yield of low- and intermediate-mass stars, its impact on a stellar population is poorly understood. Most research on binary-star stellar evolution and nucleosynthesis has focused on massive stars \citep[for example][]{Sana2012, deMink2013, DeMarco2017, Brinkman2019, Brinkman2023, Farmer2023}.

In this study, we examine how binary evolution influences low- and intermediate-mass stars and their production of C, N, O, and $s$-process elements at solar-metallicity (defined as $Z=0.015$ from \citealp{Lodders2003}). We use the binary population synthesis code \textsc{binary\_c} \citep{Izzard2004, Izzard2006, Izzard2009, Izzard2018, Izzard2023} to produce stellar grids of low- and intermediate-mass binary systems. We use \textsc{binary\_c} for its nucleosynthesis capabilities and relatively detailed AGB synthetic models compared to other population synthesis codes. We further test our models by comparing our resulting $s$-process surface abundances to those observed from Ba stars, which have surface $s$-process enhancement due to accreting material from an AGB companion \citep{Bidelman1951, McClure1983}. 

This paper is structured as follows. In Section \ref{sec:popSynth}, we describe our synthetic models, the modifications we have made to \textsc{binary\_c}, and our calibration of third dredge-up using Galactic carbon stars. In Section \ref{sec:Results}, we show results for the C, N, O, Sr, Ba, and Pb stellar population yield for populations with various binary fractions, and we discuss yield variations arising due to binary evolution. In Section \ref{sec:BaStars}, we compare our models to observations of Galactic Ba stars. In Section \ref{sec:Discussion}, we discuss the results and their uncertainties; we summarise and state our conclusions in Section \ref{sec:Conclusion}. 

\section{Binary Population Synthesis Models}
\label{sec:popSynth}

We use a modified version of \textsc{binary\_c} version 2.2.4 updated from \citet[][hereafter Paper I]{Osborn2023} interfaced with \textsc{binary\_c-python} \citep{Hendriks2023} version 1.0.0, to model our low- and intermediate-mass stellar populations. 

In Paper I, we outline the details of the previous modifications to \textsc{binary\_c}. In summary, Paper I expands the fits to the CO-core mass, third dredge-up efficiency, hot-bottom burning temperatures, and TP-AGB luminosities in \textsc{binary\_c} using models from \citet{Karakas2016} and \citet{Doherty2015}. These updates expanded the fitted AGB mass range in \textsc{binary\_c} from $1-6.5\Msun$ to $1-8\Msun$, preventing non-physical stellar evolution in the $6.5-8\Msun$ stars. In this paper, we focus on the calibration of the third dredge-up using observations of Galactic carbon stars from \citet{Abia2022} and updating to the He intershell abundance fits in \textsc{binary\_c} to the models from \citet{Karakas2016}, hereafter K16, which includes $s$-process elements.

Unless otherwise specified, our results use a synthetic grid of 1000 single-star models and 640 000 binary star models sampled according to Table \ref{tab:parameters}. To calculate the total stellar yield $y_{{\rm tot}, ij}$ of element $i$ from each model $j$, we use 

\begin{equation}
     \label{eq:Yield}
    y_{{\rm tot}, ij} =  \int ^{\tau_{\rm L}} _0 X(i,t) \frac{{\rm d}M}{ {\rm d}t}{\rm d}t,
\end{equation}

\noindent where $\tau_{\rm L}$ is the lifetime of the stellar model, $\frac{{\rm d}M}{ {\rm d}t}$ is the mass-loss rate due to stellar winds or ejection during a common envelope event, and $X(i,t)$ is the surface mass fractions of element $i$ at time $t$ \citep{Karakas2010}. We also define the net yield $y_{{\rm net}, ij}$, which describes the net production or destruction of any given element $i$, with

\begin{equation}
    \label{eq:netYield}
    y_{{\rm net}, ij} =  \int ^{\tau_{\rm L}} _0 \left[ X(i,t) - X_{\rm 0} (i)\right] \frac{{\rm d}M}{ {\rm d}t}{\rm d}t,
\end{equation}

\noindent $X_{\rm 0} (i)$ is the initial surface mass fraction of element $i$ \citep{Karakas2010}. See Paper I for more details on how we calculate the stellar yield and how we treat binary interaction.

To form a physical stellar population from our stellar grid independently of our grid-sampling distributions, we introduce a weighting factor $w_{j}$ for each stellar system $j$, based on the methodology described in \citet{Broekgaarden2019}, where

\begin{equation}
    \label{eq:Weighting}
    w_{j} = f_{\rm b} \frac{w_{\rm m}}{n} \frac{\pi(\mathbf{x}_{j})}{\xi(\mathbf{x}_{j})},
\end{equation}

\noindent where $w_{j}$ is in units per $\Msun$ of star-forming material, $f_{\rm b}$ is the binary fraction of the stellar population (we use $1-f_{\rm b}$ when weighting single-stars), $w_{\rm m}$ is a mass normalisation term describing the number of stellar systems forming per $\Msun$ of star-forming material, $n$ is the number of models sampled in the grid, $\pi(\mathbf{x}_{j})$ describes the theoretical probability distribution of initial conditions of the observed stellar-population, and $\xi(\mathbf{x}_{j})$ is the probability distribution of the initial conditions of our grid sampled in \textsc{binary\_c}. See Table \ref{tab:parameters} for the probability distribution functions applied for the theoretical and sampled $M_{\rm 1,0}$, $M_{\rm 2,0}$, and $p_{\rm 0}$. Also, see Sections 2.2 and 2.3 from Paper I for more details on calculating $w_{j}$.

We then calculate the weighted stellar yield of the stellar population,

\begin{equation}
    \label{Eq:WYield}
    y_{\rm pop, i} = \sum^{n}_{j=0} w_j \times y_{{\rm tot},ij}.
\end{equation}

Other than stellar winds, novae and supernovae can also contribute to a star's stellar yield and are significant contributors to the Universe's oxygen and iron, among many other elements (for example, see \citealp{Gehrz1998}, \citealp{Kemp2024} for novae; \citealp{Matteucci1986}, \citealp{Timmes1995}, \citealp{Limongi2018}, \citealp{Dubay2024} for supernovae). We do not include the contribution of novae or supernovae to the stellar yield in our results, as they are beyond the scope of this study. The results presented in \citet{Izzard2003} and \citet{DeDonder2004} compare the stellar yields from single and binary systems, with and without the contribution from supernovae. See also \citet{Zapartas2017}.

Throughout this paper, we refer to a \emph{standard} and \emph{modified} version of \textsc{binary\_c}. We define the standard version of \textsc{binary\_c} to be version 2.2.4 with none of the modifications outlined in Paper I or in this work. We define the modified version of \textsc{binary\_c} as the version including all of the modifications, from both Paper I and this work\footnote{The modified version of \textsc{binary\_c} is available at: \url{https://gitlab.com/binary_c/binary_c/-/tree/V2.2.4_Osb24?ref_type=heads}.}. The input physics and model parameters used in \textsc{binary\_c} are shown in Table \ref{tab:parameters}, where we also highlight the differences between the standard and modified versions of \textsc{binary\_c}.

\begin{table*}[hbt!]
\caption{The selected input physics and parameters for our \textsc{binary\_c} grids, as described in Paper I. We highlight differences between the standard and modified versions of \textsc{binary\_c}, marked here as [standard] and [modified] respectively. See Section \ref{sec:Intershell} for the details of the He intershell abundances, and Sections \ref{sec:3DUP_Method} and \ref{sec:3DUP_Results} for details on the third dredge-up parameters $\lambda_{\rm min}$ and $\Delta M_{\rm c,min}$.}

\label{tab:parameters}
\begin{tabular}{ll}
\toprule
\headrow Parameter & Setting \\
\midrule
Primary star initial mass, $M_{\rm 1,0}$, range & $0.80-8.50\Msun$ \\
$M_{\rm 1,0}$ grid-sampling probability distribution & Uniform in $M_{\rm 1,0}$ \\
$M_{\rm 1,0}$ theoretical probability distribution & \citet{Kroupa2001} normalised between $0.01-150\Msun$\\
Secondary star initial mass range & $0.1\Msun-M_{\rm 1,0}$ \\
Secondary star theoretical and grid-sampling probability distribution & Uniform in $M_{\rm 2,0}/M_{\rm 1,0}$ \\
Initial orbital period & $1.0-10^6 \, {\rm days}$ \\
Orbital period theoretical and grid-sampling probability distribution & Log-uniform in $p_{\rm 0}$\\
Metallicity, Z & 0.015 \\
Simulation time & $15 \, {\rm Gyr}$ \\
Initial eccentricity & 0.0 \\
Initial stellar rotation & 0.0 \\
Upper mass limit for AGB algorithms & $8.36 \Msun$ [modified] or $8.00\Msun$ [standard] \\
Initial chemical abundance & \citet{Lodders2003} \\ 
TP-AGB stellar wind & \citet{Vassiliadis1993} \\
Minimum third dredge-up efficiency, $\lambda_{\rm min}$ & 0.45 [modified] or 0.00 [standard] \\
Constant decrease in minimum core mass for third dredge-up, $\Delta M_{\rm c, min}$ & $-0.13\Msun$ [modified] or 0.00 [standard] \\
He intershell abundances (elements lighter then Fe) & \citet{Karakas2016} [modified] or \citet{Karakas2002} [standard] \\
He intershell abundances (elements heavier and including Fe) & \citet{Karakas2016} [modified] or \citet{Gallino1998} [standard]

\end{tabular}
\end{table*}


\subsection{Calibrating the Third Dredge-Up using Carbon-Stars}
\label{sec:3DUP_Method}

AGB stars with a surface C/O ratio by number of $\geq 1$ are called carbon stars \citep{Wallerstein1998}. The formation of carbon stars is highly dependent on the efficiency of the third dredge-up \citep{Karakas2002}. 

The \textsc{binary\_c} code models the third dredge-up using fits to the models from \citet{Karakas2002}. \textsc{Binary\_c} also allows us to adjust the operation of the third dredge-up through two parameters. It is possible to (i) lower the minimum core mass for the onset of third dredge-up (see Equation 46 in \citet{Izzard2004}) by a constant value, $\Delta M_{\rm c, min}$, and (ii) change the minimum third dredge-up efficiency, $\lambda_{\rm min}$ \citep{Marigo1996, Izzard2004_2}. We use an observed carbon-star luminosity function (CSLF, for example, see \citealp{Guandalini2013}) from solar-neighbourhood stars reported in \citet{Abia2022} to calibrate the third dredge-up parameters $\Delta M_{\rm c, min}$ and $\lambda_{\rm min}$ in our synthetic models. We use the C stars reported in \citet{Abia2022} with luminosities above the RBG tip (M$_{K_s} \leq 7 \, {\rm mag}$, see Figure 16 from \citealp{Abia2022}). 

Using the modified version of \textsc{binary\_c}, we created one grid of 100 single-star models of masses of $1-8\Msun$ for every combination of $\Delta M_{\rm c,min}$ and $\lambda_{\rm min}$ where $\Delta M_{\rm c,min}$ ranges from 0.0 to -0.2$\Msun$ with 0.01$\Msun$ increments and $\lambda_{\rm min}$ from 0.0 to 1.0 with 0.05 increments, totalling 441 stellar grids. After each thermal pulse, we utilize an approximation of the luminosity dip observed in detailed stellar models where the luminosity drops by a factor $f_{\rm L}$ \citep{Izzard2004_2},

\begin{equation}
    \label{eq:LumDip}
    f_{\rm L} = 1 - 0.5 \times {\rm min} \left[ 1, {\rm exp}\left( -3 \frac{\tau}{\tau_{\rm ip}} \right) \right],
\end{equation}

\noindent where $\tau$ is the time from the beginning of the current thermal pulse, and $\tau_{\rm ip}$ is the time between subsequent thermal pulses, known as the interpulse period.

We produced a theoretical CSLF for each stellar grid following the methodology outlined in \citet{Marigo1996, Marigo1999} and using the initial mass function from \citet{Kroupa2001}. From every AGB model, we extract the luminosities at each time step where the surface abundance ratio C/O $\geq 1$ and calculate the absolute bolometric magnitudes, $M_{\rm bol}$ \citep{Mamajek2015}. 


Finally, we bin the absolute bolometric magnitudes to 0.3-magnitude bins and produce a theoretical CSLF, which we compare to the observational CSLF. We define our best fit as the model with the lowest root-mean-squared error. To verify our fit at a higher resolution, we repeat the analysis for our best-fitting CSLF and compare it to the fits neighbouring in parameter space ($M_{\rm c,min} \pm 0.01\Msun$ and $\lambda_{\rm min} \pm 0.05$) using grids of 1000 single stars.

The C stars observed in \citet{Abia2022} have the potential to be formed by binary mass transfer \citep{Izzard2004_2}, which could motivate the use of binary models to calibrate the third dredge-up. By using C stars with M$_{K_s} \leq 7 \, {\rm mag}$, we filter out suspected extrinsic giant branch C stars from the low-luminosity tail of the CSLF. However, there is still the potential for contamination at higher bolometric luminosities. We chose not to use our binary star models for the third dredge-up calibration due to computational limitations (the binary grids would need to contain tens of thousands of models) and the additional uncertainty introduced by binary stellar evolution, such as mass transfer efficiency. Therefore, we only use single-star models to calibrate the third dredge-up, and we assume the majority of the observed AGB C stars we are fitting to are formed intrinsically. This assumption is further motivated by the results from \citet{Izzard2004_2}, which show a theoretical CSLF produced by a population of pure binaries differs considerably only at absolute bolometric magnitudes $\lesssim -4$ and mainly originates from extrinsic GB C stars. GB C stars are likely filtered out of the observational CSLF we are using.

\subsection{He Intershell Abundances}
\label{sec:Intershell}

Upon the onset of the third dredge-up, AGB models calculated using \textsc{binary\_c} instantaneously mix the products from the He intershell region into the stellar envelope, where the He intershell describes the He-rich zone between the H- and He-burning shells inside a TP-AGB star. The He intershell abundances from the standard version of \textsc{binary\_c} at solar-metallicity are fit to models presented in \citet{Gallino1998} and \citet{Karakas2002}, and are described in \citet{Bonavciv2007}. The models from \citet{Gallino1998} utilize a \iso{13}C pocket of predefined mass and \iso{13}C abundance profile to produce the $s$-process elements in AGB stars. The \iso{13}C pockets are thin layers within the He intershell rich in \iso{13}C that forms after a third dredge-up event transports protons into the He intershell (although the exact mechanism for this transportation is uncertain). The \iso{13}C burns primarily via the \iso{13}C($\alpha$, $n$)\iso{16}O reaction, which releases the neutrons required for the $s$-process \citep{Iben1982, Straniero1995}. The products of the $s$-process remain in this thin layer until the next thermal pulse, which drives the He intershell to become convective and mixes the $s$-process products throughout the He intershell.


More recent models use various methods to include the \iso{13}C pocket. For example \citet{Goriely2000}, \citet{Lugaro2012}, and K16 inject what is known as a partial mixing zone (PMZ) into the intershell at the end of each third dredge-up where the number density of the injected protons decreases monotonically from the envelope value to an arbitrary value at a predefined mass coordinate $M_{\rm PMZ}$ below the envelope. \citet{Cristallo2009, Cristallo2015} introduce an unstable convective boundary between the envelope and He intershell during the third dredge-up, allowing protons to be transported into He intershell.

At solar-metallicity, the standard version of \textsc{binary\_c} calculates the intershell abundances using an interpolation table based on fits to \citet{Karakas2002} and \citet{Gallino1998}. We update the interpolation table, which calculates He intershell abundances at solar-metallicity using fits to the abundances of 328 isotopes calculated using the same models presented by K16. For the fit, we follow a methodology similar to that described in \citet{Abate2015} that fits models from \citet{Lugaro2012} for Z = 0.0001. K16 calculates abundances for stellar models using various $M_{\rm PMZ}$ and provides a "standard" value (see Table 3 in K16), which we have used. The standard models from K16 include their $1.5\Msun$ and $1.75\Msun$ models, calculated using convective overshoot at the base of the convective envelope during the AGB. Convective overshoot is not modelled during the AGB in our \textsc{binary\_c} models. However, this does not negatively impact our stellar yields. The overshooting $1.5\Msun$ and $1.75\Msun$ models allow us to make fits to the intershell abundances of s-process elements down to $1.5\Msun$, avoiding the need to extrapolate from the $2\Msun$ K16 model. We produce an intershell abundance table fitting the He intershell abundances of the K16 solar-metallicity models by sampling the average abundance of the He intershell convective zone over the final three saved time steps of the thermal pulse. This method is valid because the He intershell is well-mixed and chemically homogeneous and uses the intershell abundances at the time of the third dredge-up. 

The standard version of \textsc{binary\_c} treats the intershell abundances independently from the third dredge-up and calculates the He intershell abundances using the metallicity, total mass at the first thermal pulse, and the number of thermal pulses experienced to that point. This method becomes problematic when stars modelled in \textsc{binary\_c} experience the first third dredge-up at a different thermal pulse number than predicted in K16. The formation of the \iso{13}C and $s$-process nucleosynthesis should begin after the conclusion of the first third dredge-up event, not after a predefined number of thermal pulses like in the standard version of \textsc{binary\_c}. To rectify this, we have produced two separate interpolation tables to calculate the He intershell abundances. The first is for the elements lighter than Fe, which calculates the He intershell abundances using the number of thermal pulses, like in the standard version of \textsc{binary\_c}. The second is for the elements heavier than and including Fe, which calculates the He intershell abundances using the number of third dredge-up events instead of the number of thermal pulses. This allows us to couple the $s$-process to the third dredge-up without further modifying the He intershell abundances of the light elements. The light element table fits all the models from K16, but our heavy element table excludes the $1.00\Msun$ and $1.25\Msun$ stars from the K16 models, as they do not experience the third dredge-up or $s$-process. For stars of mass $<1.5\Msun$ that experience the third dredge-up in our modified \textsc{binary\_c} models, the heavy element intershell abundances are calculated from the $1.5\Msun$ K16 model.


\section{Results}
\label{sec:Results}
In this section, we present our calibration of the CSLF calibration and then show the C and Ba yields from our single stars. We then show the C, N, and O yields ejected by mixed stellar populations calculated using various binary fractions. We also show the solar scaled [C/O], [N/O], and [C/N] calculated from our stellar yields to compare the limits between our single and binary stellar populations. We then examine the population yields for Sr, Ba, and Pb, focusing on Ba. Finally, we report on supernovae and the formation of black holes within our low and intermediate-mass stellar population. 

\subsection{Results of our Third Dredge-Up Calibration to the Galactic Carbon Star Luminosity Function}
\label{sec:3DUP_Results}

\begin{figure*}[hbt!]
\centering
\includegraphics[width=0.75\linewidth]{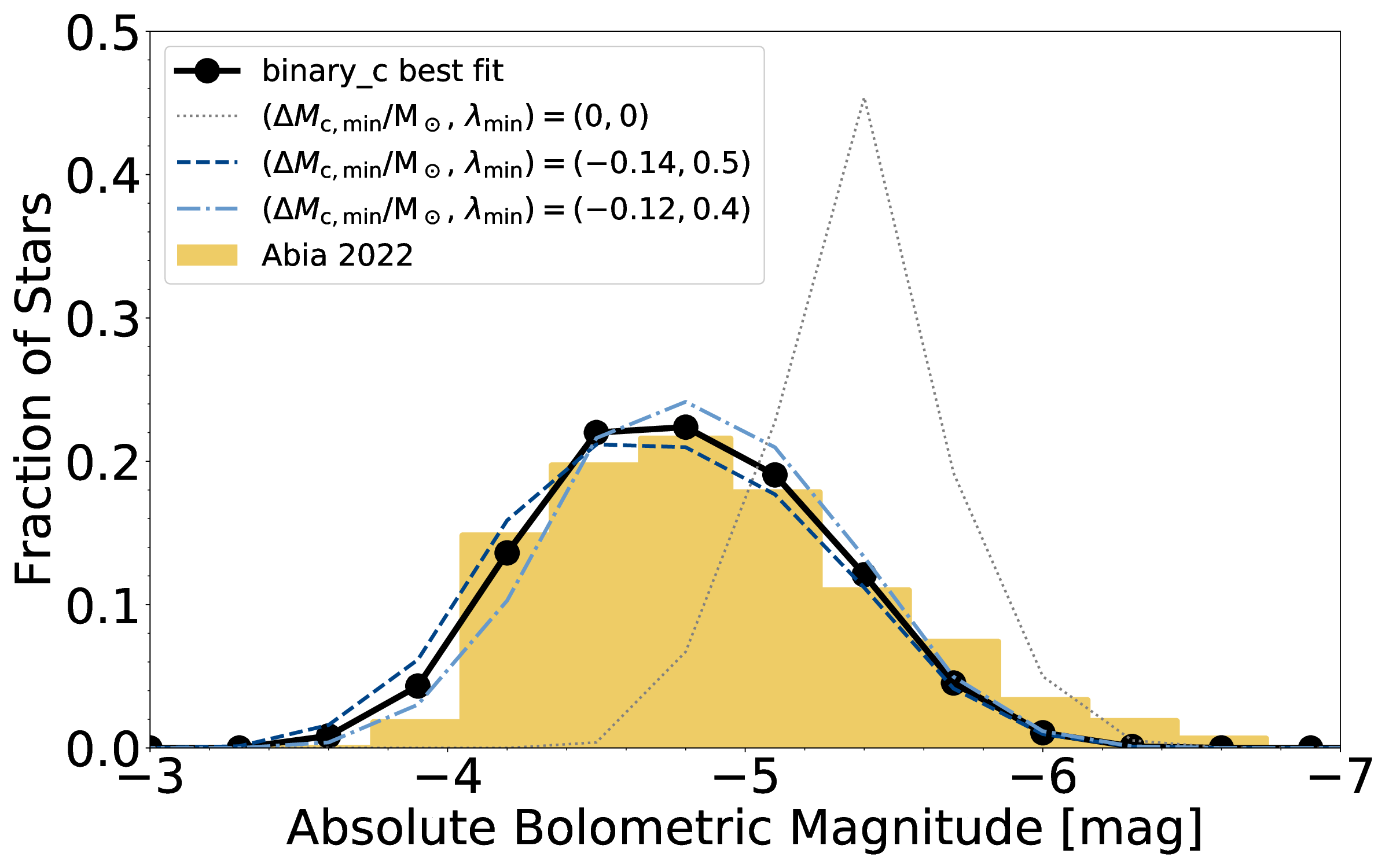}
\caption{Our best-fit to the CSLF presented in \citet{Abia2022} results when $(\Delta M_{\rm c,min}/\Msun, \lambda_{\rm min}) = (-0.13, 0.45)$. We include the results for ($M_{\rm c,min}/\Msun, \lambda_{\rm min}) = (-0.12, 0.4)$, $(0.14, 0.5)$, and $(0,0)$} for comparison.
\label{fig:CSLF_fit}
\end{figure*}

We find that $\Delta M_{\rm c, min} = -0.13\Msun$ and $\lambda_{\rm min} = 0.45$ results in the best fitting CSLF to observations from \citet{Abia2022}, as shown in Figure \ref{fig:CSLF_fit}, together with the fits for $(\Delta M_{\rm c, min}/\Msun, \lambda_{\rm min}) = (-0.12, 0.4)$, $(-0.14, 0.5)$, and $(0, 0)$ shown for comparison. Based on our binning in Figure \ref{fig:CSLF_fit}, the absolute bolometric magnitudes of our C-star population range between $-3.45$ to $-6.75$ mag and peaks at $-4.8$ mag, as observed in \citet{Abia2022}, but it also over-predicts the low-luminosity tail and under-predicts the high-luminosity tail of the observed distribution.

\begin{figure*}
\centering
\includegraphics[width=0.8\linewidth]{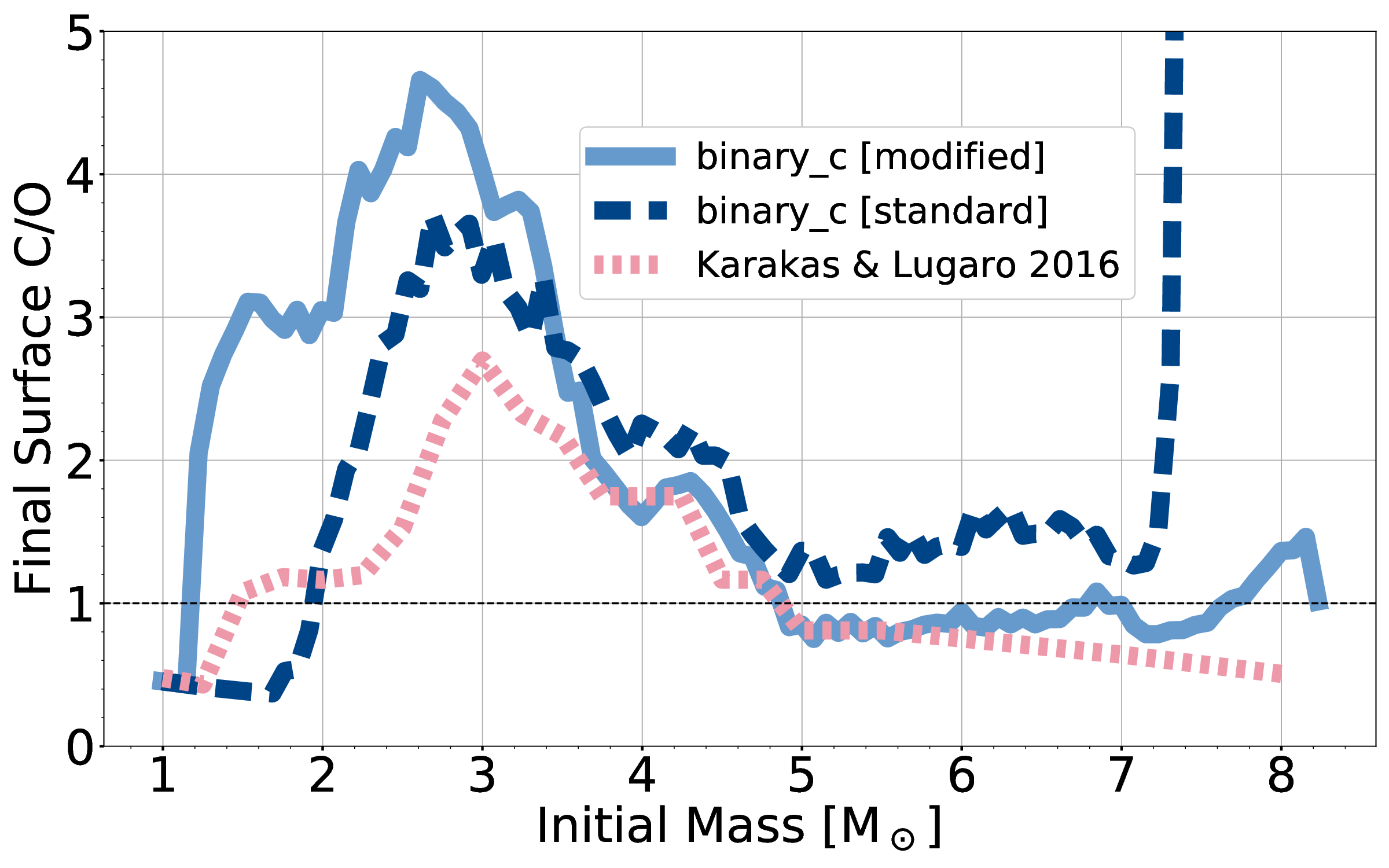}
\caption{Final surface C/O ratios of single stars from K16, and the standard and modified versions of \textsc{binary\_c}, and our modified version of \textsc{binary\_c}. The C/O ratio = 1 is marked to highlight stars that end their lives C-rich.} 
\label{fig:CORatio}
\end{figure*}

Figure \ref{fig:CORatio} shows that stars of mass $1.2-4.8\Msun$ become C-rich, with some stars of mass $\gtrsim 7\Msun$ becoming C-rich near the end of the TP-AGB after the stellar winds eject enough of the envelope for hot-bottom burning to shut down, ceasing C-destruction via the CNO cycles. Stars modelled using the standard version of \textsc{binary\_c} instead become C-rich at initial masses greater than $1.9\Msun$. Our modifications introduced in Paper I remove the unrealistic spike around $7.5\Msun$ star. Moreover, the modified \textsc{binary\_c} stellar models are more C-rich at masses $\lesssim3.5\Msun$ than the stars modelled directly by K16, due to the choice of the parameters $\Delta M_{\rm c, min}$ and $\lambda_{\rm min}$ in \textsc{binary\_c} being applied regardless of the initial stellar mass. 

A minimum C star mass of $1.2\Msun$ is low compared to other model predictions, which estimate $\sim 1.5\Msun$ \citep{Marigo2001, Karakas2014_2, Ventura2018}. Observations estimate the minimum C star mass to be $\sim1.3-1.5\Msun$, although this is uncertain due to the distances to Galactic stars \citep{Pal2021, Abia2022}. Figure \ref{fig:CORatio} shows stars of mass $\gtrsim5\Msun$ often do not become C rich when modelled using the modified version of \textsc{binary\_c}, which potentially puts too much weight on low-mass stars to produce the CSLF. Model parameters we have not explored which influence the theoretical CSLF, and therefore the third dredge-up, include the envelope mass where the third dredge-up ceases, which is set to $0.5\Msun$, the duration of hot-bottom burning, the luminosity of AGB stars, and the depth and duration of the luminosity dips approximation described in equation \ref{eq:LumDip}. Therefore, the third dredge-up remains a considerable source of uncertainty in our models. 

\subsection{Chemical Yield from Single Stars}
A natural consequence of our calibration of the third dredge-up and the updates to the He intershell abundances is the alteration of the single-star stellar yields (see Equations \ref{eq:Yield} and \ref{eq:netYield}) compared to those calculated from the standard version of \textsc{binary\_c}. To verify that the stellar yield calculated using our modified version of \textsc{binary\_c} are reasonable, we compare them to those calculated from models using the standard version of \textsc{binary\_c}, K16, and, for C, to \citet{Marigo2001}.

\begin{figure*}
\centering
\includegraphics[width=0.8\linewidth]{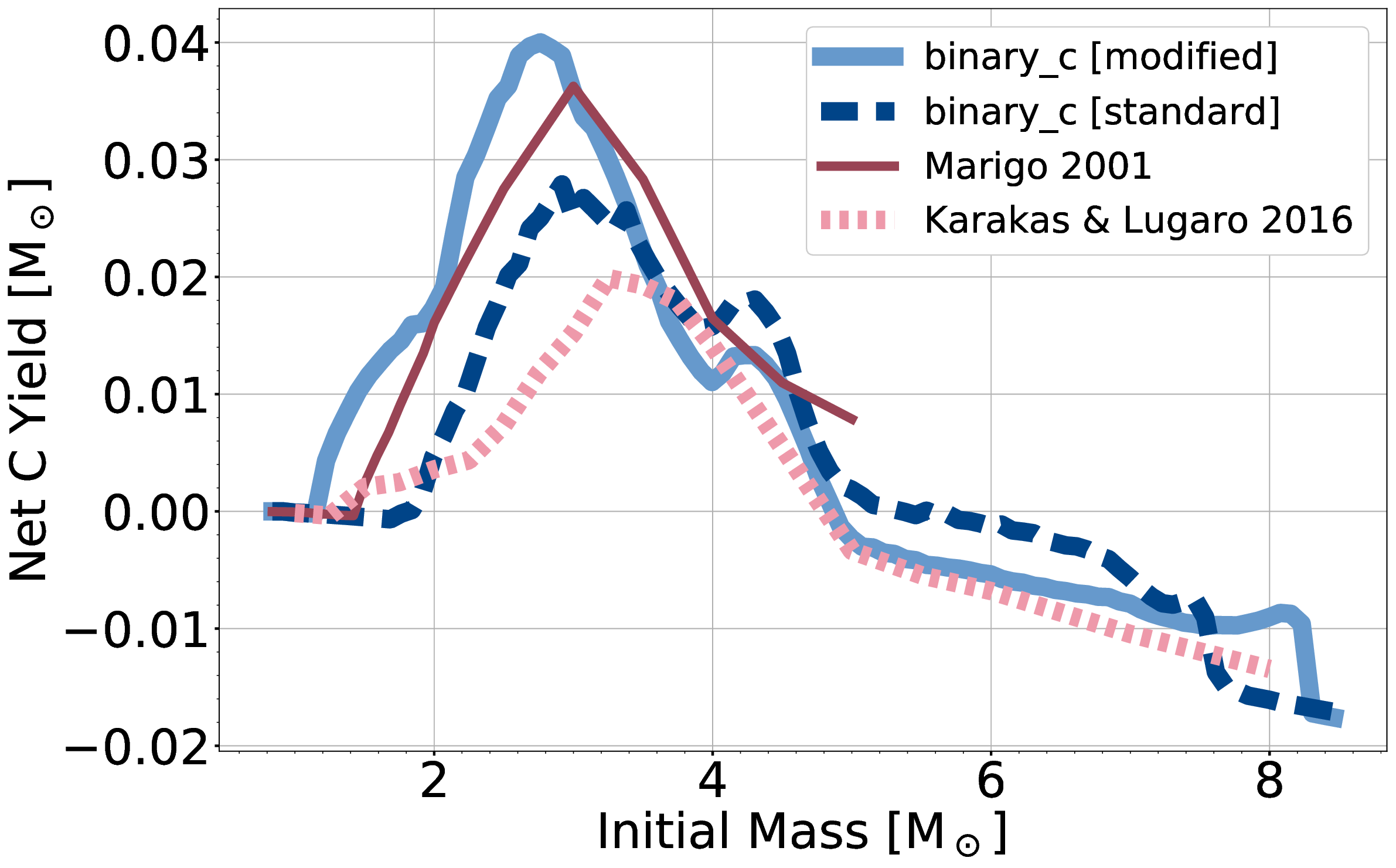}
\caption{Net C ejected our single stars as calculated from the standard and modified versions of \textsc{binary\_c}. We compare the net C yield to those calculated from K16 and \citet{Marigo2001} at solar-metallicity.}
\label{fig:SingleCNet}
\end{figure*}

We first examine the net yield of C shown in Figure \ref{fig:SingleCNet}. Here, we compare the net C yields from \citet{Marigo2001} at solar metallicity where the mixing-length parameter is 1.68. For initial masses $\lesssim 4\Msun$, the net C yield from the modified \textsc{binary\_c} models closely reflects the yield from \citet{Marigo2001}. This is likely due to \citet{Marigo2001} using the CSLF of the Large and Small Magellanic Clouds to calibrate third dredge-up in their models. Although the K16 models employ convective overshoot to force their $1.5\Msun$ and $1.75\Msun$ stars to be C-rich, they do not explicitly attempt to calibrate their models to fit any CSLF.

Our net C yield is slightly higher at masses $\lesssim 3\Msun$ than calculated in \citet{Marigo2001}, and our stars become C-rich at $1.2\Msun$, whereas in the models from \citet{Marigo2001} they become C-rich at $1.5\Msun$. This discrepancy likely arises due to the different treatments to the third dredge-up. \textsc{binary\_c} uses a variable third dredge-up efficiency, while \citet{Marigo2001} keeps the efficiency constant through the TP-AGB. Additionally, the occurrence of the third dredge-up in the models from \citet{Marigo2001} is dependent on the temperature at the base of the convective envelope, which is a dependence \textsc{binary\_c} lacks. Instead, the \textsc{binary\_c} models terminate the third dredge-up at an envelope mass of $0.5\Msun$. See \citet{Marigo1999} for the details on how the third dredge-up is treated in their models.

At masses $\gtrsim 4\Msun$, the stellar yield from our modified version of \textsc{binary\_c} more closely resembles the stellar yield from K16 than those of \citet{Marigo2001}. This is expected since our third dredge-up parameters, $\Delta M_{\rm c,min}$ and $\lambda_{\rm min}$, have very little influence at these masses as these stars enter the TP-AGB with sufficient core masses for third dredge-up. 

\begin{figure*}
\centering
\includegraphics[width=0.8\linewidth]{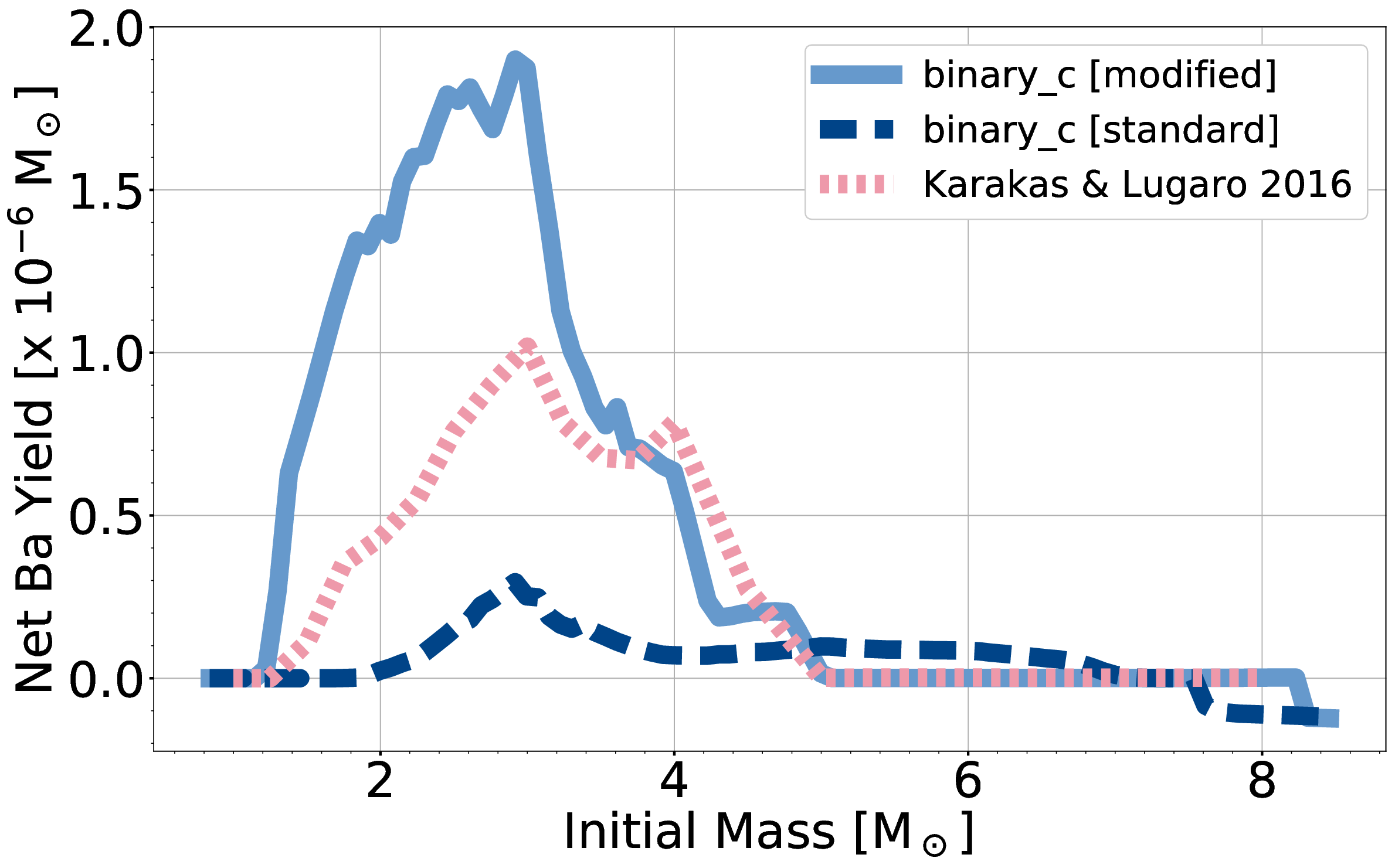}
\caption{Net Ba yield from single AGB stars calculated from the standard and modified versions of \textsc{binary\_c} compared to K16. Our modified version produces a similar Ba yield compared to K16. The standard version of \text{binary\_c} achieves a peak Ba yield of $3.3 \times 10^{-7} \Msun$ at $2.9\Msun$, which is a factor of 3.1 times lower than the peak Ba yield from K16 of $1.1 \times 10^{-6} \Msun$ at $3\Msun$. The modified version of \textsc{binary\_c} has a peak Ba yield of $1.9\times10^{-6}\Msun$, almost twice as high as the peak K16 Ba yield.}
\label{fig:SingleBa}
\end{figure*}

The elements produced by the $s$-process are affected by both the updated He intershell abundances and the new third dredge-up calibration. Figure \ref{fig:SingleBa} shows the net Ba yield from our single-star models. By construction, models from the modified \textsc{binary\_c} better agree with K16 than those calculated using the standard \textsc{binary\_c}. However, the Ba yield calculated by the modified version of \textsc{binary\_c} is higher than those from K16 in the mass range $\sim 1.3-3.8\Msun$ with a peak at $\sim3\Msun$ roughly two times higher. This is due to the third dredge-up calibration, which increases the number and efficiency of third dredge-up.

\begin{figure*}
\centering
\includegraphics[width=0.8\linewidth]{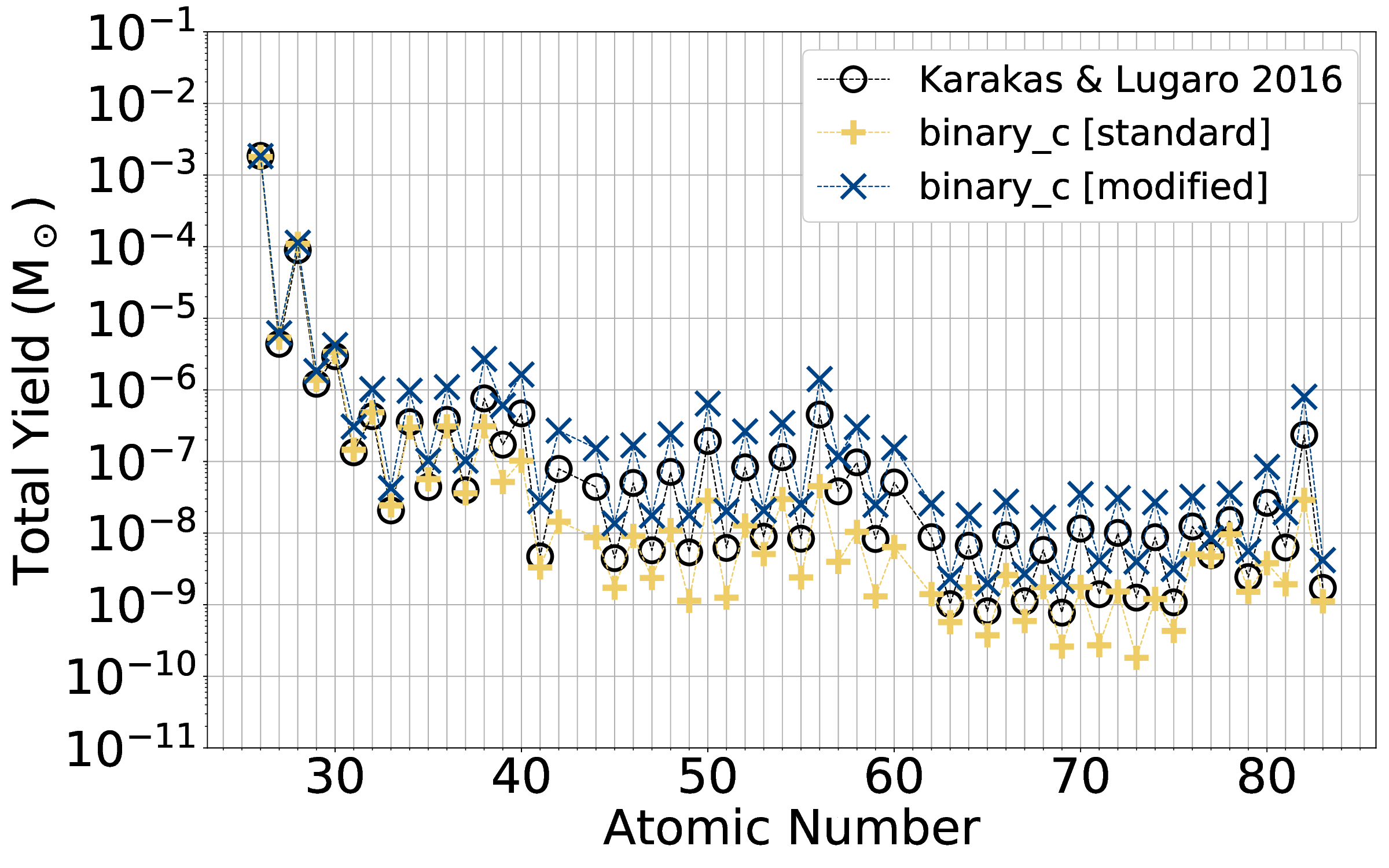}
\caption{Elemental yields ejected by a $2\Msun$ star as calculated by K16, the standard version of \textsc{binary\_c} and the modified version of \textsc{binary\_c}. We show the elemental yields for all elements from Fe to Bi, excluding radioactive Tc (not reported in K16).}
\label{fig:AllElements}
\end{figure*}

Figure \ref{fig:AllElements} shows the total yields (see Equation \ref{eq:Yield}) of all our considered elements from Fe to Bi for a $2\Msun$ star from K16, and the standard and modified \textsc{binary\_c}. Figure \ref{fig:AllElements} highlights that all yields of elements heavier than and including Ga are systemically increased in the modified \textsc{binary\_c} results compared to the yields from the $2\Msun$ K16 models. The $2\Msun$ example star shown is one of the more extreme examples of disagreement between the modified \textsc{binary\_c} models and the models from K16, even though stars of this mass experience third dredge-up in both cases. The increased yields result from our third dredge-up calibration, and this is reduced at higher masses owing to their more massive cores and thinner He-intershells. Most elemental yields calculated from modified \textsc{binary\_c} are within a factor of 3.6 times the yields from the K16 models. In contrast, the yields calculated from the standard version of \textsc{binary\_c} $2\Msun$ model (which includes fits to the He intershell to models by \citealp{Gallino1998}) show a systematic under-production compared to the $2\Msun$ K16 model for all elements including and heavier than Kr. Some key elements from the $2\Msun$ standard \textsc{binary\_c} model, such as Ba, Ce, and Pb, are under-produced by a factor of 8-10 times the $2\Msun$ K16 model. Figure \ref{fig:SingleBa} shows the standard version of \textsc{binary\_c} under-produces Ba for all masses < $4.5\Msun$. The under-production is mainly attributed to the differing treatments of the \iso{13}C pocket (see Section \ref{sec:Intershell}). 

\subsection{Stellar Population Yields and Abundances}
To investigate how binary evolution influences the stellar yield from low- and intermediate-mass populations, we first examine how the \textsc{binary\_c} stars evolve. Binary interactions, such as common envelope and Roche-lobe overflow events, may lead to the truncation of stellar evolution. Of particular interest is whether our stars experience the GB, E-AGB, and TP-AGB (at least five thermal pulses), as these evolutionary phases are the sites of the dredge-up episodes that allow low- and intermediate-stars to contribute to the chemical enrichment of the universe. Whether the TP-AGB stars have sufficient mass for hot-bottom burning when they enter the TP-AGB is also of interest, as this process influences the stellar yield of intermediate-mass stars. We use a mass of at least $5\Msun$ to indicate a hot-bottom burning star, but the \textsc{binary\_c} models show some hot-bottom burning in single stars of mass as low as about $4.5\Msun$. Binary evolution may lead to $4.5\Msun$ stars with envelopes too cool for HBB, so we use $5\Msun$ as a more conservative estimate. We show these results in Table \ref{tab:EvoStats}. Hereafter, unless otherwise specified, all results and discussion are based on calculations made using the modified version of \textsc{binary\_c} for a grid of 1000 single and 640 000 (primary stars $M_{\rm 1}: 100 \times$ secondary stars $M_{\rm 2}:80 \times $ orbital periods $ p:80$) binary stars, sampled as described in Table \ref{tab:parameters}.  


\begin{table}[hbt!]
\caption{Percentages of single and binary systems (weighted using the birth mass distribution from \citealp{Kroupa2001}) with at least one star experiencing the GB, E-AGB, and TP-AGB (at least five thermal pulses) phases and have sufficient mass ($>5\Msun$) for hot-bottom burning (HBB).}

\label{tab:EvoStats}
\begin{tabular}{lllll}
\toprule
\headrow System-type & \% GB & \% E-AGB & \% TP-AGB & \% HBB \\
\midrule
Single & 79 & 79 & 78 & 4.6 \\
Binary & 79 & 65 & 60 & 3.5 \\

\end{tabular}
\end{table}

Due to our lower mass limit of $0.8\Msun$, some stars in our stellar population do not evolve off the main sequence (MS) during the time of the $15\,$Gyr simulation. Most single-star systems that do not evolve off the MS are stars of mass $\lesssim 0.9\Msun$. As a result, only $79\%$ of our single-star population, and a similar percentage of the binary star population, enters the first giant branch. Not all of these GB stars in binary systems will experience the first dredge-up as the expansion of the stellar radius on the GB makes this phase more likely to experience Roche-lobe overflow than MS and Hertzsprung-Gap (HG) stars.

In general, binary evolution prevents stars from experiencing evolved phases. Compared to the single-star population, we find 17\% fewer stars reaching the E-AGB phase in the binary population. A further 23\% fewer enters the TP-AGB and experiences at least five thermal pulses. The contributions to the chemical enrichment of the Universe are, on average, limited rather than enhanced by binary interactions. We also find that binary systems contain 24\% fewer hot-bottom burning systems than our single-star population, mainly due to fewer systems with TP-AGB stars. 

\subsubsection{Weighted Population Yields: C, N, and O}

We first examine the stellar yield accounting for our assumed birth distributions (see Table \ref{tab:parameters}) normalised per unit of star-forming material ($\Msun/{\rm M_{\odot,SFM}}$) as described in Equation \ref{Eq:WYield}, these are hereafter referred to as weighted yield. We then examine the weighted yields of C, N, and O from our population, which are reported in Table \ref{tab:PopYields_CNO}. 

\begin{table*}[hbt!]
\caption{C, N, and O stellar population yield ejected by low- and intermediate-mass stellar populations with varying binary fractions. We also show the ratios between the C, N, and O yields produced by populations including binary systems divided by the yield produced by the population of single stars only. C is the most heavily influenced by binary evolution as the binary population (binary fraction = 1) produces 24\% less C than the population of single stars only.}

\label{tab:PopYields_CNO}
\begin{tabular}{llllllllllll}
\toprule
\headrow Element & \multicolumn{11}{c} {Weighted population yield in units of $\Msun/{\rm M_{\odot, SFM}}$ at binary fraction:} \\
\headrow & 0.0 & 0.1 & 0.2 & 0.3 & 0.4 & 0.5 & 0.6 & 0.7 & 0.8 & 0.9 & 1.0 \\
\midrule
C ($\times 10^{-3}$) & 2.4 & 2.3 & 2.2 & 2.1 & 2.1 & 2.0 & 2.0 & 1.9 & 1.9 & 1.8 & 1.8 \\
N ($\times 10^{-4}$) & 6.7 & 6.7 & 6.6 & 6.5 & 6.5 & 6.4 & 6.4 & 6.4 & 6.3 & 6.3 & 6.2 \\
O ($\times 10^{-3}$) & 1.5 & 1.5 & 1.5 & 1.5 & 1.5 & 1.5 & 1.5 & 1.5 & 1.5 & 1.5 & 1.5 \\

\headrow & \multicolumn{11}{c} {Weighted population yield ratio (population inc. binaries / population single stars only)} \\
C & 1.00 & 0.97 & 0.94 & 0.91 & 0.88 & 0.86 & 0.84 & 0.82 & 0.80 & 0.78 & 0.76 \\
N & 1.00 & 0.99 & 0.98 & 0.97 & 0.97 & 0.96 & 0.95 & 0.95 & 0.94 & 0.93 & 0.93 \\
O & 1.00 & 1.00 & 1.01 & 1.01 & 1.01 & 1.02 & 1.02 & 1.02 & 1.02 & 1.02 & 1.03 

\end{tabular}
\end{table*}

\begin{figure*}
\centering
\includegraphics[width=0.8\linewidth]{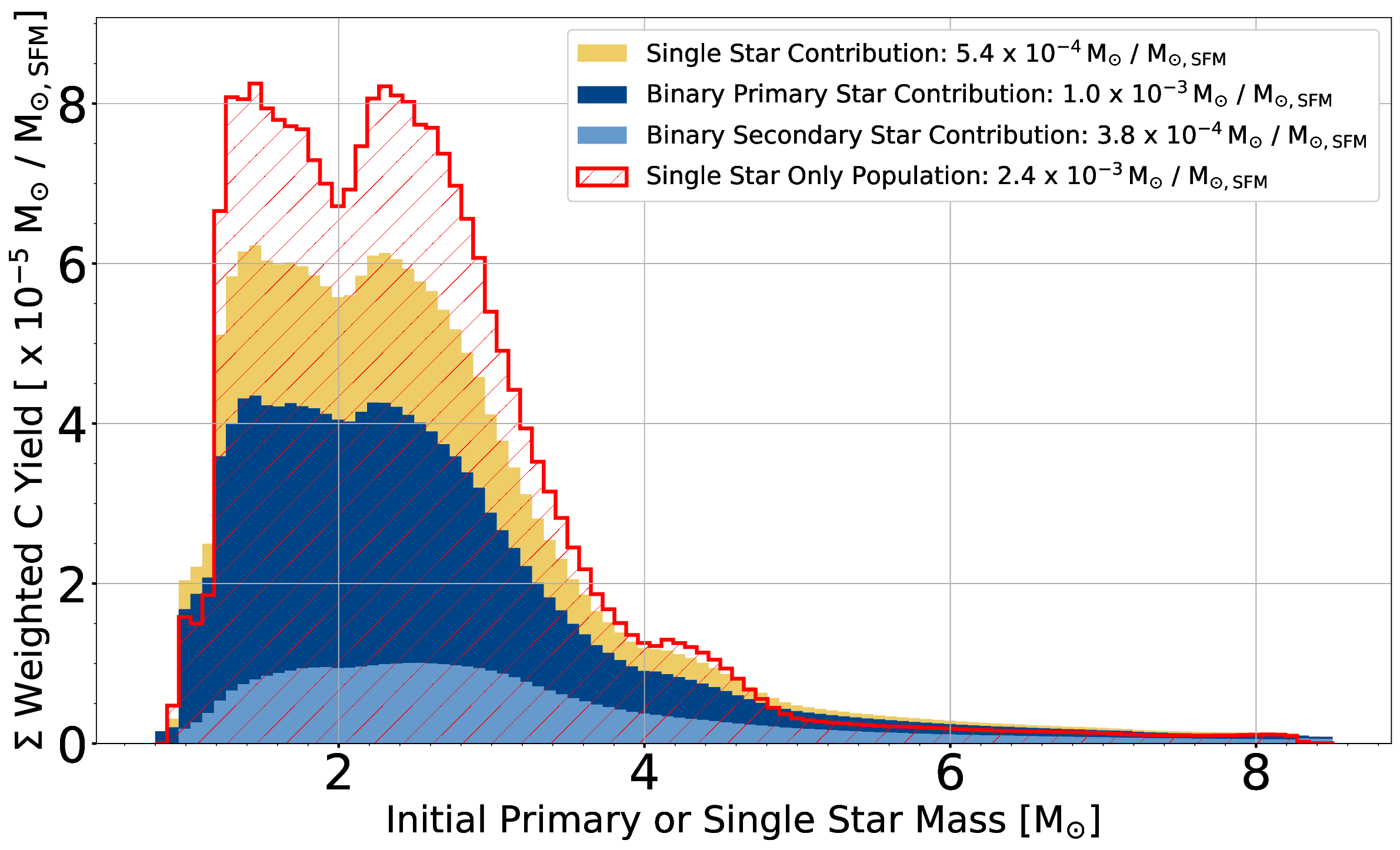}
\caption{The weighted C stellar population yield as a function of the single or primary star mass of our single star population and of a population including binaries with a 0.7 binary fraction. We sum and bin all the weighted yields based on the initial primary- or single-star mass. When including binaries, the yield contributions from the single stars, binary primary stars (and post-merger objects), and binary secondary stars are stated in the legend and stacked in the plot, with their summation totalling the yielded carbon from the population. In this case, $28\%$ of the ejected C originates from the single star portion of the population, $53\%$ from the binary primary stars, and $19\%$ from the binary secondary stars.}
\label{fig:C_Weighted}
\end{figure*}

Figure \ref{fig:C_Weighted} shows the carbon yield ejected by two stellar populations with binary fractions 0.0 and 0.7 via stellar winds as a function of the initial primary (or single) star mass. Compared to a population of single stars only, we find that including binaries results in an overall decrease in the weighted C yield. For example, we find an 18\% decrease when the binary fraction is 0.7 (see Table \ref{tab:PopYields_CNO}). C is under-produced because binary evolution sometimes truncates the TP-AGB or completely prevents the formation of a TP-AGB star (see Table \ref{tab:EvoStats}). We also find a reduction in the formation of C stars as 40\% of binary primary stars become a C star compared to 51\% in the single-star population.

Although, on average, binary evolution results in an under-production of C, some circumstances allow C overproduction in binary systems. Among low-mass (primary star mass, $M_{\rm 1,0} < 5\Msun$) binary systems, we find the largest C overproduction in stars that experience more thermal pulses than when single. This can happen through binary evolution by mass accretion onto a post-MS star, which forces the accreting star to enter the TP-AGB with a lower core mass and over-massive envelope compared to a single star of identical mass. For intermediate-mass binary systems ($5\Msun < M_{\rm 1,0} < 8.3\Msun$), mergers between He-WD and CO-WD form objects similar to R Coronae Borealis stars \citep{Clayton1996, Karakas2015, Tisserand2020} which then eject up to about $0.03\Msun$ of C, making them the objects with the highest C-overproduction in this mass range. 

The binary star population forms an R Coronae Borealis star at a rate of about $3500$ per $10^6\Msun$ of binary star-forming material. Our models estimate the average lifetime of an R Coronae Borealis star to be approximately $7 \times 10^5\,\text{yr}$, which is slightly longer than the $1-3 \times 10^5 \, \text{yr}$ estimated in other studies \citep{Saio2002, Clayton2012}. If we take the lifetime to be $1-7 \times 10^5\,\text{yr}$, and a constant Milky Way star formation rate of $2\Msun$ per year \citep{Elia2022} at solar-metallicity and a binary fraction of 0.7, we estimate there are approximately $500-3800$ R Coronae Borealis stars in the Galaxy today.

\begin{figure*}
\centering
\includegraphics[width=0.8\linewidth]{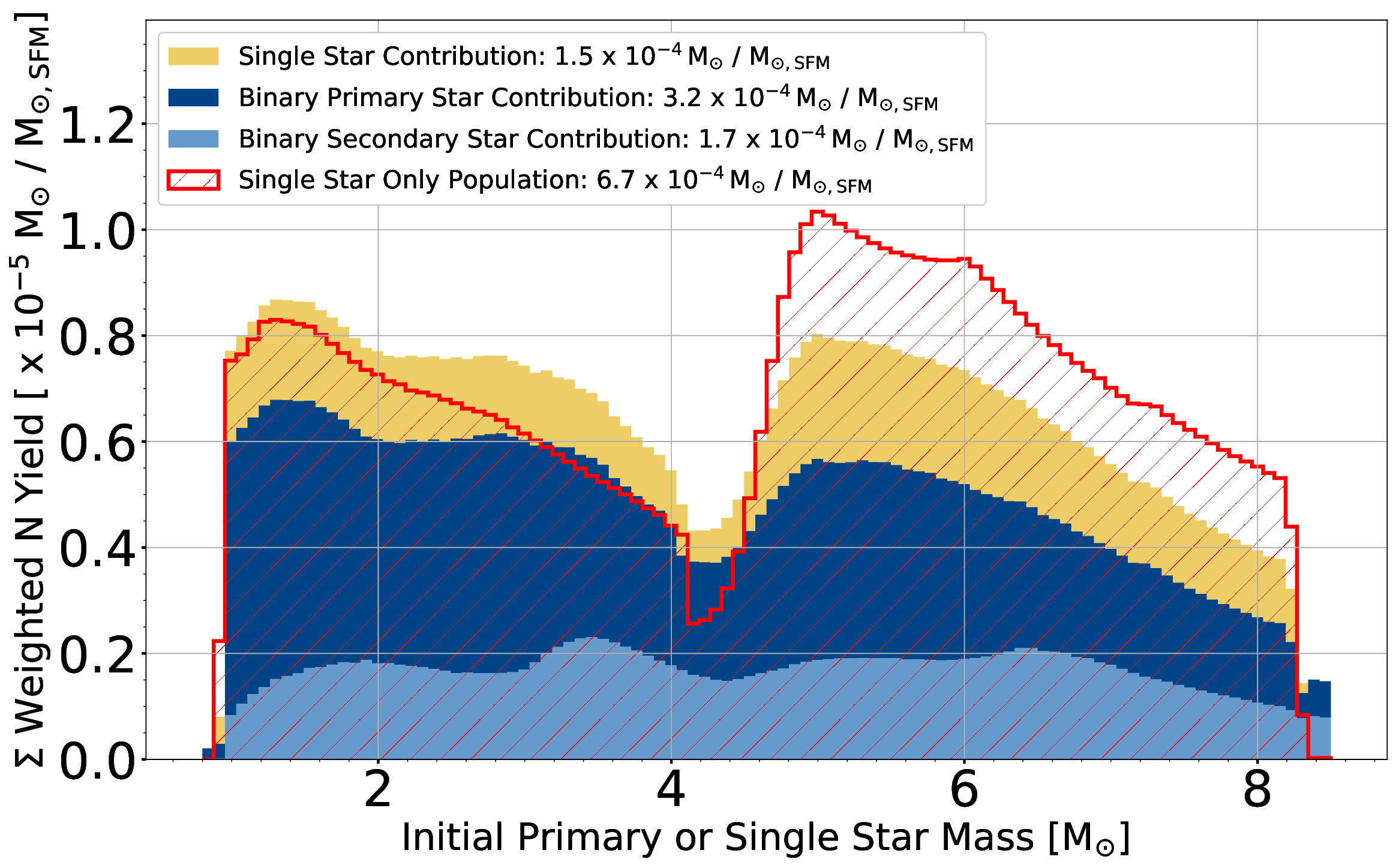}
\caption{Same as Figure \ref{fig:C_Weighted}, but for N. For the binary population, 24\% of the ejected N originates from the single star portion of the population, 50\% from the binary primary stars, and 26\% from our binary secondary stars.}
\label{fig:N_Weighted}
\end{figure*}

The nitrogen ejected via stellar winds from our low- and intermediate-mass population as a function of initial primary and single star mass is shown in Figure \ref{fig:N_Weighted}. In low-mass stars, including binaries results in a net overproduction of N despite the reduced number of AGB stars. There are two distinct evolutionary channels responsible for this. The first is mergers resulting in a star with sufficient mass for hot-bottom burning. Also, mass transfer or common envelope ejections that strip the H envelopes from stars with He-rich cores result in He-rich stars with N surface mass fractions of $\sim0.01$. Stellar winds from these He-rich stars are the second source of N overproduction in low-mass stars. Intermediate-mass stars instead, on average, experience N under-production due to binary evolution since hot-bottom burning is sometimes prevented or suppressed (see Table \ref{tab:EvoStats}). Table \ref{tab:PopYields_CNO} shows that, overall, despite the N overproduction in low-mass stars, binary evolution has little impact on the overall N production from our low- and intermediate-mass stellar population as the overproduction from the low-mass population cancels out the under-production from the intermediate-mass population.

We find little deviation of the O yield from our low- and intermediate-mass population (as shown in Table \ref{tab:PopYields_CNO}). This is expected as most of the O they synthesize remains inside the CO core. Core-collapse supernovae synthesize substantial amounts of oxygen, but we do not include their contribution. The most extreme O producers in the binary population originate from ONe-WDs merging with He-WDs and forming naked-He stars with an O mass fraction of about 0.6, similar to an R Coronae Borealis star but O-rich instead of C-rich. To our knowledge, no such objects have been observed so far, and our models predict them to be rare, with about 5 forming per $10^6\Msun$ of star-forming material. These objects survive up to approximately $3\times10^5$ years before finally ejecting their envelopes. Using this lifetime, we estimate up to 2 of these objects are present in the Galaxy today. However, the lifetimes of these objects are highly uncertain as their unique composition would likely influence mass loss.

\subsubsection{Binary System Abundance Ratios (CNO)}

Here, we investigate the statistical distributions of the solar-scaled [C/O], [C/N], and [N/O] ratios of the total material ejected into the interstellar medium and discuss the evolution of binary systems with depleted or enhanced abundance ratios compared to single stars. We calculate the ratios between elements $X$ and $Y$ (abundance by number) using,

\begin{equation}
	\label{eq:logRatio}
	[{X/Y}] = {\rm log_{10}} \left( \frac{{X}}{{Y}} \right)_{\rm {star}} - {\rm log_{10}} \left( \frac{{X}} {{Y}} \right)_{\rm {\odot}},
\end{equation}

\noindent where the solar abundances are from \citet{Lodders2003}.

\begin{figure}
\centering
\includegraphics[width=0.8\linewidth]{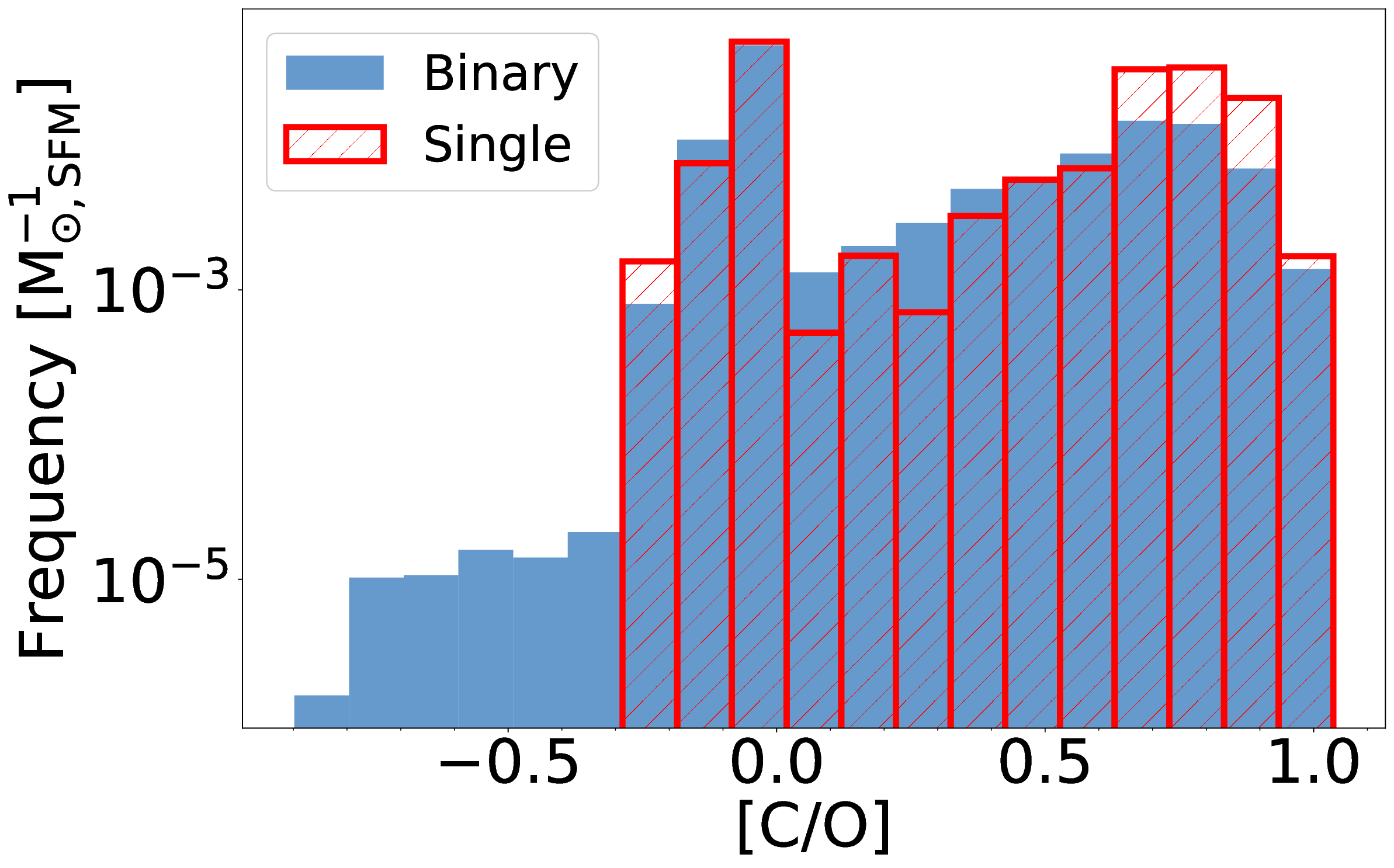}
\caption{Distribution of [C/O] ratios from our binary and single-star populations released into the interstellar medium.}
\label{fig:CO_Ratio}
\end{figure}

The [C/O] distribution ejected from both our binary and single-star populations is shown in Figure \ref{fig:CO_Ratio}. We calculate the ratios from our binary systems using the total stellar yield combined from the primary and secondary stars. This gives us the distribution of released abundances into the interstellar medium. We find that binary evolution can lead to systems with a lower [C/O] ratio than that in single stars. The lowest [C/O] in our single stars is -0.28, corresponding to an $8.28\Msun$ star, and the maximum is +0.94, corresponding to a $2.80\Msun$ star. 

\begin{figure}
\centering
\includegraphics[width=0.8\linewidth]{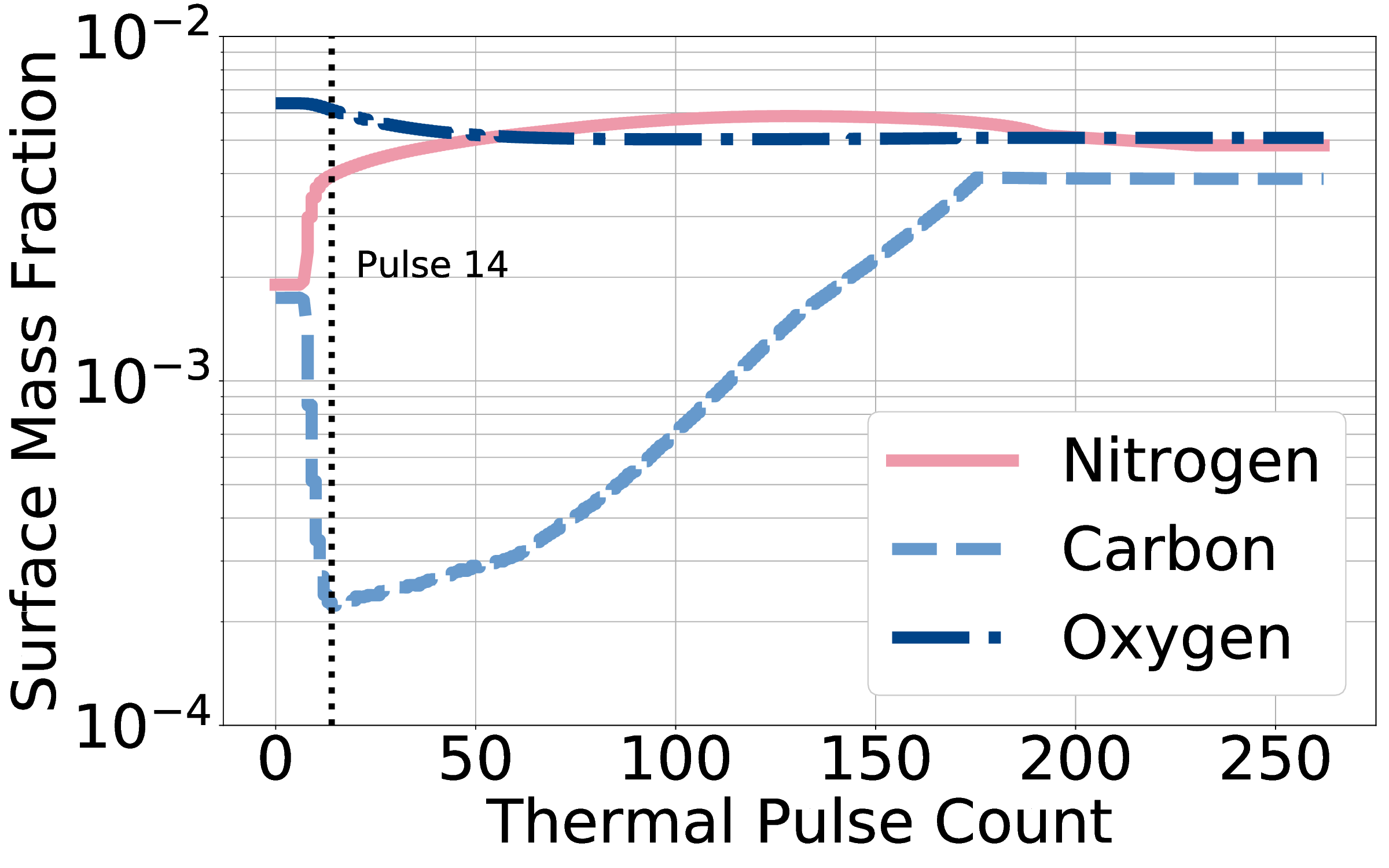}
\caption{Evolution of the surface mass fractions of C, N, and O as a function of thermal pulse count for the single star $8.23\Msun$ model. The dotted vertical line indicates thermal pulse 14 where a common envelope event truncates the stellar evolution of a binary system with $M_{\rm 1,0} = 8.23\Msun$, $M_{\rm 2,0} = 0.45\Msun$, and $p_{\rm 0} = 10.0 \, {\rm yr}$.}
\label{fig:Evol8}
\end{figure}

In our grid of binary models, the minimum [C/O] is -0.90 from the $M_{\rm 1,0} = 8.23\Msun$, $M_{\rm 2,0} = 0.45\Msun$, and $p_{\rm 0} = 10.0 \, {\rm yr}$ system, with 0.04\% of the systems born in the binary population ejecting [C/O] < -0.38 (0.1 dex lower than the minimum [C/O] ejected by the $8.23\Msun$ single star). Most of these systems are hot-bottom burning stars with low-mass companions that experience at least one common envelope event. In a single star, hot-bottom burning destroys C in the envelope, but the star recovers some surface C through third dredge-ups after mass loss shuts down hot-bottom burning. There is also some O destruction when the bottom of the convective envelope becomes hot enough to activate NO burning (see Figure \ref{fig:Evol8}). In the binary scenario, a common envelope event might interrupt the TP-AGB primary star when hot-bottom burning has destroyed a large amount of C but not a considerable amount of O. For example, in the case where $M_{\rm 1,0} = 8.23\Msun$, $M_{\rm 2,0} = 0.45\Msun$, and $p_{\rm 0} = 10.0 \, {\rm yr}$ system, the common envelope event occurs after thermal pulse 14 (as marked in Figure \ref {fig:Evol8}) and ejecting the stellar envelope when the surface [C/O] is -0.97, leaving an ONe-WD remnant. Sometimes, a second common envelope interrupts the secondary star before it enters the TP-AGB or dredges up any large amounts of C, further limiting potential C production.

\begin{figure}
\centering
\includegraphics[width=0.8\linewidth]{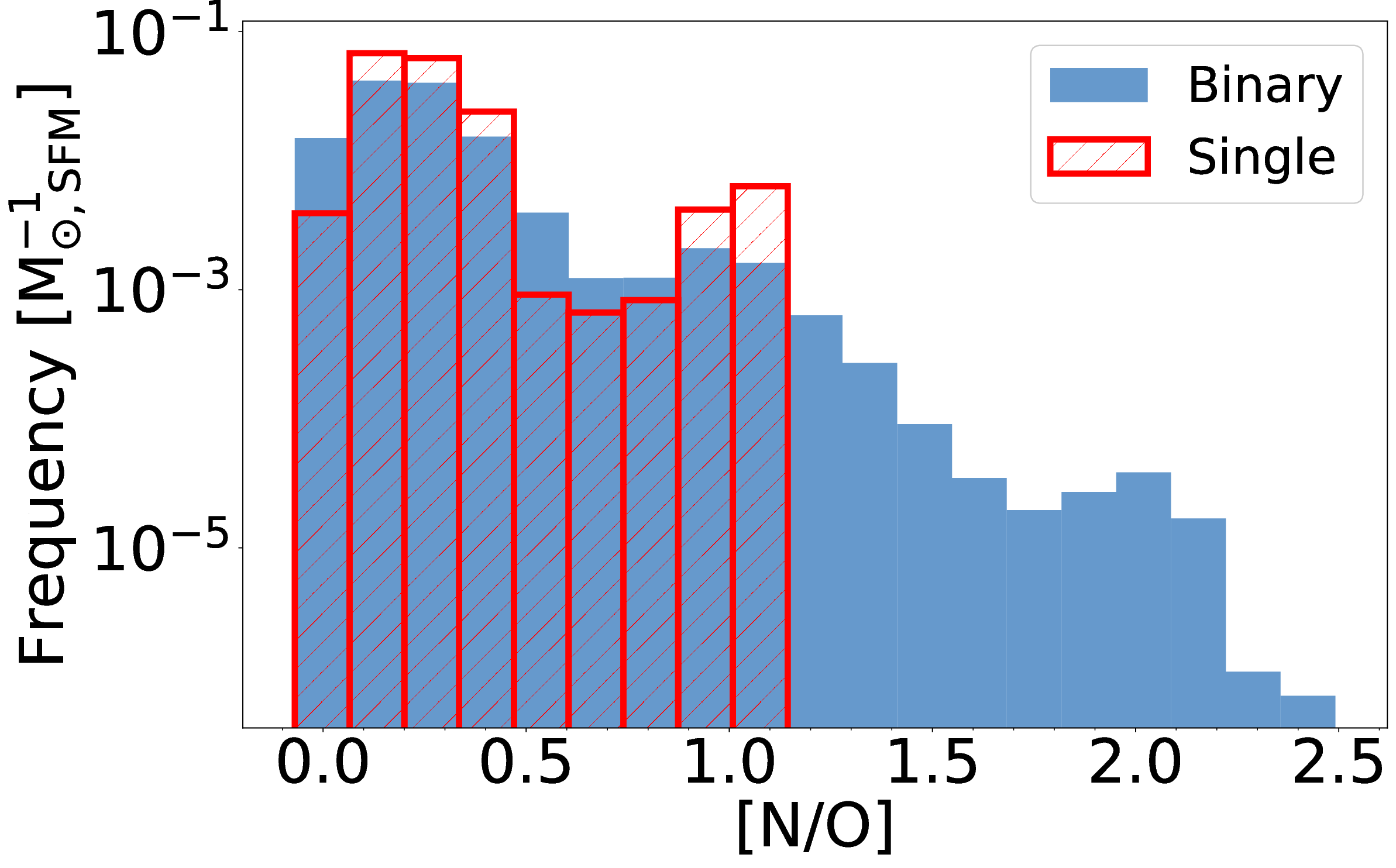}
\caption{As Figure \ref{fig:CO_Ratio}, but for [N/O].}
\label{fig:NO_Ratio}
\end{figure}

The ejected [N/O] distributions are shown in Figure \ref{fig:NO_Ratio}. The [N/O] ratios from single stars range from 0.0 (at $0.91\Msun$) to +1.1 (at $7.26\Msun$). The binary systems instead reach a higher maximum [N/O] ratio of +2.5 in the $M_{\rm 1,0} = 5.46\Msun$, $M_{\rm 2,0} = 4.28\Msun$, and $p_{\rm 0} = 0.09 \, {\rm yr}$ system, showing that binary evolution can lead to N over-production. We find 0.5\% of systems in our binary star population produce [N/O] abundance ratios $> +1.2$, which is 0.1 dex higher than the maximum [N/O] achieved by the single stars. Additionally, $0.01\%$ of the systems in the binary population achieve an [N/O] abundance ratio over +2.2, which is 1 dex higher than achieved by single stars. Most of these N-enhanced binary systems have stars that enter the TP-AGB with over-massive envelopes relative to their core masses due to stellar wind accretion or a merger with a post-MS star. These relatively massive envelopes cause the star to shrink, slowing down mass loss and consequently allowing stars to spend longer in the hot-bottom burning phase than single stars of identical mass. We discussed this in more detail in Paper I.

\begin{figure}
\centering
\includegraphics[width=0.8\linewidth]{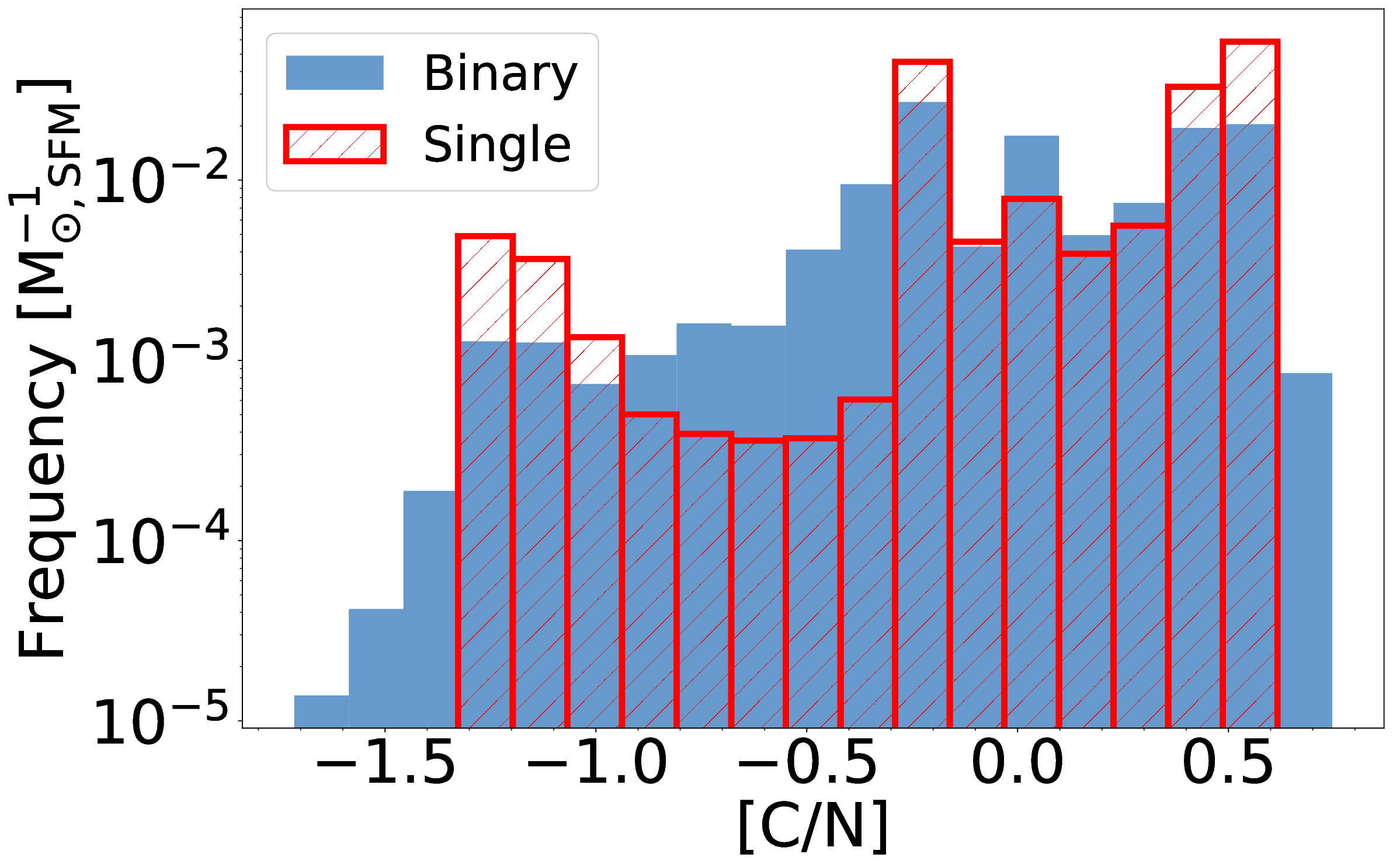}
\caption{As Figure \ref{fig:CO_Ratio}, but for [C/N].}
\label{fig:CN_Ratio}
\end{figure}

The ejected [C/N] distribution is shown in Figure \ref{fig:CN_Ratio}. The distributions from our binary- and single-star populations are similar. The minimum [C/N] is -1.7, from the $M_{\rm 1,0} = 7.61\Msun$, $M_{\rm 2,0} = 0.62\Msun$, and $p_{\rm 0} = 11.9 \, {\rm yr}$ binary system. This system evolves similarly to the $M_{\rm 1,0} = 8.23\Msun$, $M_{\rm 2,0} = 0.86\Msun$, and $p_{\rm 0} = 10.0 \, {\rm yr}$ case previously discussed. It experiences a common envelope event after the primary star experiences 25 thermal pulses, truncating the TP-AGB phase following the destruction of the carbon in its envelope. The secondary star does not evolve off the MS before the end of the simulation. The maximum [C/N] achieved by our binary systems is +0.75 from our $M_{\rm 1,0} = 1.38\Msun$, $M_{\rm 2,0} = 1.37\Msun$, and $p_{\rm 0} = 1 \, {\rm d}$. This system merges while the primary star crosses the HG and the secondary is on the MS, producing a $2.74\Msun$ star, which then evolves similarly to a single $2.74\Msun$ star but transports less N to the surface during the first dredge-up due to the star's under-massive core. 

\subsubsection{Weighted Population Yield and Binary Star Abundances of the $s$-process Elements}
Here, we present results for Sr, Ba, and Pb. These three elements are representative of the three $s$-process peaks at the magic neutron numbers 50 (Sr), 82 (Ba), and 126 (Pb). Similarly to our previous analysis of C, N, and O, we first look at how including binary stars alters the Sr, Ba, and Pb stellar yield compared with a stellar population composed of single-star systems only. We then examine the ejected [Ba/Fe] abundance ratios and discuss enhancements and depletion due to binary evolution.

\begin{figure*}
\centering
\includegraphics[width=0.8\linewidth]{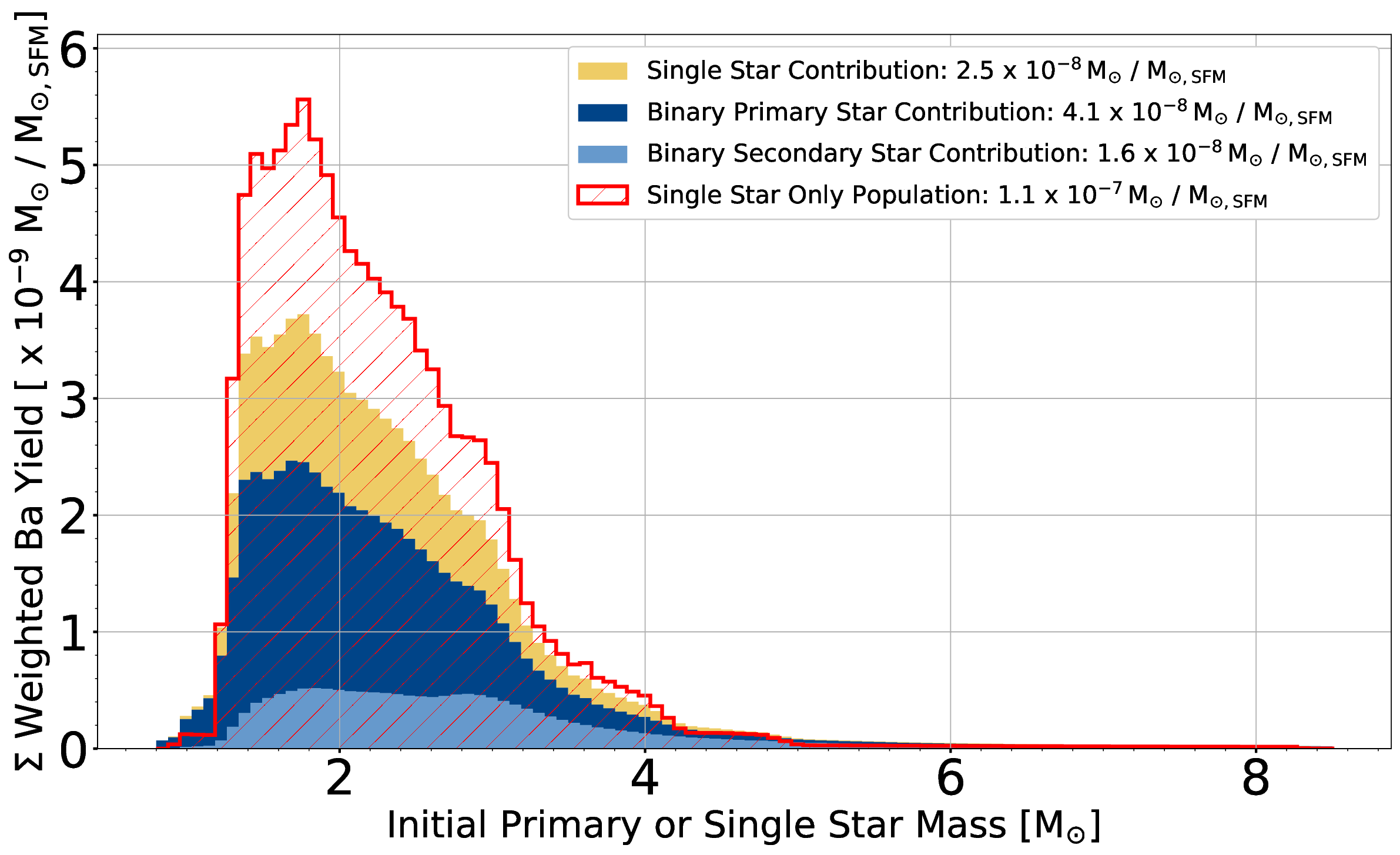}
\caption{As Figure \ref{fig:C_Weighted}, but for Ba. We find our population of only single stars ejects a weighted Ba stellar yield of $1.1 \times 10^{-7} \, \Msun/{\rm M_{\odot, SFM}}$. At a binary fraction of 0.7, our population yields $0.82 \times 10^{-7} \, \Msun/{\rm M_{\odot, SFM}}$ of Ba. From our population, including binaries, 30\% of the total ejected Ba originates from the single star portion of the population, 50\% from our binary primary stars, and 20\% from our binary secondaries.}
\label{fig:Ba_Weighted}
\end{figure*}

Table \ref{tab:PopYields_S} shows the weighted stellar yields of Sr, Ba, and Pb ejected by stellar populations with various binary fractions. Because we find that binary evolution decreases the population yield of all Sr, Ba, and Pb by about 30\%, hereafter, we will focus primarily on Ba. Figure \ref{fig:Ba_Weighted} shows the weighted population yield of Ba as a function of the initial primary and single star mass for the population of single stars only and the population with a binary fraction of 0.7. At this binary fraction, the population ejects 25\% less Ba than our population of single stars only. Most Ba under-production occurs in binary systems with initial primary masses of $\sim 1.2-4.2\Msun$. It is caused by binary evolution either truncating or preventing stars from entering the TP-AGB, similarly to the case of C.

\begin{table*}[hbt!]
\caption{Weighted Sr, Ba, and Pb stellar yield ejected by the low and intermediate-mass stellar populations of varying binary fractions. We also show the ratio of the yield produced including binaries divided by the population yield produced single stars only.}

\label{tab:PopYields_S}

\begin{tabular}{llllllllllll}
\toprule
\headrow Element & \multicolumn{11}{c} {Weighted population yield in units $\Msun/{\rm M_{\odot, SFM}}$ at binary fraction:} \\
\headrow & 0.0 & 0.1 & 0.2 & 0.3 & 0.4 & 0.5 & 0.6 & 0.7 & 0.8 & 0.9 & 1.0 \\
\midrule
Sr ($\times 10^{-7}$) & 2.3 & 2.2 & 2.1 & 2.0 & 1.9 & 1.9 & 1.8 & 1.7 & 1.7 & 1.6 & 1.6 \\
Ba ($\times 10^{-7}$) & 1.1 & 1.0 & 0.99 & 0.95 & 0.91 & 0.88 & 0.85 & 0.82 & 0.79 & 0.76 & 0.74 \\
Pb ($\times 10^{-8}$) & 5.3 & 5.1 & 4.9 & 4.7 & 4.6 & 4.4 & 4.3 & 4.1 & 4.0 & 3.9 & 3.8 \\

\headrow & \multicolumn{11}{c} {Weighted population yield ratio (population inc. binaries / population single stars only)} \\
Sr & 1.00 & 0.96 & 0.92 & 0.88 & 0.84 & 0.81 & 0.79 & 0.76 & 0.73 & 0.71 & 0.68 \\
Ba & 1.00 & 0.95 & 0.91 & 0.87 & 0.84 & 0.81 & 0.78 & 0.75 & 0.72 & 0.70 & 0.67 \\
Pb & 1.00 & 0.96 & 0.92 & 0.89 & 0.86 & 0.82 & 0.80 & 0.77 & 0.75 & 0.73 & 0.71  

\end{tabular}
\end{table*}

\begin{figure}
\centering
\includegraphics[width=0.8\linewidth]{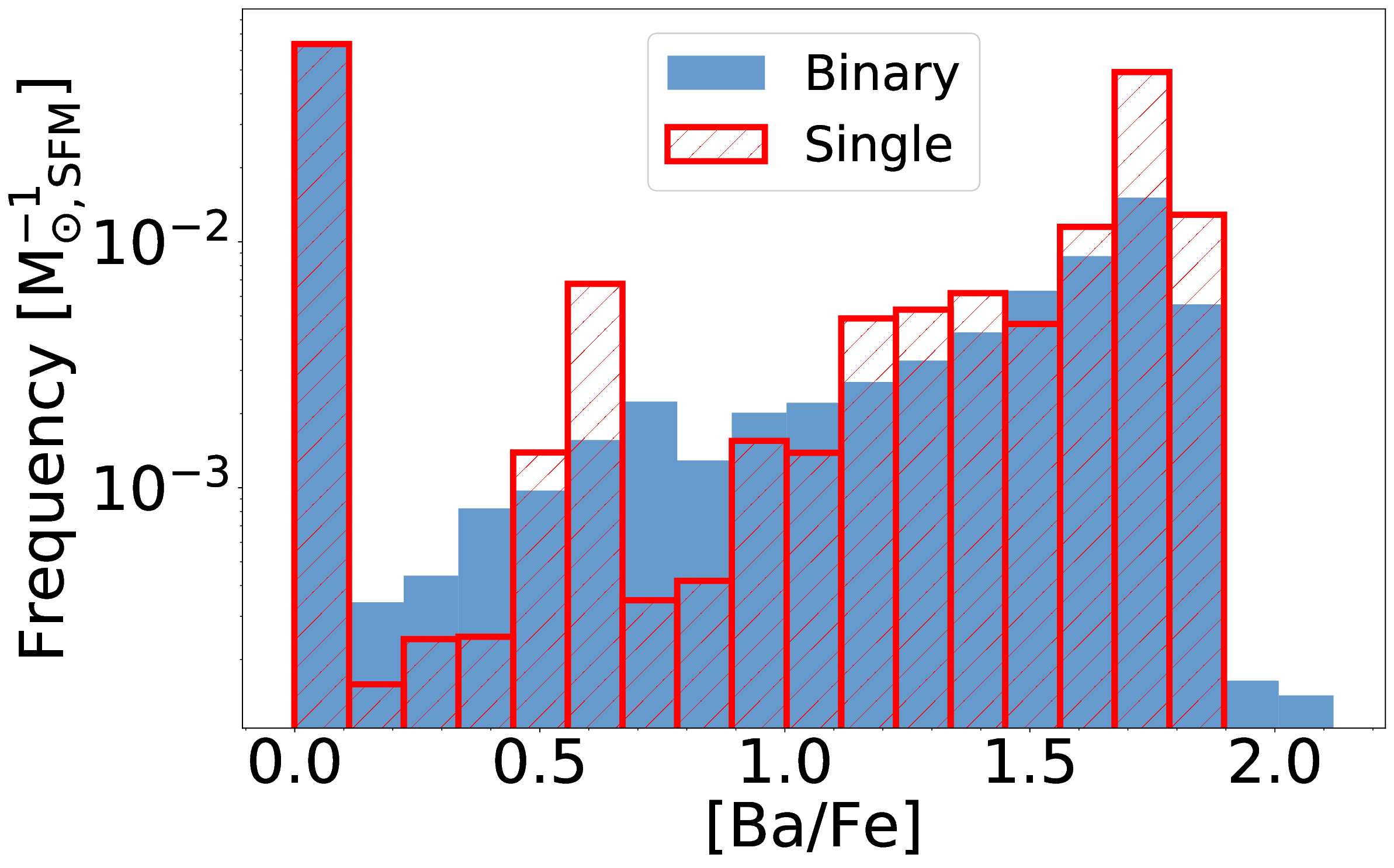}
\caption{As Figure \ref{fig:CO_Ratio}, but for [Ba/Fe]. We find that 0.2\% of the stars in our binary population have a [Ba/Fe] ratio > 1.9.}
\label{fig:BaFe_Ratio}
\end{figure}

Figure \ref{fig:BaFe_Ratio} shows the distribution (per ${\rm M_{\odot, SFM}}$) of the [Ba/Fe] abundance ratios in total material ejected into the interstellar medium by the binary- and single-star populations. As in Figures \ref{fig:CO_Ratio}, \ref{fig:NO_Ratio}, and \ref{fig:CN_Ratio}, the [Ba/Fe] for each stellar system was calculated using the total stellar yield of each system. The maximum [Ba/Fe] achieved by the single-star population is +1.8 at $1.78\Msun$. The maximum [Ba/Fe] achieved by the binary star population is +2.1 from the $M_{\rm 1,0} = 2.60\Msun$, $M_{\rm 2,0} = 1.52\Msun$, and $p_{\rm 0} = 0.03 \, {\rm yr}$ system. This system merges while the primary is on the GB and the secondary is on the MS. The merged star has a total mass of $3.75\Msun$ and a core mass of $0.55\Msun$ at its first thermal pulse. This system experiences 53 thermal pulses before transitioning into a CO-WD, which is 28 more thermal pulses and 31 more third dredge-up events than the corresponding single $3.75\Msun$ star. The extra third dredge-up events allow more Ba to be synthesised, transported to the stellar surface, and then ejected by stellar winds.

\subsection{Supernovae and Black Holes}
Mergers and mass transfer can lead to the initially low- and intermediate-mass stars gaining sufficient mass ($\gtrsim 8.3\Msun$) to end their lives via supernovae. Only 1\% of the systems in the binary population experience at least one supernova. These supernovae are mostly Type-II core collapse \citep{Limongi2018}, and Type Ib and c stripped core collapse supernovae \citep{Yoon2010}.

An interesting result from our models is the formation of black holes. Black holes are typically associated with stars of at least $20-25\Msun$ \citep{Fryer1999, Hegar2023}, yet our binary systems can only have a combined maximum mass of $17\Msun$. Within our binary star population, accretion onto, or a merger with, a neutron star forms five black holes with every $10^6\Msun$ of star-forming material (there are 321 in our grid of 640 000 models). The least massive system to form a black hole this way has the initial conditions of $M_{\rm 1} = 5.22\Msun$, $M_{\rm 2} = 5.00\Msun$, and $p_{\rm 0} = 58$ days. Stable mass transfer onto the secondary star allows it to gain sufficient mass to explode in a core-collapse supernova forming a neutron star remnant. The neutron star later collapses into a $2.24\Msun$ black hole after it merges with the CO-WD primary. However, the formation of the black hole is dependent on what is defined as the maximum neutron star mass, which we set to $2.2\Msun$ \citep{Kalogera1996, Fan2024}. 

Observations support the existence of compact objects of mass $\sim 2-5\Msun$ \citep{Abbot2020, Wyrzykowski2020}, however, their origins remain uncertain. Since our stellar population was not set up to study black hole formation, further study is required to determine the significance of low- and intermediate-mass binary systems as black hole progenitors. 

\section{Comparison to Barium Stars}
\label{sec:BaStars}

We now extend our analysis to compare our models to observations of Galactic Ba stars. Here, we use our stellar-grid of 640 000 binary stellar models as described in Section \ref{sec:popSynth}.

Barium stars are giant stars with enriched surface $s$-process abundances despite not evolving to the AGB \citep{McClure1983, Jorissen2019}. They gained their $s$-process enrichment extrinsically via mass transfer from a TP-AGB companion. In this way, Ba stars preserve the $s$-process elements from the TP-AGB companion, allowing us to calculate the $s$-process production of the progenitor TP-AGB star \citep{denHartogh2023}. We can use the observed abundances from Ba stars to test how well our models align with observations.

We identify G and K giants in our modelled binary population as in \citet{Izzard2010}. We sample the surface abundances of the secondary stars at the end of the donor star's AGB, maximising the $s$-process surface abundances of the Ba star. Since Ba lines are too strong in Ba star spectra for reliable abundance measurements, often other $s$-process peak elements are used as proxies for Ba, such as Y and La (for example, see \citealp{DeCastro2016} and \citealp{Roriz2024}). Following these authors, we define a Ba star to have an average surface abundance ratio of [Y/Fe], [La/Fe], [Ce/Fe], and [Nd/Fe] > +0.25, where we calculate the average after calculating the solar-scaled surface abundance ratios using Eq. \ref{eq:logRatio}. We define mild Ba stars as having an average $s$-process surface abundance ratio calculated between +0.25 and +1.0 dex. Strong Ba stars are defined to have an average $s$-process surface abundance ratio [s/Fe] > +1.0 dex.

We use the Ba star sample presented in \citet{Cseh2018} (derived from \citealp{DeCastro2016}) to compare to the [Ce/Y] and [Fe/H] surface abundances calculated by our models. From this sample, we only use the 75 stars with [Fe/H] of $0.00$ to $+0.05$ dex (including error bars). Additionally, we use the 12 Ba stars presented in \citet{Jorissen2019} with an [Fe/H] from -0.1 to +0.1 dex to compare the stellar masses and orbital periods of our predicted Ba star systems. For our estimate of the number of Ba stars presently in the Milky Way, we assume $10\,$Gyr of star formation at a constant star formation rate of $2\Msun/\text{yr}$ at solar-metallicity and a binary fraction of 0.7.

We find approximately 8 200 Ba stars progenitors from every $10^6\Msun$ of binary-star-forming material, and we estimate a total of $3.6 \times 10^6$ Ba stars at solar-metallicity currently in the Milky Way. 

\begin{figure*}
\centering
\includegraphics[width=0.8\linewidth]{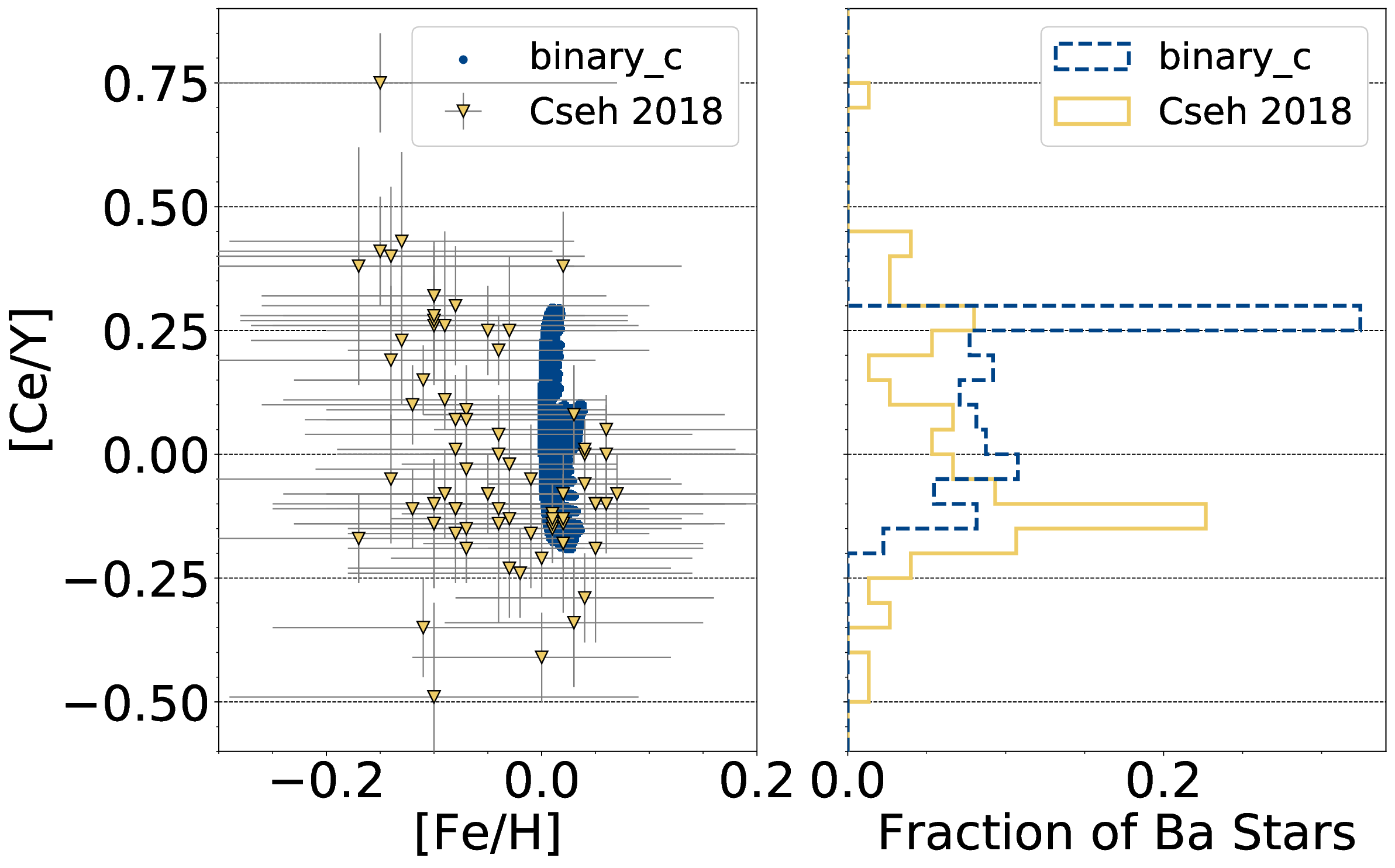}
\caption{[Ce/Y] and [Fe/H] (left) surface abundances and the [Ce/Y] distribution (right) of the predicted Ba stars compared to observed Ba stars reported in \citet{Cseh2018}. The plots share the same y-axis.}
\label{fig:CsehFreq}
\end{figure*}

Our comparison of the [Ce/Y] surface abundances to the predicted Ba stars to the data from \citet{Cseh2018} in Figure \ref{fig:CsehFreq}. The maximum predicted [Ce/Y] abundance ratio is +0.30, and the minimum is $-0.19$. Of the predicted Ba stars $40\%$ have [Ce/Y] > $+0.2$, and $60\%$ have -0.2 < [Ce/Y] < +0.2. This distribution favours Ba stars of higher surface [Ce/Y] abundance ratio than the observed distribution, which has 24\% with [Ce/Y] > +0.2, 73\% with -0.2 < [Ce/Y] < +0.4, and 3\% with [Ce/Y] < -0.4, not taking observational error into account. Restricting the star-formation time to $5\,$Gyr results in $36\%$ of Ba Stars with [Ce/Y] > $+0.2$, and $64\%$ having -0.2 < [Ce/Y] < +0.2. Additionally, because the observed minimum of [Ce/Y] $= -0.3$, we cannot reproduce these 3\% of Ba stars within $1\sigma$ of observational errors.

\begin{figure}[t!]
\centering
\includegraphics[width=0.8\linewidth]{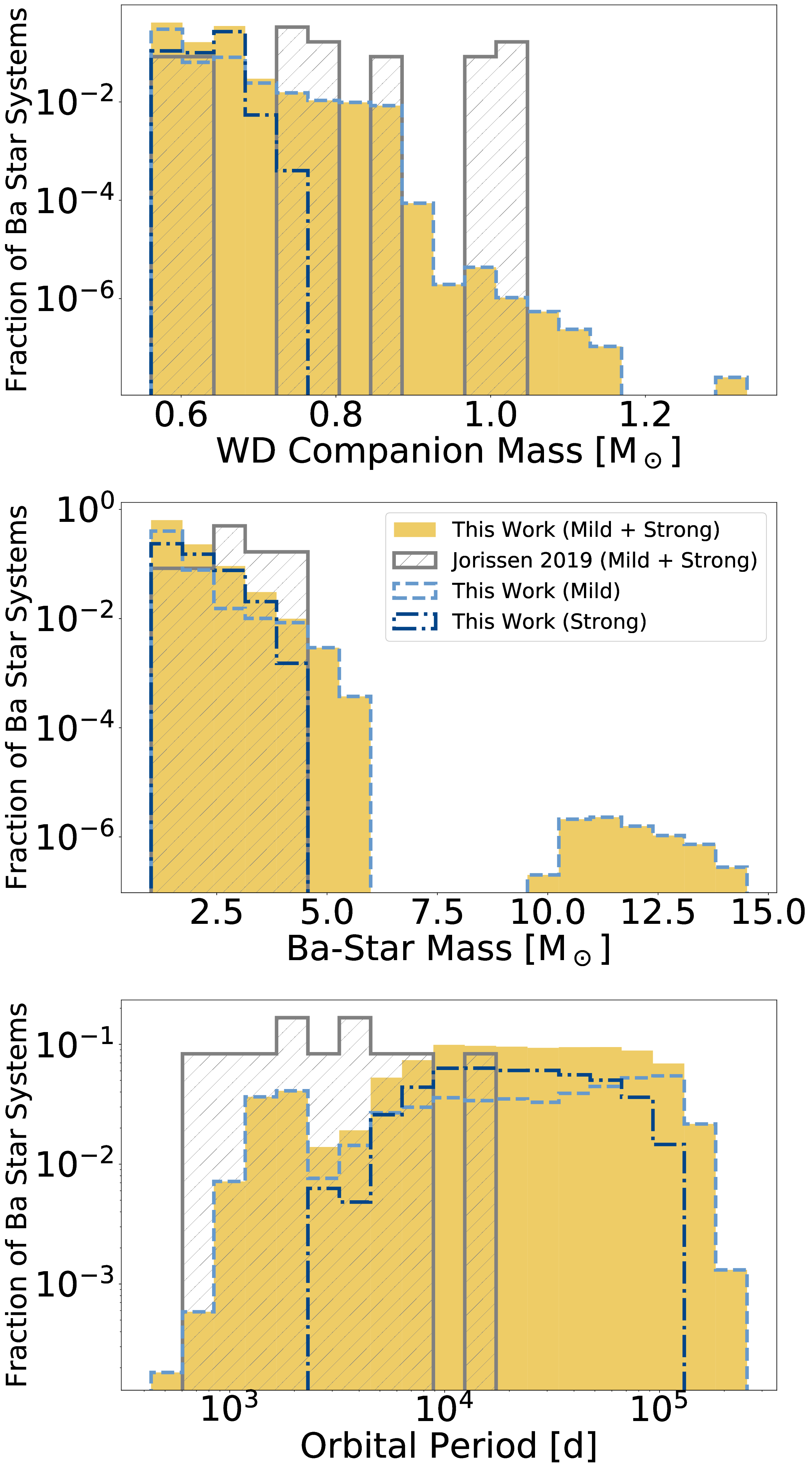}
\caption{Distributions of WD masses (top), Ba star masses (middle) and orbital periods (bottom) for predicted Ba star systems compared to observations from \citet{Jorissen2019}. The legend for all three panels is presented in the middle panel.}
\label{fig:BaProp}
\end{figure}

The masses, orbital periods, and frequencies of predicted Ba stars within our binary population are shown in Figure \ref{fig:BaProp}, compared to observations of solar-metallicity Ba stars reported in \citet{Jorissen2019}. From our predicted Ba systems, the average WD mass is $0.63\Msun$, the average Ba star mass is $1.7\Msun$, and the average orbital period is $3.6\times10^{4}$ days. \citet{Jorissen2019}, on average, observe more massive WDs and Ba stars at solar-metallicity than predicted with $0.81\Msun$ and $2.9\Msun$, respectively. The predicted orbital periods are longer than observed, with the maximum orbital period calculated in \citet{Jorissen2019} to be $1.7\times10^4$ days. Our models also estimate that 48\% observed solar-metallicity Ba stars are strong Ba stars, but there are only 2 in the sample of 12 from \citet{Jorissen2019}. 

The distribution in Figure \ref{fig:BaProp} (middle) shows the formation of a small number (0.08\% of all Ba stars) of massive Ba stars with mass $\gtrsim10\Msun$, which is not observed in \citet{Jorissen2019}. These Ba stars originate from intermediate-mass binary systems with an initial primary star mass $\gtrsim 6\Msun$ and $M_{\rm 2,0}/M_{\rm 1,0} \approx 1$. Stable mass transfer from the primary to the secondary results in a TP-AGB primary star of mass $1.71-2.25\Msun$ with a massive MS companion secondary star. Their primary stars enter the TP-AGB with massive cores compared to single stars of identical mass and experience an elevated number of thermal pulses. None of these systems experience a common envelope event. The resulting Ba star systems have WD masses between $0.95-1.33\Msun$, Ba star masses $10.8-14.9\Msun$, and orbital periods of $2971-8331$ days. These massive Ba stars fall under the mild Ba star category and are observable as Ba stars for about 1 to 3 Myr (6-16 Myr if you include their lifetime while $s$-process enriched on the MS). We estimate approximately 30 massive Ba stars are presently in the Milky Way. In this context, it is unsurprising that none have been observed, regardless of whether the channel truly occurs in nature.

\section{Discussion}
\label{sec:Discussion}
Here, we discuss the uncertainty in our model parameters and the uncertainty the Ba stars models exposed.

\subsection{Model Uncertainty}
\label{sec:ModelUncertainty}
\begin{figure*}
\centering
\includegraphics[width=0.8\linewidth]{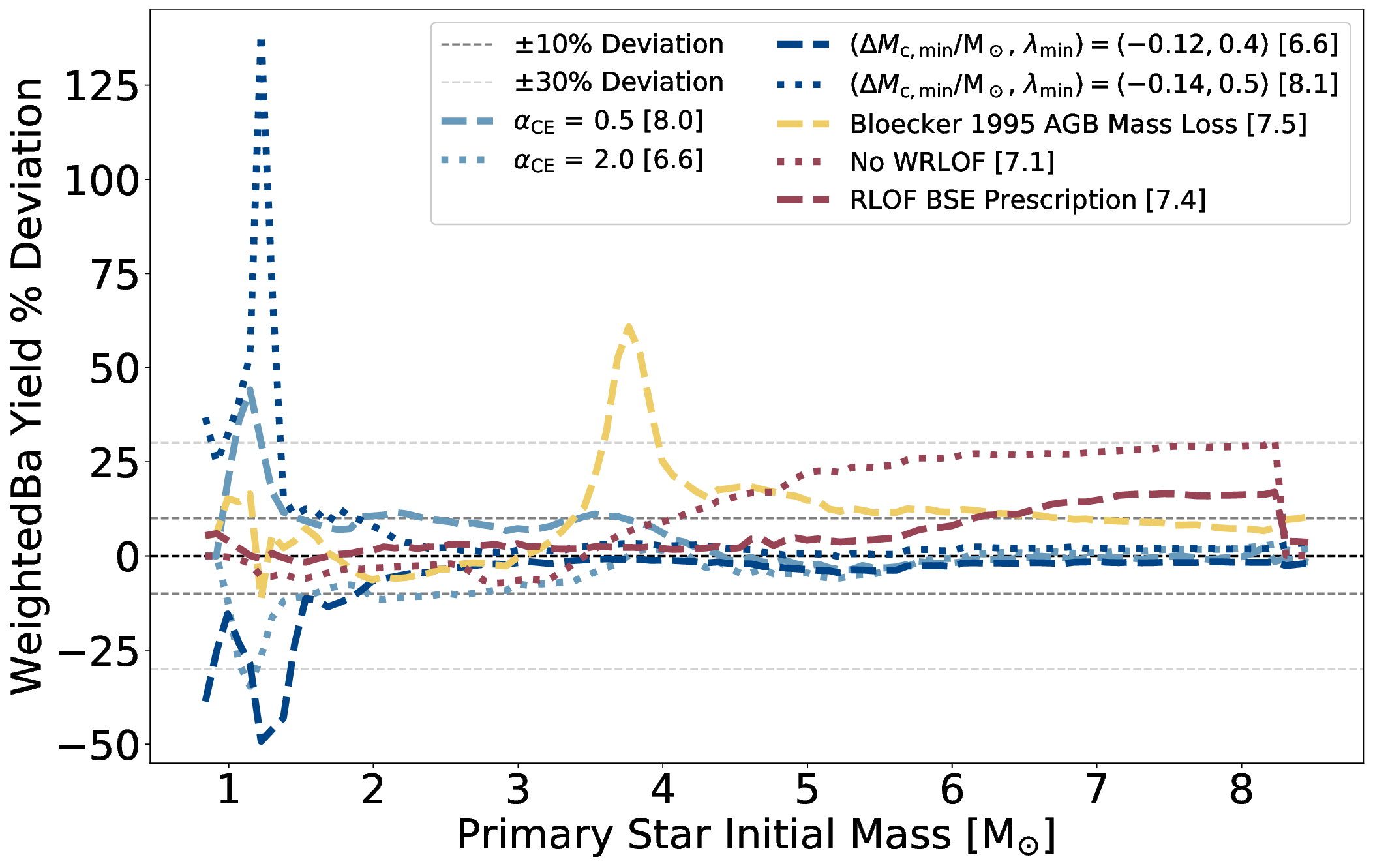}
\caption{Percentage difference in the Ba ejected by our binary star population (binary fraction is 1.0) with varying common envelope and third dredge-up parameters, mass-loss rates, the Roche-lobe overflow prescription (changed to \citealp{Hurley2002}, and notated as RLOF BSE prescription), and the wind Roche-lobe overflow (WRLOF) compared to the population calculated using the model parameters described in Table \ref{tab:parameters}. The zero-line corresponds to our models set up as shown in Table \ref{tab:parameters} where $(\Delta M_{\rm c, min}/\Msun, \lambda_{\rm min}) = ( -0.13,  0.45)$,  $\alpha_{\rm CE} = 1$, mass-loss on the TP-AGB as described in \citet{Vassiliadis1993}, Roche-lobe overflow as described in \citet{Claeys2014}, and wind Roche-lobe overflow as described in \citet{Abate2013}. The population calculated using our chosen model parameters has a total weighted Ba yield of $7.4\times 10^{-8} \,\Msun/{\rm M_{\odot, SFM}}$. In the legend, the quantities in square brackets describe the total weighted Ba yield in units $\times 10^{-8}\, \Msun/{\rm M_{\odot, SFM}}$ for each population.}
\label{fig:Ba_uncert}
\end{figure*}

AGB and binary evolution have major uncertainties related to mixing and convective boundaries, stellar winds, mass transfer, common envelope evolution, and $s$-process nucleosynthesis. To discern their impact, we computed stellar populations of binary stars calculated with grids of 640 000 stellar models ($M_{\rm 1}: 100 \times M_{\rm 2}:80 \times p:80$), varying the model parameters: common envelope efficiency $\alpha_{\rm CE}$, third dredge-up parameters $\Delta M_{\rm c, min}$ and $\lambda_{\rm min}$, TP-AGB mass-loss prescription, Roche-lobe overflow prescription, and wind Roche-Lobe overflow prescription, and compare the weighted C, N, and Ba yield to our stellar population as described in Table \ref{tab:parameters}. Our discussion will primarily focus on Ba as it shows the largest deviations. Figure \ref{fig:Ba_uncert} shows the results of the variation of these model parameters on the weighted Ba yield. 

The common envelope parameter $\alpha_{\rm CE}$ describes the fraction of orbital energy released by the stellar companion to eject the envelope of a common envelope system \citep{Hurley2002}. The higher $\alpha_{\rm CE}$, the easier it is to eject the envelope in a common envelope event. In Figure \ref{fig:Ba_uncert}, we compare the changes in the weighted Ba yield from our binary population calculated with $\alpha_{\rm CE} = 0.5$ and $2$ instead of $\alpha_{\rm CE} = 1$. $\alpha_{\rm CE}$ is among the most influential parameters influencing the Ba yield as taking $\alpha_{\rm CE} = 0.5$ results in 10\% more merger and TP-AGB progenitor systems and $\alpha_{\rm CE} = 2$ results in $16\%$ fewer mergers 8\% fewer TP-AGB progenitors than taking $\alpha_{\rm CE} = 1$. Altering $\alpha_{\rm CE}$ to 0.5 and 2 changes the total Ba yield from the binary population by $\pm 10\%$. Figure \ref{fig:Ba_uncert} shows the largest deviation in the mass range of about $1-1.2\Msun$, where mergers are essential to Ba production as these stars do not have sufficient mass to experience the third dredge-up or $s$-process nucleosynthesis when single. C and N increase by 7\% when $\alpha_{\rm CE} = 0.5$, and they decrease by 6\% and 3\% respectively when $\alpha_{\rm CE} = 2$.

When we force the third dredge-up to begin earlier and with higher minimum efficiency, $(\Delta M_{\rm c, min}/\Msun, \lambda_{\rm min}) = ( -0.14,  0.50)$, or later and with lower minimum efficiency, $(\Delta M_{\rm c, min}/\Msun, \lambda_{\rm min}) = ( -0.12,  0.4)$, than our calibration described in Section \ref{sec:3DUP_Results}, we find that stars of mass $\lesssim2\Msun$, which eject approximately 80\% of the Ba from the binary star population, have the largest deviations in the Ba yield. The largest is a 140\% deviation at $1.2\Msun$, near the low-mass boundary for the third dredge-up. The total Ba yielded by the binary populations varies by $\pm 10\%$. C and N vary by $\pm5\%$ and $-0.3\%$, respectively. The yields from the single-star populations deviate similarly. We find that the C, N and Ba yield from our binary population change by factors of 0.76, 0.93, and 0.67, respectively, with respect to their corresponding single-star populations, as in Tables \ref{tab:PopYields_CNO} and \ref{tab:PopYields_S}, for all treatments of third dredge-up tested here.

The mass-loss rate of AGB stars is a major source of uncertainty in AGB star models \citep{Karakas2016, Hofner2018}. Figure \ref{fig:Ba_uncert} shows the change in the results using the mass loss prescription described in \citet{Bloecker1995} with $\eta = 0.02$ as in \citet{Ventura2018} and \citet{Lopez2022}. This stellar wind prescription introduces a deviation of up to 60\% in the Ba yield at $3.76\Msun$, compared to models with the mass-loss described in Table \ref{tab:parameters}. This is because the single $3.76\Msun$ star experiences 24 third dredge-up events with the \citet{Bloecker1995} mass loss prescription and 20 third dredge-up events with the \citet{Vassiliadis1993} mass loss prescription. The overall C, N, and Ba weighted yields from the binary population change by $<3\%$. Additionally, when comparing our binary population yield to our single-star population yield, the yields from our binary populations are factors of 0.75, 0.92, and 0.67 of the single star population yield for C, N, and Ba respectively, which is almost identical to the result presented in Tables \ref{tab:PopYields_CNO} and \ref{tab:PopYields_S} when using winds from \citet{Vassiliadis1993}.

We investigate the effects of changing the Roche-lobe overflow prescription from the prescription described in \citet{Claeys2014} to the one described in \citet{Hurley2002}. This changes how dynamically stable mass transfer is treated during Roche-lobe overflow. In \citet{Hurley2002}, the mass transfer rate is calculated as a function of the amount the Roche-lobe is overfilled (see their Equation 58). The prescription described in \citet{Claeys2014} has an additional dependence on the stability of mass transfer (see their Equation 10), which increases the mass transfer rate on the thermal timescale. Using the Roche-lobe overflow prescription described in \citet{Hurley2002} increases the Ba yield by a maximum of 18\% at a primary mass of $7.6\Msun$. It mainly influences binary systems with initial primary mass $\gtrsim 5\Msun$ where the secondary stars are ejecting most of the Ba. The overall change in the total weighted yield for C, N, and Ba varies by $<3\%$.

Wind Roche-lobe overflow is the efficient mass transfer of stellar wind material gravitationally focused toward the accreting star \citep{Mohamed2007}. Our models use the wind Roche-lobe overflow prescription described in \citet{Abate2013}. Switching off wind Roche-lobe overflow results in all stellar wind accretion rates calculated from \citet{Bondi1944}. Switching off wind Roche-lobe overflow changes the Ba population yield by a maximum of 30\% at a primary mass of $8.23\Msun$ as fewer secondary stars accrete sufficient mass for hot-bottom burning, allowing more Ba production. For C and Ba, the variation in the total weighted yield of the population is minimal at $< 2\%$; however, for N, we find a $7\%$ decrease.

We do not study the contribution of supernovae to the stellar yield, but from the populations discussed here, we find that using $\alpha_{\rm CE} = 2$ has the largest influence on their number with a $15\%$ reduction as fewer mergers result in stars with masses $\gtrsim 8.3\Msun$. This is followed by using the Roche-lobe overflow prescription described in \citet{Hurley2002}, which reduces the number of supernovae in our population by $10\%$ originating primarily from the secondary stars as they accrete less material than when using the Roche-lobe overflow prescription described in \citet{Claeys2014}. All other variations to our model population change the number of supernovae by $<10\%$.

In summary, out of all the sources of uncertainty we investigate here, our settings for $\alpha_{\rm CE}$ and the third dredge-up introduce the highest overall uncertainty in the Ba yield from our binary population with variations of $\pm 10\%$ each. In the case of C, we find that the $\alpha_{\rm CE}$ parameter introduces the highest uncertainty, with the C yield from our binary population decreasing by 7\% when we set $\alpha_{\rm CE} = 2.0$. For N, wind Roche-lobe overflow has the most influence. We also find that although the third dredge-up treatment and the stellar winds on the TP-AGB alter the total stellar yields, they have little influence on the deviation between the yields from our single-star and binary-star populations, and it is the binary interaction which introduces the most uncertainty in this respect. Other sources of uncertainty we do not explicitly test include convective energy transport in the stellar envelope (which influences dredge-up and hot-bottom burning efficiency, for example, see \citealp{Boothroyd1988} or \citealp{Sackmann1991}), the mixing profile of the partial mixing zone \citep{Buntain2017}, and \iso{13}C pocket size.

\subsection{Barium Star Orbital Periods and Abundances}
\label{sec:BaStar_Discuss}
As described in Section \ref{sec:BaStars}, most predicted Galactic Ba stars have surface [Ce/Y] in agreement with observations (Figure \ref{fig:CsehFreq}). However, Figure \ref{fig:CsehFreq} shows we are predicting a higher fraction of Ba stars with [Ce/Y] > +0.2 than observed. 63\% of these systems had AGB companions with initial masses $<1.5\Msun$, outside the mass range of the K16 models with $s$-process stellar nucleosynthesis. Due to their colder intershells, a star of $1.2\Msun$ may experience \iso{13}C burning convectively instead of radiatively, which results in fewer free neutrons and neutron-capture reactions compared to the $1.5\Msun$ star \citepalias{Cristallo2009_2, Lugaro2012}. The K16 models do not produce \iso{13}C pockets at this mass, and therefore, the fit for \textsc{binary\_c} does not capture this. 

Additionally, the temperature of the He intershell is not calculated in \textsc{binary\_c}. An update to \textsc{binary\_c} to include He intershell temperatures and the contribution of convective \iso{13}C burning would require new fits to detailed stellar models with convective \iso{13}C burning or the use of an interpolation-based single-stellar evolution module such as METTISE \citep{Agrawal2020} or MINT \citep{Mirouh2023}; neither, at the time of publishing this work, are developed for AGB stars.

Our Ba star models also disagree in their orbital periods compared to the Ba stars observed in \citet{Jorissen2019}. The reproduction of observed orbital periods (and eccentricity, which we do not explore in this work) of Ba star systems is a known issue in binary population synthesis \citep{Bonavciv2004, Izzard2010}. Discussions in the literature surrounding the periods and eccentricity of binary systems highlight the common envelope and the origin of the energy used to eject the envelope \citep{Bonavciv2004}, the strength of tidal forces \citep{Karakas2000}, white dwarf kicks \citep{Izzard2010}, and the influence of circumbinary discs \citep{Rafikov2016, Izzard2023}. While we do not address them in this study, these mechanisms can alter the binary system's behaviour and possibly influence stellar yield, especially should the increased eccentricity allow stars to interact. 

Observations of AGB stars at solar-metallicity often have [Ba/Fe] $\lesssim$ 1.0 dex \citep{Busso2001, Abia2002, Jorissen2019}, however Figures \ref{fig:BaFe_Ratio} and \ref{fig:BaProp} suggest we should be observing many stars with [Ba/Fe] $> 1.0$ dex, although this depends on size of the partial mixing zone and the point in time of stellar evolution where the observations are recorded. Our treatment of the third dredge-up, the size of the partial mixing zone, and the number of free neutrons control the $s$-process abundances and yield. We discuss the uncertainty in our treatment of the third dredge-up in Sections \ref{sec:3DUP_Results} and \ref{sec:ModelUncertainty}. K16 discussed the uncertainties related to the partial mixing zone, especially in the $4-5\Msun$ mass range. The uncertainty in this mass range arises due to the scarcity of observational constraints and the potential for partial hot-bottom burning and hot third dredge-up \citep{Goriely2004}. All these uncertainties are inherited by the modified version of \textsc{binary\_c} and are further exacerbated in the binary models. It is even more challenging to estimate the mass of the partial mixing zone in a star that has lost or gained mass relative to a single star. For example, our models predict $\sim 30$ Galactic Ba stars of mass $\gtrsim 10\Msun$, which form from intermediate-mass binary systems (Figure \ref{fig:BaProp}). The polluting AGB stars have massive cores compared to single stars of identical mass and experience an elevated number of thermal pulses. \textsc{Binary\_c} estimates the size of \iso{13}C pocket based on the total mass of the star, which might not be appropriate for this scenario. \citet{Jorissen2019} find WD companions of solar-metallicity Ba stars of mass up to about $1\Msun$, suggesting its progenitor star was born with mass up to $7-8\Msun$, assuming a solar-metallicity progenitor \citep{Karakas2014_2}. This indicates that partially stripped intermediate-mass stars can synthesize $s$-process elements. 

\section{Conclusions}
\label{sec:Conclusion}
Binary stellar evolution is observed to shape the evolution of their stars, as shown by blue stragglers, Ba stars, and the shapes of planetary nebulae \citep{DeMarco2009, Jones2017}. We use a modified version of the binary population synthesis code \textsc{binary\_c} to explore how binary evolution shapes the evolution of low- and intermediate-mass stars and their stellar yield at solar-metallicity. We update how \textsc{binary\_c} handles the $s$-process by fitting the intershell abundances based on the models presented in \citet{Karakas2016} and coupling the production of $s$-process elements to the third dredge-up. We also calibrate the third dredge-up based on the Galactic carbon-star luminosity function presented in \citet{Abia2022}.

We evolve a grid of 640 000 low- and intermediate-mass binary systems and weighted them based on their birth probability to build theoretical stellar populations. We find that 60\% of our binary population has at least one star entering the TP-AGB over the life of the simulation, a reduction compared to the 78\% of single stars entering the TP-AGB. The consequence is that the third dredge-up, active only during the TP-AGB, is required to transport C and the products of the $s$-process to the stellar surface. In a low- and intermediate-mass stellar population with a binary fraction of 0.7, we find a $20-25$\% reduction in the C, Sr, Ba, and Pb yields compared to a population of single stars only. In contrast, binary evolution has little influence on the N and O yielded by our stellar population (excluding novae and supernovae). However, we find rare cases where binary evolution leads to abundance ratios calculated from the stellar yields of [N/O], [C/O], and [Ba/Fe], with the largest deviations in the [N/O] ratios, as shown in Figure \ref{fig:NO_Ratio}. 

Within the binary population, $\sim$8 200 Ba stars form from every $10^6\Msun$ of binary-star-forming material. Comparing the surface abundances of our Ba star models to the observed Ba stars, we find that the [Ce/Y] abundances are in agreement. However, we predict a higher fraction of Ba stars with [Ce/Y] > +0.2 dex than observed. Additionally, the [Ba/Fe] abundances ejected by the single and binary stellar populations show relatively high frequencies of stars with [Ba/Fe] > +1.5 dex, which are not observed. This suggests our models are over-efficient in producing $s$-process elements.

Our treatment of the third dredge-up, the common envelope parameter $\alpha_{\rm CE}$, and the efficiency of stellar wind accretion all introduce considerable uncertainty in our stellar yields. This is most apparent in our Ba yield, where our treatment of the third dredge-up introduces a similar uncertainty to the uncertainty introduced by binary evolution. However, when comparing the change in the stellar yield between the single and binary-star populations, the treatment of the third dredge-up has little impact.

Our updated He intershell abundance table includes 328 isotopes and elements up to and including \iso{210}Po. Combined with our updates presented in Paper I, this introduces a wealth of potential new data to analyse. For example, the abundances of Galactic planetary nebulae reflect only the abundances of the stellar envelopes immediately before ejection \citep{Werner2006}, but many of them have observed abundances that are not sufficiently explained by single star models \citep{Pottasch2010, Henry2018, Wesson2018}. We will use our models to attempt to explain the evolutions of chemically peculiar Galactic planetary nebulae. In this work, we did not discuss the elements Na, F, Al, Mg, or their isotopes. These elements can be synthesized in AGB \citep{Cristallo2011, Karakas2016} and super-AGB ($M\gtrsim7\Msun$) stars \citep{Siess2010, Doherty2015}. Previous studies of these elements mainly focused on globular cluster anomalies \citep{Siess2010,Ventura2011,Doherty2014}, and they did not consider AGB binary influence. Therefore, Na, F, Al, and Mg should be explored in future work, focusing on metal-poor systems.

\begin{acknowledgement}
We acknowledge C. Abia for kindly sharing his AGB CSLF data from \citet{Abia2022}. Thank you to the anonymous referee for their helpful comments.
\end{acknowledgement}

\paragraph{Funding Statement}

ZO acknowledges this research was supported by an Australian Government Research Training Program (RTP) Scholarship and the ASA Student Travel Assistance Scheme. AIK, DK, and ZO were supported by the Australian Research Council Centre of Excellence for All Sky Astrophysics in 3 Dimensions (ASTRO 3D) through project number CE170100013. RGI is funded by STFC grants ST/Y002350/1, ST/L003910/1, and ST/R000603/1 as part of the BRIDGCE UK network. DK acknowledges funding support from the Australian Research Council Discovery Project DP240101150. ML was supported by the Lend\"ulet Program LP2023-10 of the Hungarian Academy of Sciences and the NKFIH excellence grant TKP2021-NKTA-64. ZO and ML also thank the EU ChETEC-INFRA project (G. A. no. 101008324).

\paragraph{Competing Interests}
None

\paragraph{Data Availability Statement}
Data can be made available upon reasonable request to the corresponding authors.

\printendnotes

\printbibliography

@ARTICLE{Abate2013,
       author = {{Abate}, C. and {Pols}, O.~R. and {Izzard}, R.~G. and {Mohamed}, S.~S. and {de Mink}, S.~E.},
        title = "{Wind Roche-lobe overflow: Application to carbon-enhanced metal-poor stars}",
      journal = {A\&A},
         year = 2013,
        month = apr,
       volume = {552},
          eid = {A26},
        pages = {A26},
          doi = {10.1051/0004-6361/201220007},
archivePrefix = {arXiv},
       eprint = {1302.4441},
 primaryClass = {astro-ph.SR},
       adsurl = {https://ui.adsabs.harvard.edu/abs/2013A&A...552A..26A}
}

@ARTICLE{Abate2015,
       author = {{Abate}, C. and {Pols}, O.~R. and {Karakas}, A.~I. and {Izzard}, R.~G.},
        title = "{Carbon-enhanced metal-poor stars: a window on AGB nucleosynthesis and binary evolution. I. Detailed analysis of 15 binary stars with known orbital periods}",
      journal = {A\&A},
         year = 2015,
        month = apr,
       volume = {576},
          eid = {A118},
        pages = {A118},
          doi = {10.1051/0004-6361/201424739}
}

@ARTICLE{Abbot2020,
       author = {{Abbott}, R. and {Abbott}, T.~D. and {Abraham}, S. and {Acernese}, F. and {Ackley}, K. and {Adams}, C. and {Adhikari}, R.~X. and {Adya}, V.~B. and {Affeldt}, C. and {Agathos}, M. and {Agatsuma}, K. and {Aggarwal}, N. and {Aguiar}, O.~D. and {Aich}, A. and {Aiello}, L. and {Ain}, A. and {Ajith}, P. and {Akcay}, S. and {Allen}, G. and {Allocca}, A. and {Altin}, P.~A. and {Amato}, A. and {Anand}, S. and {Ananyeva}, A. and {Anderson}, S.~B. and {Anderson}, W.~G. and {Angelova}, S.~V. and {Ansoldi}, S. and {Antier}, S. and {Appert}, S. and {Arai}, K. and {Araya}, M.~C. and {Areeda}, J.~S. and {Ar{\`e}ne}, M. and {Arnaud}, N. and {Aronson}, S.~M. and {Arun}, K.~G. and {Asali}, Y. and {Ascenzi}, S. and {Ashton}, G. and {Aston}, S.~M. and {Astone}, P. and {Aubin}, F. and {Aufmuth}, P. and {AultONeal}, K. and {Austin}, C. and {Avendano}, V. and {Babak}, S. and {Bacon}, P. and {Badaracco}, F. and {Bader}, M.~K.~M. and {Bae}, S. and {Baer}, A.~M. and {Baird}, J. and {Baldaccini}, F. and {Ballardin}, G. and {Ballmer}, S.~W. and {Bals}, A. and {Balsamo}, A. and {Baltus}, G. and {Banagiri}, S. and {Bankar}, D. and {Bankar}, R.~S. and {Barayoga}, J.~C. and {Barbieri}, C. and {Barish}, B.~C. and {Barker}, D. and {Barkett}, K. and {Barneo}, P. and {Barone}, F. and {Barr}, B. and {Barsotti}, L. and {Barsuglia}, M. and {Barta}, D. and {Bartlett}, J. and {Bartos}, I. and {Bassiri}, R. and {Basti}, A. and {Bawaj}, M. and {Bayley}, J.~C. and {Bazzan}, M. and {B{\'e}csy}, B. and {Bejger}, M. and {Belahcene}, I. and {Bell}, A.~S. and {Beniwal}, D. and {Benjamin}, M.~G. and {Benkel}, R. and {Bentley}, J.~D. and {Bergamin}, F. and {Berger}, B.~K. and {Bergmann}, G. and {Bernuzzi}, S. and {Berry}, C.~P.~L. and {Bersanetti}, D. and {Bertolini}, A. and {Betzwieser}, J. and {Bhandare}, R. and {Bhandari}, A.~V. and {Bidler}, J. and {Biggs}, E. and {Bilenko}, I.~A. and {Billingsley}, G. and {Birney}, R. and {Birnholtz}, O. and {Biscans}, S. and {Bischi}, M. and {Biscoveanu}, S. and {Bisht}, A. and {Bissenbayeva}, G. and {Bitossi}, M. and {Bizouard}, M.~A. and {Blackburn}, J.~K. and {Blackman}, J. and {Blair}, C.~D. and {Blair}, D.~G. and {Blair}, R.~M. and {Bobba}, F. and {Bode}, N. and {Boer}, M. and {Boetzel}, Y. and {Bogaert}, G. and {Bondu}, F. and {Bonilla}, E. and {Bonnand}, R. and {Booker}, P. and {Boom}, B.~A. and {Bork}, R. and {Boschi}, V. and {Bose}, S. and {Bossilkov}, V. and {Bosveld}, J. and {Bouffanais}, Y. and {Bozzi}, A. and {Bradaschia}, C. and {Brady}, P.~R. and {Bramley}, A. and {Branchesi}, M. and {Brau}, J.~E. and {Breschi}, M. and {Briant}, T. and {Briggs}, J.~H. and {Brighenti}, F. and {Brillet}, A. and {Brinkmann}, M. and {Brito}, R. and {Brockill}, P. and {Brooks}, A.~F. and {Brooks}, J. and {Brown}, D.~D. and {Brunett}, S. and {Bruno}, G. and {Bruntz}, R. and {Buikema}, A. and {Bulik}, T. and {Bulten}, H.~J. and {Buonanno}, A. and {Buskulic}, D. and {Byer}, R.~L. and {Cabero}, M. and {Cadonati}, L. and {Cagnoli}, G. and {Cahillane}, C. and {Bustillo}, J. Calder{\'o}n and {Callaghan}, J.~D. and {Callister}, T.~A. and {Calloni}, E. and {Camp}, J.~B. and {Canepa}, M. and {Cannon}, K.~C. and {Cao}, H. and {Cao}, J. and {Carapella}, G. and {Carbognani}, F. and {Caride}, S. and {Carney}, M.~F. and {Carullo}, G. and {Diaz}, J. Casanueva and {Casentini}, C. and {Casta{\~n}eda}, J. and {Caudill}, S. and {Cavagli{\`a}}, M. and {Cavalier}, F. and {Cavalieri}, R. and {Cella}, G. and {Cerd{\'a}-Dur{\'a}n}, P. and {Cesarini}, E. and {Chaibi}, O. and {Chakravarti}, K. and {Chan}, C. and {Chan}, M. and {Chao}, S. and {Charlton}, P. and {Chase}, E.~A. and {Chassande-Mottin}, E. and {Chatterjee}, D. and {Chaturvedi}, M. and {Chatziioannou}, K. and {Chen}, H.~Y. and {Chen}, X. and {Chen}, Y. and {Cheng}, H. -P. and {Cheong}, C.~K. and {Chia}, H.~Y. and {Chiadini}, F. and {Chierici}, R. and {Chincarini}, A. and {Chiummo}, A. and {Cho}, G. and {Cho}, H.~S. and {Cho}, M. and {Christensen}, N. and {Chu}, Q. and {Chua}, S. and {Chung}, K.~W. and {Chung}, S. and {Ciani}, G. and {Ciecielag}, P. and {Cie{\'s}lar}, M. and {Ciobanu}, A.~A. and {Ciolfi}, R. and {Cipriano}, F. and {Cirone}, A. and {Clara}, F. and {Clark}, J.~A. and {Clearwater}, P. and {Clesse}, S. and {Cleva}, F. and {Coccia}, E. and {Cohadon}, P. -F. and {Cohen}, D. and {Colleoni}, M. and {Collette}, C.~G. and {Collins}, C. and {Colpi}, M. and {Constancio}, M., Jr. and {Conti}, L. and {Cooper}, S.~J. and {Corban}, P. and {Corbitt}, T.~R. and {Cordero-Carri{\'o}n}, I. and {Corezzi}, S. and {Corley}, K.~R. and {Cornish}, N. and {Corre}, D. and {Corsi}, A. and {Cortese}, S. and {Costa}, C.~A. and {Cotesta}, R. and {Coughlin}, M.~W. and {Coughlin}, S.~B. and {Coulon}, J. -P. and {Countryman}, S.~T. and {Couvares}, P. and {Covas}, P.~B. and {Coward}, D.~M. and {Cowart}, M.~J. and {Coyne}, D.~C. and {Coyne}, R. and {Creighton}, J.~D.~E. and {Creighton}, T.~D. and {Cripe}, J. and {Croquette}, M. and {Crowder}, S.~G. and {Cudell}, J. -R. and {Cullen}, T.~J. and {Cumming}, A. and {Cummings}, R. and {Cunningham}, L. and {Cuoco}, E. and {Curylo}, M. and {Canton}, T. Dal and {D{\'a}lya}, G. and {Dana}, A. and {Daneshgaran-Bajastani}, L.~M. and {D'Angelo}, B. and {Danilishin}, S.~L. and {D'Antonio}, S. and {Danzmann}, K. and {Darsow-Fromm}, C. and {Dasgupta}, A. and {Datrier}, L.~E.~H. and {Dattilo}, V. and {Dave}, I. and {Davier}, M. and {Davies}, G.~S. and {Davis}, D. and {Daw}, E.~J. and {DeBra}, D. and {Deenadayalan}, M. and {Degallaix}, J. and {De Laurentis}, M. and {Del{\'e}glise}, S. and {Delfavero}, M. and {De Lillo}, N. and {Del Pozzo}, W. and {DeMarchi}, L.~M. and {D'Emilio}, V. and {Demos}, N. and {Dent}, T. and {De Pietri}, R. and {De Rosa}, R. and {De Rossi}, C. and {DeSalvo}, R. and {de Varona}, O. and {Dhurandhar}, S. and {D{\'\i}az}, M.~C. and {Diaz-Ortiz}, M., Jr. and {Dietrich}, T. and {Di Fiore}, L. and {Di Fronzo}, C. and {Di Giorgio}, C. and {Di Giovanni}, F. and {Di Giovanni}, M. and {Di Girolamo}, T. and {Di Lieto}, A. and {Ding}, B. and {Di Pace}, S. and {Di Palma}, I. and {Di Renzo}, F. and {Divakarla}, A.~K. and {Dmitriev}, A. and {Doctor}, Z. and {Donovan}, F. and {Dooley}, K.~L. and {Doravari}, S. and {Dorrington}, I. and {Downes}, T.~P. and {Drago}, M. and {Driggers}, J.~C. and {Du}, Z. and {Ducoin}, J. -G. and {Dupej}, P. and {Durante}, O. and {D'Urso}, D. and {Dwyer}, S.~E. and {Easter}, P.~J. and {Eddolls}, G. and {Edelman}, B. and {Edo}, T.~B. and {Edy}, O. and {Effler}, A. and {Ehrens}, P. and {Eichholz}, J. and {Eikenberry}, S.~S. and {Eisenmann}, M. and {Eisenstein}, R.~A. and {Ejlli}, A. and {Errico}, L. and {Essick}, R.~C. and {Estelles}, H. and {Estevez}, D. and {Etienne}, Z.~B. and {Etzel}, T. and {Evans}, M. and {Evans}, T.~M. and {Ewing}, B.~E. and {Fafone}, V. and {Fairhurst}, S. and {Fan}, X. and {Farinon}, S. and {Farr}, B. and {Farr}, W.~M. and {Fauchon-Jones}, E.~J. and {Favata}, M. and {Fays}, M. and {Fazio}, M. and {Feicht}, J. and {Fejer}, M.~M. and {Feng}, F. and {Fenyvesi}, E. and {Ferguson}, D.~L. and {Fernandez-Galiana}, A. and {Ferrante}, I. and {Ferreira}, E.~C. and {Ferreira}, T.~A. and {Fidecaro}, F. and {Fiori}, I. and {Fiorucci}, D. and {Fishbach}, M. and {Fisher}, R.~P. and {Fittipaldi}, R. and {Fitz-Axen}, M. and {Fiumara}, V. and {Flaminio}, R. and {Floden}, E. and {Flynn}, E. and {Fong}, H. and {Font}, J.~A. and {Forsyth}, P.~W.~F. and {Fournier}, J. -D. and {Frasca}, S. and {Frasconi}, F. and {Frei}, Z. and {Freise}, A. and {Frey}, R. and {Frey}, V. and {Fritschel}, P. and {Frolov}, V.~V. and {Fronz{\`e}}, G. and {Fulda}, P. and {Fyffe}, M. and {Gabbard}, H.~A. and {Gadre}, B.~U. and {Gaebel}, S.~M. and {Gair}, J.~R. and {Galaudage}, S. and {Ganapathy}, D. and {Ganguly}, A. and {Gaonkar}, S.~G. and {Garc{\'\i}a-Quir{\'o}s}, C. and {Garufi}, F. and {Gateley}, B. and {Gaudio}, S. and {Gayathri}, V. and {Gemme}, G. and {Genin}, E. and {Gennai}, A. and {George}, D. and {George}, J. and {Gergely}, L. and {Ghonge}, S. and {Ghosh}, Abhirup and {Ghosh}, Archisman and {Ghosh}, S. and {Giacomazzo}, B. and {Giaime}, J.~A. and {Giardina}, K.~D. and {Gibson}, D.~R. and {Gier}, C. and {Gill}, K. and {Glanzer}, J. and {Gniesmer}, J. and {Godwin}, P. and {Goetz}, E. and {Goetz}, R. and {Gohlke}, N. and {Goncharov}, B. and {Gonz{\'a}lez}, G. and {Gopakumar}, A. and {Gossan}, S.~E. and {Gosselin}, M. and {Gouaty}, R. and {Grace}, B. and {Grado}, A. and {Granata}, M. and {Grant}, A. and {Gras}, S. and {Grassia}, P. and {Gray}, C. and {Gray}, R. and {Greco}, G. and {Green}, A.~C. and {Green}, R. and {Gretarsson}, E.~M. and {Griggs}, H.~L. and {Grignani}, G. and {Grimaldi}, A. and {Grimm}, S.~J. and {Grote}, H. and {Grunewald}, S. and {Gruning}, P. and {Guidi}, G.~M. and {Guimaraes}, A.~R. and {Guix{\'e}}, G. and {Gulati}, H.~K. and {Guo}, Y. and {Gupta}, A. and {Gupta}, Anchal and {Gupta}, P. and {Gustafson}, E.~K. and {Gustafson}, R. and {Haegel}, L. and {Halim}, O. and {Hall}, E.~D. and {Hamilton}, E.~Z. and {Hammond}, G. and {Haney}, M. and {Hanke}, M.~M. and {Hanks}, J. and {Hanna}, C. and {Hannam}, M.~D. and {Hannuksela}, O.~A. and {Hansen}, T.~J. and {Hanson}, J. and {Harder}, T. and {Hardwick}, T. and {Haris}, K. and {Harms}, J. and {Harry}, G.~M. and {Harry}, I.~W. and {Hasskew}, R.~K. and {Haster}, C. -J. and {Haughian}, K. and {Hayes}, F.~J. and {Healy}, J. and {Heidmann}, A. and {Heintze}, M.~C. and {Heinze}, J. and {Heitmann}, H. and {Hellman}, F. and {Hello}, P. and {Hemming}, G. and {Hendry}, M. and {Heng}, I.~S. and {Hennes}, E. and {Hennig}, J. and {Heurs}, M. and {Hild}, S. and {Hinderer}, T. and {Hoback}, S.~Y. and {Hochheim}, S. and {Hofgard}, E. and {Hofman}, D. and {Holgado}, A.~M. and {Holland}, N.~A. and {Holt}, K. and {Holz}, D.~E. and {Hopkins}, P. and {Horst}, C. and {Hough}, J. and {Howell}, E.~J. and {Hoy}, C.~G. and {Huang}, Y. and {H{\"u}bner}, M.~T. and {Huerta}, E.~A. and {Huet}, D. and {Hughey}, B. and {Hui}, V. and {Husa}, S. and {Huttner}, S.~H. and {Huxford}, R. and {Huynh-Dinh}, T. and {Idzkowski}, B. and {Iess}, A. and {Inchauspe}, H. and {Ingram}, C. and {Intini}, G. and {Isac}, J. -M. and {Isi}, M. and {Iyer}, B.~R. and {Jacqmin}, T. and {Jadhav}, S.~J. and {Jadhav}, S.~P. and {James}, A.~L. and {Jani}, K. and {Janthalur}, N.~N. and {Jaranowski}, P. and {Jariwala}, D. and {Jaume}, R. and {Jenkins}, A.~C. and {Jiang}, J. and {Johns}, G.~R. and {Johnson-McDaniel}, N.~K. and {Jones}, A.~W. and {Jones}, D.~I. and {Jones}, J.~D. and {Jones}, P. and {Jones}, R. and {Jonker}, R.~J.~G. and {Ju}, L. and {Junker}, J. and {Kalaghatgi}, C.~V. and {Kalogera}, V. and {Kamai}, B. and {Kandhasamy}, S. and {Kang}, G. and {Kanner}, J.~B. and {Kapadia}, S.~J. and {Karki}, S. and {Kashyap}, R. and {Kasprzack}, M. and {Kastaun}, W. and {Katsanevas}, S. and {Katsavounidis}, E. and {Katzman}, W. and {Kaufer}, S. and {Kawabe}, K. and {K{\'e}f{\'e}lian}, F. and {Keitel}, D. and {Keivani}, A. and {Kennedy}, R. and {Key}, J.~S. and {Khadka}, S. and {Khalili}, F.~Y. and {Khan}, I. and {Khan}, S. and {Khan}, Z.~A. and {Khazanov}, E.~A. and {Khetan}, N. and {Khursheed}, M. and {Kijbunchoo}, N. and {Kim}, Chunglee and {Kim}, G.~J. and {Kim}, J.~C. and {Kim}, K. and {Kim}, W. and {Kim}, W.~S. and {Kim}, Y. -M. and {Kimball}, C. and {King}, P.~J. and {Kinley-Hanlon}, M. and {Kirchhoff}, R. and {Kissel}, J.~S. and {Kleybolte}, L. and {Klimenko}, S. and {Knowles}, T.~D. and {Knyazev}, E. and {Koch}, P. and {Koehlenbeck}, S.~M. and {Koekoek}, G. and {Koley}, S. and {Kondrashov}, V. and {Kontos}, A. and {Koper}, N. and {Korobko}, M. and {Korth}, W.~Z. and {Kovalam}, M. and {Kozak}, D.~B. and {Kringel}, V. and {Krishnendu}, N.~V. and {Kr{\'o}lak}, A. and {Krupinski}, N. and {Kuehn}, G. and {Kumar}, A. and {Kumar}, P. and {Kumar}, Rahul and {Kumar}, Rakesh and {Kumar}, S. and {Kuo}, L. and {Kutynia}, A. and {Lackey}, B.~D. and {Laghi}, D. and {Lalande}, E. and {Lam}, T.~L. and {Lamberts}, A. and {Landry}, M. and {Landry}, P. and {Lane}, B.~B. and {Lang}, R.~N. and {Lange}, J. and {Lantz}, B. and {Lanza}, R.~K. and {La Rosa}, I. and {Lartaux-Vollard}, A. and {Lasky}, P.~D. and {Laxen}, M. and {Lazzarini}, A. and {Lazzaro}, C. and {Leaci}, P. and {Leavey}, S. and {Lecoeuche}, Y.~K. and {Lee}, C.~H. and {Lee}, H.~M. and {Lee}, H.~W. and {Lee}, J. and {Lee}, K. and {Lehmann}, J. and {Leroy}, N. and {Letendre}, N. and {Levin}, Y. and {Li}, A.~K.~Y. and {Li}, J. and {li}, K. and {Li}, T.~G.~F. and {Li}, X. and {Linde}, F. and {Linker}, S.~D. and {Linley}, J.~N. and {Littenberg}, T.~B. and {Liu}, J. and {Liu}, X. and {Llorens-Monteagudo}, M. and {Lo}, R.~K.~L. and {Lockwood}, A. and {London}, L.~T. and {Longo}, A. and {Lorenzini}, M. and {Loriette}, V. and {Lormand}, M. and {Losurdo}, G. and {Lough}, J.~D. and {Lousto}, C.~O. and {Lovelace}, G. and {L{\"u}ck}, H. and {Lumaca}, D. and {Lundgren}, A.~P. and {Ma}, Y. and {Macas}, R. and {Macfoy}, S. and {MacInnis}, M. and {Macleod}, D.~M. and {MacMillan}, I.~A.~O. and {Macquet}, A. and {Hernandez}, I. Maga{\~n}a and {Maga{\~n}a-Sandoval}, F. and {Magee}, R.~M. and {Majorana}, E. and {Maksimovic}, I. and {Malik}, A. and {Man}, N. and {Mandic}, V. and {Mangano}, V. and {Mansell}, G.~L. and {Manske}, M. and {Mantovani}, M. and {Mapelli}, M. and {Marchesoni}, F. and {Marion}, F. and {M{\'a}rka}, S. and {M{\'a}rka}, Z. and {Markakis}, C. and {Markosyan}, A.~S. and {Markowitz}, A. and {Maros}, E. and {Marquina}, A. and {Marsat}, S. and {Martelli}, F. and {Martin}, I.~W. and {Martin}, R.~M. and {Martinez}, V. and {Martynov}, D.~V. and {Masalehdan}, H. and {Mason}, K. and {Massera}, E. and {Masserot}, A. and {Massinger}, T.~J. and {Masso-Reid}, M. and {Mastrogiovanni}, S. and {Matas}, A. and {Matichard}, F. and {Mavalvala}, N. and {Maynard}, E. and {McCann}, J.~J. and {McCarthy}, R. and {McClelland}, D.~E. and {McCormick}, S. and {McCuller}, L. and {McGuire}, S.~C. and {McIsaac}, C. and {McIver}, J. and {McManus}, D.~J. and {McRae}, T. and {McWilliams}, S.~T. and {Meacher}, D. and {Meadors}, G.~D. and {Mehmet}, M. and {Mehta}, A.~K. and {Villa}, E. Mejuto and {Melatos}, A. and {Mendell}, G. and {Mercer}, R.~A. and {Mereni}, L. and {Merfeld}, K. and {Merilh}, E.~L. and {Merritt}, J.~D. and {Merzougui}, M. and {Meshkov}, S. and {Messenger}, C. and {Messick}, C. and {Metzdorff}, R. and {Meyers}, P.~M. and {Meylahn}, F. and {Mhaske}, A. and {Miani}, A. and {Miao}, H. and {Michaloliakos}, I. and {Michel}, C. and {Middleton}, H. and {Milano}, L. and {Miller}, A.~L. and {Millhouse}, M. and {Mills}, J.~C. and {Milotti}, E. and {Milovich-Goff}, M.~C. and {Minazzoli}, O. and {Minenkov}, Y. and {Mishkin}, A. and {Mishra}, C. and {Mistry}, T. and {Mitra}, S. and {Mitrofanov}, V.~P. and {Mitselmakher}, G. and {Mittleman}, R. and {Mo}, G. and {Mogushi}, K. and {Mohapatra}, S.~R.~P. and {Mohite}, S.~R. and {Molina-Ruiz}, M. and {Mondin}, M. and {Montani}, M. and {Moore}, C.~J. and {Moraru}, D. and {Morawski}, F. and {Moreno}, G. and {Morisaki}, S. and {Mours}, B. and {Mow-Lowry}, C.~M. and {Mozzon}, S. and {Muciaccia}, F. and {Mukherjee}, Arunava and {Mukherjee}, D. and {Mukherjee}, S. and {Mukherjee}, Subroto and {Mukund}, N. and {Mullavey}, A. and {Munch}, J. and {Mu{\~n}iz}, E.~A. and {Murray}, P.~G. and {Nagar}, A. and {Nardecchia}, I. and {Naticchioni}, L. and {Nayak}, R.~K. and {Neil}, B.~F. and {Neilson}, J. and {Nelemans}, G. and {Nelson}, T.~J.~N. and {Nery}, M. and {Neunzert}, A. and {Ng}, K.~Y. and {Ng}, S. and {Nguyen}, C. and {Nguyen}, P. and {Nichols}, D. and {Nichols}, S.~A. and {Nissanke}, S. and {Nocera}, F. and {Noh}, M. and {North}, C. and {Nothard}, D. and {Nuttall}, L.~K. and {Oberling}, J. and {O'Brien}, B.~D. and {Oganesyan}, G. and {Ogin}, G.~H. and {Oh}, J.~J. and {Oh}, S.~H. and {Ohme}, F. and {Ohta}, H. and {Okada}, M.~A. and {Oliver}, M. and {Olivetto}, C. and {Oppermann}, P. and {Oram}, Richard J. and {O'Reilly}, B. and {Ormiston}, R.~G. and {Ortega}, L.~F. and {O'Shaughnessy}, R. and {Ossokine}, S. and {Osthelder}, C. and {Ottaway}, D.~J. and {Overmier}, H. and {Owen}, B.~J. and {Pace}, A.~E. and {Pagano}, G. and {Page}, M.~A. and {Pagliaroli}, G. and {Pai}, A. and {Pai}, S.~A. and {Palamos}, J.~R. and {Palashov}, O. and {Palomba}, C. and {Pan}, H. and {Panda}, P.~K. and {Pang}, P.~T.~H. and {Pankow}, C. and {Pannarale}, F. and {Pant}, B.~C. and {Paoletti}, F. and {Paoli}, A. and {Parida}, A. and {Parker}, W. and {Pascucci}, D. and {Pasqualetti}, A. and {Passaquieti}, R. and {Passuello}, D. and {Patricelli}, B. and {Payne}, E. and {Pearlstone}, B.~L. and {Pechsiri}, T.~C. and {Pedersen}, A.~J. and {Pedraza}, M. and {Pele}, A. and {Penn}, S. and {Perego}, A. and {Perez}, C.~J. and {P{\'e}rigois}, C. and {Perreca}, A. and {Perri{\`e}s}, S. and {Petermann}, J. and {Pfeiffer}, H.~P. and {Phelps}, M. and {Phukon}, K.~S. and {Piccinni}, O.~J. and {Pichot}, M. and {Piendibene}, M. and {Piergiovanni}, F. and {Pierro}, V. and {Pillant}, G. and {Pinard}, L. and {Pinto}, I.~M. and {Piotrzkowski}, K. and {Pirello}, M. and {Pitkin}, M. and {Plastino}, W. and {Poggiani}, R. and {Pong}, D.~Y.~T. and {Ponrathnam}, S. and {Popolizio}, P. and {Porter}, E.~K. and {Powell}, J. and {Prajapati}, A.~K. and {Prasai}, K. and {Prasanna}, R. and {Pratten}, G. and {Prestegard}, T. and {Principe}, M. and {Prodi}, G.~A. and {Prokhorov}, L. and {Punturo}, M. and {Puppo}, P. and {P{\"u}rrer}, M. and {Qi}, H. and {Quetschke}, V. and {Quinonez}, P.~J. and {Raab}, F.~J. and {Raaijmakers}, G. and {Radkins}, H. and {Radulesco}, N. and {Raffai}, P. and {Rafferty}, H. and {Raja}, S. and {Rajan}, C. and {Rajbhandari}, B. and {Rakhmanov}, M. and {Ramirez}, K.~E. and {Ramos-Buades}, A. and {Rana}, Javed and {Rao}, K. and {Rapagnani}, P. and {Raymond}, V. and {Razzano}, M. and {Read}, J. and {Regimbau}, T. and {Rei}, L. and {Reid}, S. and {Reitze}, D.~H. and {Rettegno}, P. and {Ricci}, F. and {Richardson}, C.~J. and {Richardson}, J.~W. and {Ricker}, P.~M. and {Riemenschneider}, G. and {Riles}, K. and {Rizzo}, M. and {Robertson}, N.~A. and {Robinet}, F. and {Rocchi}, A. and {Rodriguez-Soto}, R.~D. and {Rolland}, L. and {Rollins}, J.~G. and {Roma}, V.~J. and {Romanelli}, M. and {Romano}, R. and {Romel}, C.~L. and {Romero-Shaw}, I.~M. and {Romie}, J.~H. and {Rose}, C.~A. and {Rose}, D. and {Rose}, K. and {Rosi{\'n}ska}, D. and {Rosofsky}, S.~G. and {Ross}, M.~P. and {Rowan}, S. and {Rowlinson}, S.~J. and {Roy}, P.~K. and {Roy}, Santosh and {Roy}, Soumen and {Ruggi}, P. and {Rutins}, G. and {Ryan}, K. and {Sachdev}, S. and {Sadecki}, T. and {Sakellariadou}, M. and {Salafia}, O.~S. and {Salconi}, L. and {Saleem}, M. and {Salemi}, F. and {Samajdar}, A. and {Sanchez}, E.~J. and {Sanchez}, L.~E. and {Sanchis-Gual}, N. and {Sanders}, J.~R. and {Santiago}, K.~A. and {Santos}, E. and {Sarin}, N. and {Sassolas}, B. and {Sathyaprakash}, B.~S. and {Sauter}, O. and {Savage}, R.~L. and {Savant}, V. and {Sawant}, D. and {Sayah}, S. and {Schaetzl}, D. and {Schale}, P. and {Scheel}, M. and {Scheuer}, J. and {Schmidt}, P. and {Schnabel}, R. and {Schofield}, R.~M.~S. and {Sch{\"o}nbeck}, A. and {Schreiber}, E. and {Schulte}, B.~W. and {Schutz}, B.~F. and {Schwarm}, O. and {Schwartz}, E. and {Scott}, J. and {Scott}, S.~M. and {Seidel}, E. and {Sellers}, D. and {Sengupta}, A.~S. and {Sennett}, N. and {Sentenac}, D. and {Sequino}, V. and {Sergeev}, A. and {Setyawati}, Y. and {Shaddock}, D.~A. and {Shaffer}, T. and {Shahriar}, M.~S. and {Sharma}, A. and {Sharma}, P. and {Shawhan}, P. and {Shen}, H. and {Shikauchi}, M. and {Shink}, R. and {Shoemaker}, D.~H. and {Shoemaker}, D.~M. and {Shukla}, K. and {ShyamSundar}, S. and {Siellez}, K. and {Sieniawska}, M. and {Sigg}, D. and {Singer}, L.~P. and {Singh}, D. and {Singh}, N. and {Singha}, A. and {Singhal}, A. and {Sintes}, A.~M. and {Sipala}, V. and {Skliris}, V. and {Slagmolen}, B.~J.~J. and {Slaven-Blair}, T.~J. and {Smetana}, J. and {Smith}, J.~R. and {Smith}, R.~J.~E. and {Somala}, S. and {Son}, E.~J. and {Soni}, S. and {Sorazu}, B. and {Sordini}, V. and {Sorrentino}, F. and {Souradeep}, T. and {Sowell}, E. and {Spencer}, A.~P. and {Spera}, M. and {Srivastava}, A.~K. and {Srivastava}, V. and {Staats}, K. and {Stachie}, C. and {Standke}, M. and {Steer}, D.~A. and {Steinhoff}, J. and {Steinke}, M. and {Steinlechner}, J. and {Steinlechner}, S. and {Steinmeyer}, D. and {Stevenson}, S. and {Stocks}, D. and {Stops}, D.~J. and {Stover}, M. and {Strain}, K.~A. and {Stratta}, G. and {Strunk}, A. and {Sturani}, R. and {Stuver}, A.~L. and {Sudhagar}, S. and {Sudhir}, V. and {Summerscales}, T.~Z. and {Sun}, L. and {Sunil}, S. and {Sur}, A. and {Suresh}, J. and {Sutton}, P.~J. and {Swinkels}, B.~L. and {Szczepa{\'n}czyk}, M.~J. and {Tacca}, M. and {Tait}, S.~C. and {Talbot}, C. and {Tanasijczuk}, A.~J. and {Tanner}, D.~B. and {Tao}, D. and {T{\'a}pai}, M. and {Tapia}, A. and {San Martin}, E.~N. Tapia and {Tasson}, J.~D. and {Taylor}, R. and {Tenorio}, R. and {Terkowski}, L. and {Thirugnanasambandam}, M.~P. and {Thomas}, M. and {Thomas}, P. and {Thompson}, J.~E. and {Thondapu}, S.~R. and {Thorne}, K.~A. and {Thrane}, E. and {Tinsman}, C.~L. and {Saravanan}, T.~R. and {Tiwari}, Shubhanshu and {Tiwari}, S. and {Tiwari}, V. and {Toland}, K. and {Tonelli}, M. and {Tornasi}, Z. and {Torres-Forn{\'e}}, A. and {Torrie}, C.~I. and {Tosta e Melo}, I. and {T{\"o}yr{\"a}}, D. and {Trail}, E.~A. and {Travasso}, F. and {Traylor}, G. and {Tringali}, M.~C. and {Tripathee}, A. and {Trovato}, A. and {Trudeau}, R.~J. and {Tsang}, K.~W. and {Tse}, M. and {Tso}, R. and {Tsukada}, L. and {Tsuna}, D. and {Tsutsui}, T. and {Turconi}, M. and {Ubhi}, A.~S. and {Ueno}, K. and {Ugolini}, D. and {Unnikrishnan}, C.~S. and {Urban}, A.~L. and {Usman}, S.~A. and {Utina}, A.~C. and {Vahlbruch}, H. and {Vajente}, G. and {Valdes}, G. and {Valentini}, M. and {van Bakel}, N. and {van Beuzekom}, M. and {van den Brand}, J.~F.~J. and {Van Den Broeck}, C. and {Vander-Hyde}, D.~C. and {van der Schaaf}, L. and {Van Heijningen}, J.~V. and {van Veggel}, A.~A. and {Vardaro}, M. and {Varma}, V. and {Vass}, S. and {Vas{\'u}th}, M. and {Vecchio}, A. and {Vedovato}, G. and {Veitch}, J. and {Veitch}, P.~J. and {Venkateswara}, K. and {Venugopalan}, G. and {Verkindt}, D. and {Veske}, D. and {Vetrano}, F. and {Vicer{\'e}}, A. and {Viets}, A.~D. and {Vinciguerra}, S. and {Vine}, D.~J. and {Vinet}, J. -Y. and {Vitale}, S. and {Vivanco}, Francisco Hernandez and {Vo}, T. and {Vocca}, H. and {Vorvick}, C. and {Vyatchanin}, S.~P. and {Wade}, A.~R. and {Wade}, L.~E. and {Wade}, M. and {Walet}, R. and {Walker}, M. and {Wallace}, G.~S. and {Wallace}, L. and {Walsh}, S. and {Wang}, J.~Z. and {Wang}, S. and {Wang}, W.~H. and {Ward}, R.~L. and {Warden}, Z.~A. and {Warner}, J. and {Was}, M. and {Watchi}, J. and {Weaver}, B. and {Wei}, L. -W. and {Weinert}, M. and {Weinstein}, A.~J. and {Weiss}, R. and {Wellmann}, F. and {Wen}, L. and {We{\ss}els}, P. and {Westhouse}, J.~W. and {Wette}, K. and {Whelan}, J.~T. and {Whiting}, B.~F. and {Whittle}, C. and {Wilken}, D.~M. and {Williams}, D. and {Willis}, J.~L. and {Willke}, B. and {Winkler}, W. and {Wipf}, C.~C. and {Wittel}, H. and {Woan}, G. and {Woehler}, J. and {Wofford}, J.~K. and {Wong}, C. and {Wright}, J.~L. and {Wu}, D.~S. and {Wysocki}, D.~M. and {Xiao}, L. and {Yamamoto}, H. and {Yang}, L. and {Yang}, Y. and {Yang}, Z. and {Yap}, M.~J. and {Yazback}, M. and {Yeeles}, D.~W. and {Yu}, Hang and {Yu}, Haocun and {Yuen}, S.~H.~R. and {Zadro{\.z}ny}, A.~K. and {Zadro{\.z}ny}, A. and {Zanolin}, M. and {Zelenova}, T. and {Zendri}, J. -P. and {Zevin}, M. and {Zhang}, J. and {Zhang}, L. and {Zhang}, T. and {Zhao}, C. and {Zhao}, G. and {Zhou}, M. and {Zhou}, Z. and {Zhu}, X.~J. and {Zimmerman}, A.~B. and {Zucker}, M.~E. and {Zweizig}, J. and {LIGO Scientific Collaboration} and {Virgo Collaboration}},
        title = "{GW190814: Gravitational Waves from the Coalescence of a 23 Solar Mass Black Hole with a 2.6 Solar Mass Compact Object}",
      journal = {ApJL},
         year = 2020,
        month = jun,
       volume = {896},
       number = {2},
          eid = {L44},
        pages = {L44},
          doi = {10.3847/2041-8213/ab960f}
}

@ARTICLE{Abia2002,
       author = {{Abia}, C. and {Dom{\'i}nguez}, I. and {Gallino}, R. and {Busso}, M. and {Masera}, S. and {Straniero}, O. and {de Laverny}, P. and {Plez}, B. and {Isern}, J.},
        title = "{s-Process Nucleosynthesis in Carbon Stars}",
      journal = {ApJ},
         year = 2002,
        month = nov,
       volume = {579},
       number = {2},
        pages = {817-831},
          doi = {10.1086/342924}
}

@ARTICLE{Abia2022,
       author = {{Abia}, C. and {de Laverny}, P. and {Romero-G{\'o}mez}, M. and {Figueras}, F.},
        title = "{Characterisation of Galactic carbon stars and related stars from Gaia EDR3}",
      journal = {A\&A},
         year = 2022,
        month = aug,
       volume = {664},
          eid = {A45},
        pages = {A45},
          doi = {10.1051/0004-6361/202243595}
}

@ARTICLE{Agrawal2020,
       author = {{Agrawal}, Poojan and {Hurley}, Jarrod and {Stevenson}, Simon and {Sz{\'e}csi}, Dorottya and {Flynn}, Chris},
        title = "{The fates of massive stars: exploring uncertainties in stellar evolution with METISSE}",
      journal = {MNRAS},
         year = 2020,
        month = oct,
       volume = {497},
       number = {4},
        pages = {4549-4564},
          doi = {10.1093/mnras/staa2264}
}

@ARTICLE{Bidelman1951,
       author = {{Bidelman}, William P. and {Keenan}, Philip C.},
        title = "{The Ba II Stars.}",
      journal = {ApJ},
         year = 1951,
        month = nov,
       volume = {114},
        pages = {473},
          doi = {10.1086/145488}
}

@ARTICLE{Bloecker1995,
       author = {{Bloecker}, T.},
        title = "{Stellar evolution of low and intermediate-mass stars. I. Mass loss on the AGB and its consequences for stellar evolution.}",
      journal = {A\&A},
         year = 1995,
        month = may,
       volume = {297},
        pages = {727},
       adsurl = {https://ui.adsabs.harvard.edu/abs/1995A&A...297..727B}
}

@ARTICLE{Bensby2006,
       author = {{Bensby}, T. and {Feltzing}, S.},
        title = "{The origin and chemical evolution of carbon in the Galactic thin and thick discs$^{*}$}",
      journal = {MNRAS},
         year = 2006,
        month = apr,
       volume = {367},
       number = {3},
        pages = {1181-1193},
          doi = {10.1111/j.1365-2966.2006.10037.x}
}

@ARTICLE{Bonavciv2004,
       author = {{Bona{\v{c}}i{\'c} Marinovi{\'c}}, A.~A. and {Pols}, O.~R.},
        title = "{Barium star population synthesis with an improved TP-AGB model}",
      journal = {Mem. S.A.It.},
         year = 2004,
        month = jan,
       volume = {75},
        pages = {760},
       adsurl = {https://ui.adsabs.harvard.edu/abs/2004MmSAI..75..760B}
}

@ARTICLE{Bonavciv2007,
       author = {{Bona{\v{c}}i{\'c} Marinovi{\'c}}, A. and {Izzard}, R.~G. and {Lugaro}, M. and {Pols}, O.~R.},
        title = "{The s-process in stellar population synthesis: a new approach to understanding AGB stars}",
      journal = {A\&A},
         year = 2007,
        month = jul,
       volume = {469},
       number = {3},
        pages = {1013-1025},
          doi = {10.1051/0004-6361:20066861}
}

@ARTICLE{Brinkman2019,
       author = {{Brinkman}, H.~E. and {Doherty}, C.~L. and {Pols}, O.~R. and {Li}, E.~T. and {C{\^o}t{\'e}}, B. and {Lugaro}, M.},
        title = "{Aluminium-26 from Massive Binary Stars. I. Nonrotating Models}",
      journal = {ApJ},
         year = 2019,
        month = oct,
       volume = {884},
       number = {1},
          eid = {38},
        pages = {38},
          doi = {10.3847/1538-4357/ab40ae},
archivePrefix = {arXiv},
       eprint = {1909.04433},
 primaryClass = {astro-ph.SR},
       adsurl = {https://ui.adsabs.harvard.edu/abs/2019ApJ...884...38B}
}

@ARTICLE{Bondi1944,
       author = {{Bondi}, H. and {Hoyle}, F.},
        title = "{On the mechanism of accretion by stars}",
      journal = {MNRAS},
         year = 1944,
        month = jan,
       volume = {104},
        pages = {273},
          doi = {10.1093/mnras/104.5.273}
}

@ARTICLE{Boothroyd1995,
       author = {{Boothroyd}, Arnold I. and {Sackmann}, I. -Juliana and {Wasserburg}, G.~J.},
        title = "{Hot Bottom Burning in Asymptotic Giant Branch Stars and Its Effect on Oxygen Isotopic Abundances}",
      journal = {ApJ},
         year = 1995,
        month = mar,
       volume = {442},
        pages = {L21},
          doi = {10.1086/187806},
       adsurl = {https://ui.adsabs.harvard.edu/abs/1995ApJ...442L..21B}
}

@ARTICLE{Boothroyd1988,
       author = {{Boothroyd}, Arnold I. and {Sackmann}, I. -Juliana},
        title = "{Low-Mass Stars. IV. The Production of Carbon Stars}",
      journal = {ApJ},
         year = 1988,
        month = may,
       volume = {328},
        pages = {671},
          doi = {10.1086/166324}
}

@ARTICLE{Broekgaarden2019,
       author = {{Broekgaarden}, Floor S. and {Justham}, Stephen and {de Mink}, Selma E. and {Gair}, Jonathan and {Mandel}, Ilya and {Stevenson}, Simon and {Barrett}, Jim W. and {Vigna-G{\'o}mez}, Alejandro and {Neijssel}, Coenraad J.},
        title = "{STROOPWAFEL: simulating rare outcomes from astrophysical populations, with application to gravitational-wave sources}",
      journal = {MNRAS},
         year = 2019,
        month = dec,
       volume = {490},
       number = {4},
        pages = {5228-5248},
          doi = {10.1093/mnras/stz2558}
}

@ARTICLE{Brinkman2023,
       author = {{Brinkman}, H.~E. and {Doherty}, Carolyn and {Pignatari}, Marco and {Pols}, Onno and {Lugaro}, Maria},
        title = "{Aluminium-26 from Massive Binary Stars. III. Binary Stars up to Core Collapse and Their Impact on the Early Solar System}",
      journal = {ApJ},
         year = 2023,
        month = jul,
       volume = {951},
       number = {2},
          eid = {110},
        pages = {110},
          doi = {10.3847/1538-4357/acd7ea}
}

@ARTICLE{Buntain2017,
       author = {{Buntain}, J.~F. and {Doherty}, C.~L. and {Lugaro}, M. and {Lattanzio}, J.~C. and {Stancliffe}, R.~J. and {Karakas}, A.~I.},
        title = "{Partial mixing and the formation of $^{13}$C pockets in AGB stars: effects on the s-process elements}",
      journal = {MNRAS},
         year = 2017,
        month = oct,
       volume = {471},
       number = {1},
        pages = {824-838},
          doi = {10.1093/mnras/stx1502}
}

@ARTICLE{Busso2001,
       author = {{Busso}, Maurizio and {Gallino}, Roberto and {Lambert}, David L. and {Travaglio}, Claudia and {Smith}, Verne V.},
        title = "{Nucleosynthesis and Mixing on the Asymptotic Giant Branch. III. Predicted and Observed s-Process Abundances}",
      journal = {ApJ},
         year = 2001,
        month = aug,
       volume = {557},
       number = {2},
        pages = {802-821},
          doi = {10.1086/322258}
}

@ARTICLE{Cinquegrana2022_2,
       author = {{Karakas}, Amanda I. and {Cinquegrana}, Giulia and {Joyce}, Meridith},
        title = "{The most metal-rich asymptotic giant branch stars}",
      journal = {MNRAS},
         year = 2022,
        month = jan,
       volume = {509},
       number = {3},
        pages = {4430-4447},
          doi = {10.1093/mnras/stab3205}
}

@ARTICLE{Claeys2014,
       author = {{Claeys}, J.~S.~W. and {Pols}, O.~R. and {Izzard}, R.~G. and {Vink}, J. and {Verbunt}, F.~W.~M.},
        title = "{Theoretical uncertainties of the Type Ia supernova rate}",
      journal = {A\&A},
         year = 2014,
        month = mar,
       volume = {563},
          eid = {A83},
        pages = {A83},
          doi = {10.1051/0004-6361/201322714},
archivePrefix = {arXiv},
       eprint = {1401.2895},
 primaryClass = {astro-ph.SR},
       adsurl = {https://ui.adsabs.harvard.edu/abs/2014A&A...563A..83C}
}

@ARTICLE{Clayton1996,
       author = {{Clayton}, Geoffrey C.},
        title = "{The R Coronae Borealis Stars}",
      journal = {PASP},
         year = 1996,
        month = mar,
       volume = {108},
        pages = {225},
          doi = {10.1086/133715}
}

@ARTICLE{Clayton2012,
       author = {{Clayton}, G.~C.},
        title = "{What Are the R Coronae Borealis Stars?}",
      journal = {JAVSO},
         year = 2012,
        month = jun,
       volume = {40},
       number = {1},
        pages = {539},
          doi = {10.48550/arXiv.1206.3448}
}

@ARTICLE{Cristallo2009,
       author = {{Cristallo}, S. and {Straniero}, O. and {Gallino}, R. and {Piersanti}, L. and {Dom{\'i}nguez}, I. and {Lederer}, M.~T.},
        title = "{Evolution, Nucleosynthesis, and Yields of Low-Mass Asymptotic Giant Branch Stars at Different Metallicities}",
      journal = {ApJ},
         year = 2009,
        month = may,
       volume = {696},
       number = {1},
        pages = {797-820},
          doi = {10.1088/0004-637X/696/1/797}
}

@ARTICLE{Cristallo2009_2,
       author = {{Cristallo}, S. and {Piersanti}, L. and {Straniero}, O. and {Gallino}, R. and {Dom{\'\i}nguez}, I. and {K{\"a}ppeler}, F.},
        title = "{Asymptotic-Giant-Branch Models at Very Low Metallicity}",
      journal = {PASA},
         year = 2009,
        month = aug,
       volume = {26},
       number = {3},
        pages = {139-144},
          doi = {10.1071/AS09003}
}

@ARTICLE{Cristallo2011,
       author = {{Cristallo}, S. and {Piersanti}, L. and {Straniero}, O. and {Gallino}, R. and {Dom{\'\i}nguez}, I. and {Abia}, C. and {Di Rico}, G. and {Quintini}, M. and {Bisterzo}, S.},
        title = "{Evolution, Nucleosynthesis, and Yields of Low-mass Asymptotic Giant Branch Stars at Different Metallicities. II. The FRUITY Database}",
      journal = {ApJS},
         year = 2011,
        month = dec,
       volume = {197},
       number = {2},
          eid = {17},
        pages = {17},
          doi = {10.1088/0067-0049/197/2/17}
}

@ARTICLE{Cristallo2015,
       author = {{Cristallo}, S. and {Straniero}, O. and {Piersanti}, L. and {Gobrecht}, D.},
        title = "{Evolution, Nucleosynthesis, and Yields of AGB Stars at Different Metallicities. III. Intermediate-mass Models, Revised Low-mass Models, and the ph-FRUITY Interface}",
      journal = {ApJs},
         year = 2015,
        month = aug,
       volume = {219},
       number = {2},
          eid = {40},
        pages = {40},
          doi = {10.1088/0067-0049/219/2/40}
}

@ARTICLE{Cseh2018,
       author = {{Cseh}, B. and {Lugaro}, M. and {D'Orazi}, V. and {de Castro}, D.~B. and {Pereira}, C.~B. and {Karakas}, A.~I. and {Moln{\'a}r}, L. and {Plachy}, E. and {Szab{\'o}}, R. and {Pignatari}, M. and {Cristallo}, S.},
        title = "{The s process in AGB stars as constrained by a large sample of barium stars}",
      journal = {A\&A},
         year = 2018,
        month = dec,
       volume = {620},
          eid = {A146},
        pages = {A146},
          doi = {10.1051/0004-6361/201834079}
}

@ARTICLE{DeCastro2016,
       author = {{de Castro}, D.~B. and {Pereira}, C.~B. and {Roig}, F. and {Jilinski}, E. and {Drake}, N.~A. and {Chavero}, C. and {Sales Silva}, J.~V.},
        title = "{Chemical abundances and kinematics of barium stars}",
      journal = {MNRAS},
         year = 2016,
        month = jul,
       volume = {459},
       number = {4},
        pages = {4299-4324},
          doi = {10.1093/mnras/stw815}
}

@ARTICLE{DeDonder2004,
       author = {{De Donder}, Erwin and {Vanbeveren}, Dany},
        title = "{The influence of binaries on galactic chemical evolution}",
      journal = {NewAR},
         year = 2004,
        month = sep,
       volume = {48},
       number = {10},
        pages = {861-975},
          doi = {10.1016/j.newar.2004.07.001}
}

@ARTICLE{deMink2013,
       author = {{de Mink}, S.~E. and {Langer}, N. and {Izzard}, R.~G. and {Sana}, H. and {de Koter}, A.},
        title = "{The Rotation Rates of Massive Stars: The Role of Binary Interaction through Tides, Mass Transfer, and Mergers}",
      journal = {ApJ},
         year = 2013,
        month = feb,
       volume = {764},
       number = {2},
          eid = {166},
        pages = {166}
}

@ARTICLE{DeMarco2009,
       author = {{De Marco}, Orsola},
        title = "{The Origin and Shaping of Planetary Nebulae: Putting the Binary Hypothesis to the Test}",
      journal = {PASP},
         year = 2009,
        month = apr,
       volume = {121},
       number = {878},
        pages = {316},
          doi = {10.1086/597765}
}

@ARTICLE{DeMarco2017,
       author = {{De Marco}, Orsola and {Izzard}, Robert G.},
        title = "{Dawes Review 6: The Impact of Companions on Stellar Evolution}",
      journal = {PASA},
         year = 2017,
        month = jan,
       volume = {34},
          eid = {e001},
        pages = {e001},
          doi = {10.1017/pasa.2016.52}
}

@ARTICLE{DeMarco2022,
       author = {{De Marco}, Orsola and {Akashi}, Muhammad and {Akras}, Stavros and {Alcolea}, Javier and {Aleman}, Isabel and {Amram}, Philippe and {Balick}, Bruce and {De Beck}, Elvire and {Blackman}, Eric G. and {Boffin}, Henri M.~J. and {Boumis}, Panos and {Bublitz}, Jesse and {Bucciarelli}, Beatrice and {Bujarrabal}, Valentin and {Cami}, Jan and {Chornay}, Nicholas and {Chu}, You-Hua and {Corradi}, Romano L.~M. and {Frank}, Adam and {Garc{\'\i}a-Hern{\'a}ndez}, D.~A. and {Garc{\'\i}a-Rojas}, Jorge and {Garc{\'\i}a-Segura}, Guillermo and {G{\'o}mez-Llanos}, Veronica and {Gon{\c{c}}alves}, Denise R. and {Guerrero}, Mart{\'\i}n A. and {Jones}, David and {Karakas}, Amanda I. and {Kastner}, Joel H. and {Kwok}, Sun and {Lykou}, Foteini and {Manchado}, Arturo and {Matsuura}, Mikako and {McDonald}, Iain and {Miszalski}, Brent and {Mohamed}, Shazrene S. and {Monreal-Ibero}, Ana and {Monteiro}, Hektor and {Montez}, Rodolfo and {Baez}, Paula Moraga and {Morisset}, Christophe and {Nordhaus}, Jason and {Mendes de Oliveira}, Claudia and {Osborn}, Zara and {Otsuka}, Masaaki and {Parker}, Quentin A. and {Peeters}, Els and {Quint}, Bruno C. and {Quintana-Lacaci}, Guillermo and {Redman}, Matt and {Ruiter}, Ashley J. and {Sabin}, Laurence and {Sahai}, Raghvendra and {Contreras}, Carmen S{\'a}nchez and {Santander-Garc{\'\i}a}, Miguel and {Seitenzahl}, Ivo and {Soker}, Noam and {Speck}, Angela K. and {Stanghellini}, Letizia and {Steffen}, Wolfgang and {Toal{\'a}}, Jes{\'u}s A. and {Ueta}, Toshiya and {Van de Steene}, Griet and {Van Winckel}, Hans and {Ventura}, Paolo and {Villaver}, Eva and {Vlemmings}, Wouter and {Walsh}, Jeremy R. and {Wesson}, Roger and {Zijlstra}, Albert A.},
        title = "{The messy death of a multiple star system and the resulting planetary nebula as observed by JWST}",
      journal = {Nature Astronomy},
         year = 2022,
        month = dec,
       volume = {6},
        pages = {1421-1432},
          doi = {10.1038/s41550-022-01845-2}
}

@ARTICLE{denHartogh2023,
       author = {{den Hartogh}, J.~W. and {Yag{\"u}e L{\'o}pez}, A. and {Cseh}, B. and {Pignatari}, M. and {Vil{\'a}gos}, B. and {Roriz}, M.~P. and {Pereira}, C.~B. and {Drake}, N.~A. and {Junqueira}, S. and {Lugaro}, M.},
        title = "{Barium stars as tracers of s-process nucleosynthesis in AGB stars. II. Using machine learning techniques on 169 stars}",
      journal = {A\&A},
         year = 2023,
        month = apr,
       volume = {672},
          eid = {A143},
        pages = {A143},
          doi = {10.1051/0004-6361/202244189}
}

@ARTICLE{Doherty2014,
       author = {{Doherty}, Carolyn L. and {Gil-Pons}, Pilar and {Lau}, Herbert H.~B. and {Lattanzio}, John C. and {Siess}, Lionel},
        title = "{Super and massive AGB stars - II. Nucleosynthesis and yields - Z = 0.02, 0.008 and 0.004}",
      journal = {MNRAS},
         year = 2014,
        month = jan,
       volume = {437},
       number = {1},
        pages = {195-214},
          doi = {10.1093/mnras/stt1877},
archivePrefix = {arXiv},
       eprint = {1310.2614},
 primaryClass = {astro-ph.SR},
       adsurl = {https://ui.adsabs.harvard.edu/abs/2014MNRAS.437..195D}
}

@ARTICLE{Doherty2015,
       author = {{Doherty}, Carolyn L. and {Gil-Pons}, Pilar and {Siess}, Lionel and {Lattanzio}, John C. and {Lau}, Herbert H.~B.},
        title = "{Super- and massive AGB stars - IV. Final fates - initial-to-final mass relation}",
      journal = {MNRAS},
         year = 2015,
        month = jan,
       volume = {446},
       number = {3},
        pages = {2599-2612},
          doi = {10.1093/mnras/stu2180},
archivePrefix = {arXiv},
       eprint = {1410.5431},
 primaryClass = {astro-ph.SR},
       adsurl = {https://ui.adsabs.harvard.edu/abs/2015MNRAS.446.2599D}
}

@ARTICLE{Dubay2024,
       author = {{Dubay}, Liam O. and {Johnson}, Jennifer A. and {Johnson}, James W.},
        title = "{Galactic Chemical Evolution Models Favor an Extended Type Ia Supernova Delay-Time Distribution}",
      journal = {arXiv e-prints},
         year = 2024,
        month = apr,
          eid = {arXiv:2404.08059},
        pages = {arXiv:2404.08059},
          doi = {10.48550/arXiv.2404.08059}
}

@ARTICLE{Elia2022,
       author = {{Elia}, D. and {Molinari}, S. and {Schisano}, E. and {Soler}, J.~D. and {Merello}, M. and {Russeil}, D. and {Veneziani}, M. and {Zavagno}, A. and {Noriega-Crespo}, A. and {Olmi}, L. and {Benedettini}, M. and {Hennebelle}, P. and {Klessen}, R.~S. and {Leurini}, S. and {Paladini}, R. and {Pezzuto}, S. and {Traficante}, A. and {Eden}, D.~J. and {Martin}, P.~G. and {Sormani}, M. and {Coletta}, A. and {Colman}, T. and {Plume}, R. and {Maruccia}, Y. and {Mininni}, C. and {Liu}, S.~J.},
        title = "{The Star Formation Rate of the Milky Way as Seen by Herschel}",
      journal = {ApJ},
         year = 2022,
        month = dec,
       volume = {941},
       number = {2},
          eid = {162},
        pages = {162},
          doi = {10.3847/1538-4357/aca27d}
}

@ARTICLE{Eggleton1983,
       author = {{Eggleton}, P.~P.},
        title = "{Aproximations to the radii of Roche lobes.}",
      journal = {ApJ},
         year = 1983,
        month = may,
       volume = {268},
        pages = {368-369},
          doi = {10.1086/160960},
       adsurl = {https://ui.adsabs.harvard.edu/abs/1983ApJ...268..368E}
}

@ARTICLE{Fan2024,
       author = {{Fan}, Yi-Zhong and {Han}, Ming-Zhe and {Jiang}, Jin-Liang and {Shao}, Dong-Sheng and {Tang}, Shao-Peng},
        title = "{Maximum gravitational mass M$_{TOV}$=2.2 5$_{-0.07}$$^{+0.08}$M$_{{\ensuremath{\odot}}}$ inferred at about 3\% precision with multimessenger data of neutron stars}",
      journal = {PhysRevD},
         year = 2024,
        month = feb,
       volume = {109},
       number = {4},
          eid = {043052},
        pages = {043052},
          doi = {10.1103/PhysRevD.109.043052}
}

@ARTICLE{Farmer2023,
       author = {{Farmer}, R. and {Laplace}, E. and {Ma}, Jing-ze and {de Mink}, S.~E. and {Justham}, S.},
        title = "{Nucleosynthesis of Binary-stripped Stars}",
      journal = {ApJ},
         year = 2023,
        month = may,
       volume = {948},
       number = {2},
          eid = {111},
        pages = {111},
          doi = {10.3847/1538-4357/acc315}
}

@ARTICLE{Fryer1999,
       author = {{Fryer}, Chris L.},
        title = "{Mass Limits For Black Hole Formation}",
      journal = {ApJ},
         year = 1999,
        month = sep,
       volume = {522},
       number = {1},
        pages = {413-418},
          doi = {10.1086/307647}
}

@ARTICLE{Gallino1998,
       author = {{Gallino}, Roberto and {Arlandini}, Claudio and {Busso}, Maurizio and {Lugaro}, Maria and {Travaglio}, Claudia and {Straniero}, Oscar and {Chieffi}, Alessandro and {Limongi}, Marco},
        title = "{Evolution and Nucleosynthesis in Low-Mass Asymptotic Giant Branch Stars. II. Neutron Capture and the S-Process}",
      journal = {ApJ},
         year = 1998,
        month = apr,
       volume = {497},
       number = {1},
        pages = {388-403},
          doi = {10.1086/305437},
       adsurl = {https://ui.adsabs.harvard.edu/abs/1998ApJ...497..388G}
}

@ARTICLE{Gehrz1998,
       author = {{Gehrz}, Robert D. and {Truran}, James W. and {Williams}, Robert E. and {Starrfield}, Sumner},
        title = "{Nucleosynthesis in Classical Novae and Its Contribution to the Interstellar Medium}",
      journal = {PASP},
         year = 1998,
        month = jan,
       volume = {110},
       number = {743},
        pages = {3-26},
          doi = {10.1086/316107}
}

@ARTICLE{Goriely2000,
       author = {{Goriely}, S. and {Mowlavi}, N.},
        title = "{Neutron-capture nucleosynthesis in AGB stars}",
      journal = {A\&A},
         year = 2000,
        month = oct,
       volume = {362},
        pages = {599-614},
       adsurl = {https://ui.adsabs.harvard.edu/abs/2000A&A...362..599G}
}

@ARTICLE{Goriely2004,
       author = {{Goriely}, S. and {Siess}, L.},
        title = "{S-process in hot AGB stars: A complex interplay between diffusive mixing and nuclear burning}",
      journal = {A\&A},
         year = 2004,
        month = jul,
       volume = {421},
        pages = {L25-L28},
          doi = {10.1051/0004-6361:20040184}
}

@ARTICLE{Guandalini2013,
       author = {{Guandalini}, R. and {Cristallo}, S.},
        title = "{Luminosities of carbon-rich asymptotic giant branch stars in the Milky Way}",
      journal = {A\&A},
         year = 2013,
        month = jul,
       volume = {555},
          eid = {A120},
        pages = {A120},
          doi = {10.1051/0004-6361/201321225}
}

@ARTICLE{Hegar2023,
       author = {{Heger}, Alexander and {M{\"u}ller}, Bernhard and {Mandel}, Ilya},
        title = "{Black holes as the end state of stellar evolution: Theory and simulations}",
      journal = {arXiv e-prints},
         year = 2023,
        month = apr,
          eid = {arXiv:2304.09350},
        pages = {arXiv:2304.09350},
          doi = {10.48550/arXiv.2304.09350}
}

@ARTICLE{Hendriks2023,
       author = {{Hendriks}, D. and {Izzard}, R.},
        title = "{binary\_c-python: A Python-based stellar population synthesis tool and interface to binary\_c}",
      journal = {The Journal of Open Source Software},
         year = 2023,
        month = may,
       volume = {8},
       number = {85},
          eid = {4642},
        pages = {4642},
          doi = {10.21105/joss.04642}
}

@ARTICLE{Henry2018,
       author = {{Henry}, R.~B.~C. and {Stephenson}, B.~G. and {Miller Bertolami}, M.~M. and {Kwitter}, K.~B. and {Balick}, B.},
        title = "{On the production of He, C, and N by low- and intermediate-mass stars: a comparison of observed and model-predicted planetary nebula abundances}",
      journal = {MNRAS},
         year = 2018,
        month = jan,
       volume = {473},
       number = {1},
        pages = {241-260},
          doi = {10.1093/mnras/stx2286}
}

@ARTICLE{Herwig2005,
       author = {{Herwig}, Falk},
        title = "{Evolution of Asymptotic Giant Branch Stars}",
      journal = {ARA\&A},
         year = 2005,
        month = sep,
       volume = {43},
       number = {1},
        pages = {435-479},
          doi = {10.1146/annurev.astro.43.072103.150600},
       adsurl = {https://ui.adsabs.harvard.edu/abs/2005ARA&A..43..435H}
}

@ARTICLE{Hofner2018,
       author = {{H{\"o}fner}, Susanne and {Olofsson}, Hans},
        title = "{Mass loss of stars on the asymptotic giant branch. Mechanisms, models and measurements}",
      journal = {A\&Ar},
         year = 2018,
        month = jan,
       volume = {26},
       number = {1},
          eid = {1},
        pages = {1},
          doi = {10.1007/s00159-017-0106-5}
}

@ARTICLE{Hurley2002,
       author = {{Hurley}, Jarrod R. and {Tout}, Christopher A. and {Pols}, Onno R.},
        title = "{Evolution of binary stars and the effect of tides on binary populations}",
      journal = {MNRAS},
         year = 2002,
        month = feb,
       volume = {329},
       number = {4},
        pages = {897-928},
          doi = {10.1046/j.1365-8711.2002.05038.x},
archivePrefix = {arXiv},
       eprint = {astro-ph/0201220},
 primaryClass = {astro-ph},
       adsurl = {https://ui.adsabs.harvard.edu/abs/2002MNRAS.329..897H}
}

@ARTICLE{Iben1982,
       author = {{Iben}, I., Jr. and {Renzini}, A.},
        title = "{On the formation of carbon star characteristics and the production of neutron-rich isotopes in asymptotic giant branch stars of small core mass}",
      journal = {ApJL},
         year = 1982,
        month = dec,
       volume = {263},
        pages = {L23-L27},
          doi = {10.1086/183916}
}

@ARTICLE{Iben1991,
       author = {{Iben}, Icko, Jr.},
        title = "{Single and Binary Star Evolution}",
      journal = {ApJs},
         year = 1991,
        month = may,
       volume = {76},
        pages = {55},
          doi = {10.1086/191565},
       adsurl = {https://ui.adsabs.harvard.edu/abs/1991ApJS...76...55I}
}

@ARTICLE{Izzard2003,
       author = {{Izzard}, R.~G. and {Tout}, C~A.},
        title = "{Nucleosynthesis in Binary Populations}",
      journal = {PASA},
         year = 2003,
        month = jan,
       volume = {20},
       number = {4},
        pages = {345-350},
          doi = {10.1071/AS03026}
}

@ARTICLE{Izzard2004,
       author = {{Izzard}, R.~G. and {Tout}, C~A. and {Karakas}, A~I. and {Pols}, O~R.},
        title = "{A new synthetic model for asymptotic giant branch stars}",
      journal = {MNRAS},
         year = 2004,
        month = may,
       volume = {350},
       number = {2},
        pages = {407-426},
          doi = {10.1111/j.1365-2966.2004.07446.x}
}

@ARTICLE{Izzard2004_2,
       author = {{Izzard}, R.~G. and {Tout}, C~A.},
        title = "{A binary origin for low-luminosity carbon stars}",
      journal = {MNRAS},
         year = 2004,
        month = may,
       volume = {350},
       number = {1},
        pages = {L1-L4},
          doi = {10.1111/j.1365-2966.2004.07466.x}
}

@ARTICLE{Izzard2006,
       author = {{Izzard}, R.~G. and {Dray}, L.~M. and {Karakas}, A.~I. and {Lugaro}, M. and {Tout}, C.~A.},
        title = "{Population nucleosynthesis in single and binary stars. I. Model}",
      journal = {A\&A},
         year = 2006,
        month = dec,
       volume = {460},
       number = {2},
        pages = {565-572},
          doi = {10.1051/0004-6361:20066129},
       adsurl = {https://ui.adsabs.harvard.edu/abs/2006A&A...460..565I}
}

@ARTICLE{Izzard2009,
       author = {{Izzard}, R.~G. and {Glebbeek}, E. and {Stancliffe}, R.~J. and {Pols}, O.~R.},
        title = "{Population synthesis of binary carbon-enhanced metal-poor stars}",
      journal = {A\&A},
         year = 2009,
        month = dec,
       volume = {508},
       number = {3},
        pages = {1359-1374},
          doi = {10.1051/0004-6361/200912827},
archivePrefix = {arXiv},
       eprint = {0910.2158},
 primaryClass = {astro-ph.SR},
       adsurl = {https://ui.adsabs.harvard.edu/abs/2009A&A...508.1359I}
}

@ARTICLE{Izzard2010,
       author = {{Izzard}, R.~G. and {Dermine}, T. and {Church}, R.~P.},
        title = "{White-dwarf kicks and implications for barium stars}",
      journal = {A\&A},
         year = 2010,
        month = nov,
       volume = {523},
          eid = {A10},
        pages = {A10},
          doi = {10.1051/0004-6361/201015254}
}

@ARTICLE{Izzard2018,
       author = {{Izzard}, R.~G. and {Preece}, H. and {Jofre}, P. and {Halabi}, G.~M. and {Masseron}, T. and {Tout}, C.~A.},
        title = "{Binary stars in the Galactic thick disc}",
      journal = {MNRAS},
         year = 2018,
        month = jan,
       volume = {473},
       number = {3},
        pages = {2984-2999},
          doi = {10.1093/mnras/stx2355}
}

@ARTICLE{Izzard2023,
       author = {{Izzard}, Robert G. and {Jermyn}, Adam S.},
        title = "{Circumbinary discs for stellar population models}",
      journal = {MNRAS},
         year = 2023,
        month = may,
       volume = {521},
       number = {1},
        pages = {35-50},
          doi = {10.1093/mnras/stac2899}
}

@ARTICLE{Jones2012,
       author = {{Jones}, D. and {Mitchell}, D.~L. and {Lloyd}, M. and {Pollacco}, D. and {O'Brien}, T.~J. and {Meaburn}, J. and {Vaytet}, N.~M.~H.},
        title = "{The morphology and kinematics of the Fine Ring Nebula, planetary nebula Sp 1, and the shaping influence of its binary central star}",
      journal = {MNRAS},
         year = 2012,
        month = mar,
       volume = {420},
       number = {3},
        pages = {2271-2279},
          doi = {10.1111/j.1365-2966.2011.20192.x}
}

@ARTICLE{Jones2017,
       author = {{Jones}, David and {Boffin}, Henri M.~J.},
        title = "{Binary stars as the key to understanding planetary nebulae}",
      journal = {Nature Astronomy},
         year = 2017,
        month = may,
       volume = {1},
          eid = {0117},
        pages = {0117},
          doi = {10.1038/s41550-017-0117}
}

@ARTICLE{Jorissen2019,
       author = {{Jorissen}, A. and {Boffin}, H.~M.~J. and {Karinkuzhi}, D. and {Van Eck}, S. and {Escorza}, A. and {Shetye}, S. and {Van Winckel}, H.},
        title = "{Barium and related stars, and their white-dwarf companions. I. Giant stars}",
      journal = {A\&A},
         year = 2019,
        month = jun,
       volume = {626},
          eid = {A127},
        pages = {A127},
          doi = {10.1051/0004-6361/201834630}
}

@ARTICLE{Kalogera1996,
       author = {{Kalogera}, Vassiliki and {Baym}, Gordon},
        title = "{The Maximum Mass of a Neutron Star}",
      journal = {ApJL},
         year = 1996,
        month = oct,
       volume = {470},
        pages = {L61},
          doi = {10.1086/310296}
}

@ARTICLE{Kamath2015,
       author = {{Kamath}, D. and {Wood}, P.~R. and {Van Winckel}, H.},
        title = "{Optically visible post-AGB stars, post-RGB stars and young stellar objects in the Large Magellanic Cloud}",
      journal = {MNRAS},
         year = 2015,
        month = dec,
       volume = {454},
       number = {2},
        pages = {1468-1502},
          doi = {10.1093/mnras/stv1202}
}

@ARTICLE{Karakas2000,
       author = {{Karakas}, A.~I. and {Tout}, C~A. and {Lattanzio}, J~C.},
        title = "{The eccentricities of the barium stars}",
      journal = {MNRAS},
         year = 2000,
        month = aug,
       volume = {316},
       number = {3},
        pages = {689-698},
          doi = {10.1046/j.1365-8711.2000.03561.x}
}

@ARTICLE{Karakas2002,
       author = {{Karakas}, A.~I. and {Lattanzio}, J.~C. and {Pols}, O.~R.},
        title = "{Parameterising the Third Dredge-up in Asymptotic Giant Branch Stars}",
      journal = {PASA},
         year = 2002,
        month = jan,
       volume = {19},
       number = {4},
        pages = {515-526},
          doi = {10.1071/AS02013},
archivePrefix = {arXiv},
       eprint = {astro-ph/0210058},
 primaryClass = {astro-ph},
       adsurl = {https://ui.adsabs.harvard.edu/abs/2002PASA...19..515K}
}

@ARTICLE{Karakas2010,
       author = {{Karakas}, A.~I.},
        title = "{Updated stellar yields from asymptotic giant branch models}",
      journal = {MNRAS},
         year = 2010,
        month = apr,
       volume = {403},
       number = {3},
        pages = {1413-1425},
          doi = {10.1111/j.1365-2966.2009.16198.x},
archivePrefix = {arXiv},
       eprint = {0912.2142},
 primaryClass = {astro-ph.SR},
       adsurl = {https://ui.adsabs.harvard.edu/abs/2010MNRAS.403.1413K}
}

@ARTICLE{Karakas2014,
       author = {{Karakas}, A.~I. and {Lattanzio}, J~C.},
        title = "{The Dawes Review 2: Nucleosynthesis and Stellar Yields of Low- and Intermediate-Mass Single Stars}",
      journal = {PASA},
         year = 2014,
        month = jul,
       volume = {31},
          eid = {e030},
        pages = {e030},
          doi = {10.1017/pasa.2014.21},
archivePrefix = {arXiv},
       eprint = {1405.0062},
 primaryClass = {astro-ph.SR},
       adsurl = {https://ui.adsabs.harvard.edu/abs/2014PASA...31...30K}
}

@ARTICLE{Karakas2014_2,
       author = {{Karakas}, A.~I.},
        title = "{Helium enrichment and carbon-star production in metal-rich populations}",
      journal = {MNRAS},
         year = 2014,
        month = nov,
       volume = {445},
       number = {1},
        pages = {347-358},
          doi = {10.1093/mnras/stu1727}
}

@ARTICLE{Karakas2015,
       author = {{Karakas}, A.~I. and {Ruiter}, A~J. and {Hampel}, M.},
        title = "{R Coronae Borealis Stars Are Viable Factories of Pre-solar Grains}",
      journal = {ApJ},
          year = 2015,
        month = aug,
       volume = {809},
       number = {2},
          eid = {184},
        pages = {184},
          doi = {10.1088/0004-637X/809/2/184}
}

@ARTICLE{Karakas2016,
       author = {{Karakas}, A.~I. and {Lugaro}, M.},
        title = "{Stellar Yields from Metal-rich Asymptotic Giant Branch Models}",
      journal = {ApJ},
         year = 2016,
        month = jul,
       volume = {825},
       number = {1},
          eid = {26},
        pages = {26},
          doi = {10.3847/0004-637X/825/1/26},
archivePrefix = {arXiv},
       eprint = {1604.02178},
 primaryClass = {astro-ph.SR},
       adsurl = {https://ui.adsabs.harvard.edu/abs/2016ApJ...825...26K}
}

@ARTICLE{Kemp2024,
       author = {{Kemp}, Alex J. and {Karakas}, Amanda I. and {Casey}, Andrew R. and {Cote}, Benoit and {Izzard}, Robert G. and {Osborn}, Zara},
        title = "{Nova contributions to the chemical evolution of the Milky Way}",
      journal = {arXiv e-prints},
         year = 2024,
        month = jul,
          eid = {arXiv:2407.18718},
        pages = {arXiv:2407.18718},
          doi = {10.48550/arXiv.2407.18718}
}

@ARTICLE{Kobayashi2020,
       author = {{Kobayashi}, Chiaki and {Karakas}, Amanda I. and {Lugaro}, Maria},
        title = "{The Origin of Elements from Carbon to Uranium}",
      journal = {ApJ},
         year = 2020,
        month = sep,
       volume = {900},
       number = {2},
          eid = {179},
        pages = {179},
          doi = {10.3847/1538-4357/abae65}
}

@ARTICLE{Kroupa2001,
       author = {{Kroupa}, Pavel},
        title = "{On the variation of the initial mass function}",
      journal = {MNRAS},
         year = 2001,
        month = apr,
       volume = {322},
       number = {2},
        pages = {231-246},
          doi = {10.1046/j.1365-8711.2001.04022.x}
}

@ARTICLE{Limongi2018,
       author = {{Limongi}, Marco and {Chieffi}, Alessandro},
        title = "{Presupernova Evolution and Explosive Nucleosynthesis of Rotating Massive Stars in the Metallicity Range -3 {\ensuremath{\leq}} [Fe/H] {\ensuremath{\leq}} 0}",
      journal = {ApJs},
         year = 2018,
        month = jul,
       volume = {237},
       number = {1},
          eid = {13},
        pages = {13}
}

@ARTICLE{Lodders2003,
       author = {{Lodders}, Katharina},
        title = "{Solar System Abundances and Condensation Temperatures of the Elements}",
      journal = {ApJ},
         year = 2003,
        month = jul,
       volume = {591},
       number = {2},
        pages = {1220-1247},
          doi = {10.1086/375492}
}

@ARTICLE{Lopez2022,
       author = {{Yag{\"u}e L{\'o}pez}, A. and {Garc{\'i}a-Hern{\'a}ndez}, D.~A. and {Ventura}, P. and {Doherty}, C.~L. and {den Hartogh}, J.~W. and {Jones}, S.~W. and {Lugaro}, M.},
        title = "{First models of the s process in AGB stars of solar metallicity for the stellar evolutionary code ATON with a novel stable explicit numerical solver}",
      journal = {A\&A},
         year = 2022,
        month = jan,
       volume = {657},
          eid = {A28},
        pages = {A28},
          doi = {10.1051/0004-6361/202039318}
}

@ARTICLE{Lugaro2012,
       author = {{Lugaro}, Maria and {Karakas}, Amanda I. and {Stancliffe}, Richard J. and {Rijs}, Carlos},
        title = "{The s-process in Asymptotic Giant Branch Stars of Low Metallicity and the Composition of Carbon-enhanced Metal-poor Stars}",
      journal = {ApJ},
         year = 2012,
        month = mar,
       volume = {747},
       number = {1},
          eid = {2},
        pages = {2},
          doi = {10.1088/0004-637X/747/1/2}
}

@ARTICLE{Mamajek2015,
       author = {{Mamajek}, E.~E. and {Torres}, G. and {Prsa}, A. and {Harmanec}, P. and {Asplund}, M. and {Bennett}, P.~D. and {Capitaine}, N. and {Christensen-Dalsgaard}, J. and {Depagne}, E. and {Folkner}, W.~M. and {Haberreiter}, M. and {Hekker}, S. and {Hilton}, J.~L. and {Kostov}, V. and {Kurtz}, D.~W. and {Laskar}, J. and {Mason}, B.~D. and {Milone}, E.~F. and {Montgomery}, M.~M. and {Richards}, M.~T. and {Schou}, J. and {Stewart}, S.~G.},
        title = "{IAU 2015 Resolution B2 on Recommended Zero Points for the Absolute and Apparent Bolometric Magnitude Scales}",
      journal = {arXiv e-prints},
         year = 2015,
        month = oct,
          eid = {arXiv:1510.06262},
        pages = {arXiv:1510.06262},
          doi = {10.48550/arXiv.1510.06262},
archivePrefix = {arXiv},
       eprint = {1510.06262},
 primaryClass = {astro-ph.SR}
}

@ARTICLE{Matteucci1986,
       author = {{Matteucci}, F. and {Greggio}, L.},
        title = "{Relative roles of type I and II supernovae in the chemical enrichment of the interstellar gas}",
      journal = {A\&A},
         year = 1986,
        month = jan,
       volume = {154},
       number = {1-2},
        pages = {279-287},
       adsurl = {https://ui.adsabs.harvard.edu/abs/1986A&A...154..279M}
}

@ARTICLE{Marigo1996,
       author = {{Marigo}, P. and {Bressan}, A. and {Chiosi}, C.},
        title = "{The TP-AGB phase: a new model.}",
      journal = {A\&A},
         year = 1996,
        month = sep,
       volume = {313},
        pages = {545-564},
       adsurl = {https://ui.adsabs.harvard.edu/abs/1996A&A...313..545M}
}

@ARTICLE{Marigo1999,
       author = {{Marigo}, Paola and {Girardi}, L{\'e}o and {Bressan}, Alessandro},
        title = "{The third dredge-up and the carbon star luminosity functions in the Magellanic Clouds}",
      journal = {A\&A},
         year = 1999,
        month = apr,
       volume = {344},
        pages = {123-142},
          doi = {10.48550/arXiv.astro-ph/9901235}
}

@ARTICLE{Marigo2001,
       author = {{Marigo}, P.},
        title = "{Chemical yields from low- and intermediate-mass stars: Model predictions and basic observational constraints}",
      journal = {A\&A},
         year = 2001,
        month = apr,
       volume = {370},
        pages = {194-217},
          doi = {10.1051/0004-6361:20000247}
}

@ARTICLE{McClure1983,
       author = {{McClure}, R.~D.},
        title = "{The binary nature of the barium stars. II. Velocities, binary frequency, and preliminary orbits.}",
      journal = {ApJ},
         year = 1983,
        month = may,
       volume = {268},
        pages = {264-273},
          doi = {10.1086/160951}
}

@ARTICLE{Mirouh2023,
       author = {{Mirouh}, Giovanni M. and {Hendriks}, David D. and {Dykes}, Sophie and {Moe}, Maxwell and {Izzard}, Robert G.},
        title = "{Detailed equilibrium and dynamical tides: impact on circularization and synchronization in open clusters}",
      journal = {MNRAS},
         year = 2023,
        month = sep,
       volume = {524},
       number = {3},
        pages = {3978-3999},
          doi = {10.1093/mnras/stad2048}
}

@ARTICLE{Moe2017,
       author = {{Moe}, Maxwell and {Di Stefano}, Rosanne},
        title = "{Mind Your Ps and Qs: The Interrelation between Period (P) and Mass-ratio (Q) Distributions of Binary Stars}",
      journal = {ApJs},
         year = 2017,
        month = jun,
       volume = {230},
       number = {2},
          eid = {15},
        pages = {15},
          doi = {10.3847/1538-4365/aa6fb6},
archivePrefix = {arXiv},
       eprint = {1606.05347},
 primaryClass = {astro-ph.SR},
       adsurl = {https://ui.adsabs.harvard.edu/abs/2017ApJS..230...15M}
}

@INPROCEEDINGS{Mohamed2007,
       author = {{Mohamed}, S. and {Podsiadlowski}, Ph.},
        title = "{Wind Roche-Lobe Overflow: a New Mass-Transfer Mode for Wide Binaries}",
    booktitle = {15th European Workshop on White Dwarfs},
         year = 2007,
       editor = {{Napiwotzki}, R. and {Burleigh}, M.~R.},
       series = {Astronomical Society of the Pacific Conference Series},
       volume = {372},
        month = sep,
        pages = {397},
       adsurl = {https://ui.adsabs.harvard.edu/abs/2007ASPC..372..397M}
}

@article{Osborn2023,
    author = {{Osborn}, Zara and {Karakas}, Amanda I and {Kemp}, Alex J and {Izzard}, Robert G},
    title = "{Aluminium-26 production in low- and intermediate-mass binary systems}",
    journal = {MNRAS},
    pages = {stad3174},
    year = {2023},
    month = {10},
    issn = {0035-8711},
    doi = {10.1093/mnras/stad3174}
}

@ARTICLE{Pal2021,
       author = {{Pal}, Tathagata and {Worthey}, G.},
        title = "{The frequency by mass of Galactic carbon stars inferred from Gaia measurements of star cluster members}",
      journal = {MNRAS},
         year = 2021,
        month = sep,
       volume = {506},
       number = {3},
        pages = {3669-3677},
          doi = {10.1093/mnras/stab1967}
}

@ARTICLE{Pottasch2010,
       author = {{Pottasch}, S.~R. and {Bernard-Salas}, J.},
        title = "{Planetary nebulae abundances and stellar evolution II}",
      journal = {A\&A},
         year = 2010,
        month = jul,
       volume = {517},
          eid = {A95},
        pages = {A95},
          doi = {10.1051/0004-6361/201014009}
}

@ARTICLE{Prantzos2020,
       author = {{Prantzos}, N. and {Abia}, C. and {Cristallo}, S. and {Limongi}, M. and {Chieffi}, A.},
        title = "{Chemical evolution with rotating massive star yields II. A new assessment of the solar s- and r-process components}",
      journal = {MNRAS},
         year = 2020,
        month = jan,
       volume = {491},
       number = {2},
        pages = {1832-1850},
          doi = {10.1093/mnras/stz3154}
}

@ARTICLE{Rafikov2016,
       author = {{Rafikov}, Roman R.},
        title = "{On the Eccentricity Excitation in Post-main-sequence Binaries}",
      journal = {ApJ},
         year = 2016,
        month = oct,
       volume = {830},
       number = {1},
          eid = {8},
        pages = {8},
          doi = {10.3847/0004-637X/830/1/8}
}

@ARTICLE{Raghavan2010,
       author = {{Raghavan}, Deepak and {McAlister}, Harold A. and {Henry}, Todd J. and {Latham}, David W. and {Marcy}, Geoffrey W. and {Mason}, Brian D. and {Gies}, Douglas R. and {White}, Russel J. and {ten Brummelaar}, Theo A.},
        title = "{A Survey of Stellar Families: Multiplicity of Solar-type Stars}",
      journal = {ApJs},
         year = 2010,
        month = sep,
       volume = {190},
       number = {1},
        pages = {1-42},
          doi = {10.1088/0067-0049/190/1/1}
}

@ARTICLE{Renda2004,
       author = {{Renda}, Agostino and {Fenner}, Yeshe and {Gibson}, Brad K. and {Karakas}, Amanda I. and {Lattanzio}, John C. and {Campbell}, Simon and {Chieffi}, Alessandro and {Cunha}, Katia and {Smith}, Verne V.},
        title = "{On the origin of fluorine in the Milky Way}",
      journal = {MNRAS},
         year = 2004,
        month = oct,
       volume = {354},
       number = {2},
        pages = {575-580},
          doi = {10.1111/j.1365-2966.2004.08215.x}
}

@ARTICLE{Roriz2024,
       author = {{Roriz}, M.~P. and {Holanda}, N. and {da Concei{\c{c}}{\~a}o}, L.~V. and {Junqueira}, S. and {Drake}, N.~A. and {Sonally}, A. and {Pereira}, C.~B.},
        title = "{High-resolution Spectroscopic Analysis of Four Unevolved Barium Stars}",
      journal = {AJ},
         year = 2024,
        month = apr,
       volume = {167},
       number = {4},
          eid = {184},
        pages = {184},
          doi = {10.3847/1538-3881/ad29f2}
}

@ARTICLE{Sackmann1991,
       author = {{Sackmann}, I. -Juliana and {Boothroyd}, Arnold I.},
        title = "{Mixing Length and Opacity Effects: Deep Convective Envelopes on the Asymptotic Giant Branch}",
      journal = {ApJ},
         year = 1991,
        month = jan,
       volume = {366},
        pages = {529},
          doi = {10.1086/169587}
}

@ARTICLE{Saio2002,
       author = {{Saio}, Hideyuki and {Jeffery}, C. Simon},
        title = "{Merged binary white dwarf evolution: rapidly accreting carbon-oxygen white dwarfs and the progeny of extreme helium stars}",
      journal = {MNRAS},
         year = 2002,
        month = jun,
       volume = {333},
       number = {1},
        pages = {121-132},
          doi = {10.1046/j.1365-8711.2002.05384.x}
}

@ARTICLE{Sana2012,
       author = {{Sana}, H. and {de Mink}, S.~E. and {de Koter}, A. and {Langer}, N. and {Evans}, C.~J. and {Gieles}, M. and {Gosset}, E. and {Izzard}, R.~G. and {Le Bouquin}, J. -B. and {Schneider}, F.~R.~N.},
        title = "{Binary Interaction Dominates the Evolution of Massive Stars}",
      journal = {Science},
         year = 2012,
        month = jul,
       volume = {337},
       number = {6093},
        pages = {444},
          doi = {10.1126/science.1223344}
}

@ARTICLE{Siess2010,
       author = {{Siess}, L.},
        title = "{Evolution of massive AGB stars. III. the thermally pulsing super-AGB phase}",
      journal = {A\&A},
         year = 2010,
        month = mar,
       volume = {512},
          eid = {A10},
        pages = {A10},
          doi = {10.1051/0004-6361/200913556},
       adsurl = {https://ui.adsabs.harvard.edu/abs/2010A&A...512A..10S}
}

@ARTICLE{Straniero1995,
       author = {{Straniero}, O. and {Gallino}, R. and {Busso}, M. and {Chiefei}, A. and {Raiteri}, C.~M. and {Limongi}, M. and {Salaris}, M.},
        title = "{Radiative 13C Burning in Asymptotic Giant Branch Stars and s-Processing}",
      journal = {ApJL},
         year = 1995,
        month = feb,
       volume = {440},
        pages = {L85},
          doi = {10.1086/187767}
}

@ARTICLE{Timmes1995,
       author = {{Timmes}, F.~X. and {Woosley}, S.~E. and {Hartmann}, D.~H. and {Hoffman}, R.~D. and {Weaver}, T.~A. and {Matteucci}, F.},
        title = "{26Al and 60Fe from Supernova Explosions}",
      journal = {ApJ},
         year = 1995,
        month = aug,
       volume = {449},
        pages = {204},
          doi = {10.1086/176046},
archivePrefix = {arXiv},
       eprint = {astro-ph/9503120},
 primaryClass = {astro-ph},
       adsurl = {https://ui.adsabs.harvard.edu/abs/1995ApJ...449..204T}
}

@ARTICLE{Tisserand2020,
       author = {{Tisserand}, P. and {Clayton}, G.~C. and {Bessell}, M.~S. and {Welch}, D.~L. and {Kamath}, D. and {Wood}, P.~R. and {Wils}, P. and {Wyrzykowski}, {\L}. and {Mr{\'o}z}, P. and {Udalski}, A.},
        title = "{A plethora of new R Coronae Borealis stars discovered from a dedicated spectroscopic follow-up survey}",
      journal = {A\&A},
         year = 2020,
        month = mar,
       volume = {635},
          eid = {A14},
        pages = {A14},
          doi = {10.1051/0004-6361/201834410}
}

@ARTICLE{Vangioni2018,
       author = {{Vangioni}, Elisabeth and {Dvorkin}, Irina and {Olive}, Keith A. and {Dubois}, Yohan and {Molaro}, Paolo and {Petitjean}, Patrick and {Silk}, Joe and {Kimm}, Taysun},
        title = "{Cosmological evolution of the nitrogen abundance}",
      journal = {MNRAS},
         year = 2018,
        month = jun,
       volume = {477},
       number = {1},
        pages = {56-66},
          doi = {10.1093/mnras/sty559}
}

@ARTICLE{Vassiliadis1993,
       author = {{Vassiliadis}, E. and {Wood}, P.~R.},
        title = "{Evolution of Low- and Intermediate-Mass Stars to the End of the Asymptotic Giant Branch with Mass Loss}",
      journal = {ApJ},
         year = 1993,
        month = aug,
       volume = {413},
        pages = {641},
          doi = {10.1086/173033},
       adsurl = {https://ui.adsabs.harvard.edu/abs/1993ApJ...413..641V}
}

@ARTICLE{Ventura2011,
       author = {{Ventura}, P. and {Carini}, R. and {D'Antona}, F.},
        title = "{A deep insight into the Mg-Al-Si nucleosynthesis in massive asymptotic giant branch and super-asymptotic giant branch stars}",
      journal = {MNRAS},
         year = 2011,
        month = aug,
       volume = {415},
       number = {4},
        pages = {3865-3871},
          doi = {10.1111/j.1365-2966.2011.18997.x},
archivePrefix = {arXiv},
       eprint = {1105.0603},
 primaryClass = {astro-ph.SR},
       adsurl = {https://ui.adsabs.harvard.edu/abs/2011MNRAS.415.3865V}
}

@ARTICLE{Ventura2018,
       author = {{Ventura}, P. and {Karakas}, A. and {Dell'Agli}, F. and {Garc{\'i}a-Hern{\'a}ndez}, D.~A. and {Guzman-Ramirez}, L.},
        title = "{Gas and dust from solar metallicity AGB stars}",
      journal = {MNRAS},
         year = 2018,
        month = apr,
       volume = {475},
       number = {2},
        pages = {2282-2305},
          doi = {10.1093/mnras/stx3338}
}

@ARTICLE{Ventura2020,
       author = {{Ventura}, P. and {Dell'Agli}, F. and {Lugaro}, M. and {Romano}, D. and {Tailo}, M. and {Yag{\"u}e}, A.},
        title = "{Gas and dust from metal-rich AGB stars}",
      journal = {A\&A},
         year = 2020,
        month = sep,
       volume = {641},
          eid = {A103},
        pages = {A103},
          doi = {10.1051/0004-6361/202038289}
}

@ARTICLE{Wallerstein1998,
       author = {{Wallerstein}, George and {Knapp}, Gillian R.},
        title = "{Carbon Stars}",
      journal = {ARA\&A},
         year = 1998,
        month = jan,
       volume = {36},
        pages = {369-434},
          doi = {10.1146/annurev.astro.36.1.369}
}

@ARTICLE{Werner2006,
       author = {{Werner}, Klaus and {Herwig}, Falk},
        title = "{The Elemental Abundances in Bare Planetary Nebula Central Stars and the Shell Burning in AGB Stars}",
      journal = {PASP},
         year = 2006,
        month = feb,
       volume = {118},
       number = {840},
        pages = {183-204},
          doi = {10.1086/500443}
}

@ARTICLE{Wesson2018,
       author = {{Wesson}, R. and {Jones}, D. and {Garc{\'\i}a-Rojas}, J. and {Boffin}, H.~M.~J. and {Corradi}, R.~L.~M.},
        title = "{Confirmation of the link between central star binarity and extreme abundance discrepancy factors in planetary nebulae}",
      journal = {MNRAS},
         year = 2018,
        month = nov,
       volume = {480},
       number = {4},
        pages = {4589-4613},
          doi = {10.1093/mnras/sty1871}
}

@ARTICLE{Wyrzykowski2020,
       author = {{Wyrzykowski}, {\L}ukasz and {Mandel}, Ilya},
        title = "{Constraining the masses of microlensing black holes and the mass gap with Gaia DR2}",
      journal = {A\&A},
         year = 2020,
        month = apr,
       volume = {636},
          eid = {A20},
        pages = {A20},
          doi = {10.1051/0004-6361/201935842}
}

@ARTICLE{Yoon2010,
       author = {{Yoon}, S. -C. and {Woosley}, S.~E. and {Langer}, N.},
        title = "{Type Ib/c Supernovae in Binary Systems. I. Evolution and Properties of the Progenitor Stars}",
      journal = {ApJ},
         year = 2010,
        month = dec,
       volume = {725},
       number = {1},
        pages = {940-954},
          doi = {10.1088/0004-637X/725/1/940}
}

@ARTICLE{Zapartas2017,
       author = {{Zapartas}, E. and {de Mink}, S.~E. and {Izzard}, R.~G. and {Yoon}, S. -C. and {Badenes}, C. and {G{\"o}tberg}, Y. and {de Koter}, A. and {Neijssel}, C.~J. and {Renzo}, M. and {Schootemeijer}, A. and {Shrotriya}, T.~S.},
        title = "{Delay-time distribution of core-collapse supernovae with late events resulting from binary interaction}",
      journal = {A\&A},
         year = 2017,
        month = may,
       volume = {601},
          eid = {A29},
        pages = {A29},
          doi = {10.1051/0004-6361/201629685}
}




\end{document}